\documentclass[useAMS]{mn2e}

\usepackage{psfig,epsfig,supertabular,wrapfig,placeins}
\usepackage{graphicx,amsmath,amssymb,subfigure,multirow,units}
\usepackage{marvosym}
\usepackage{graphics}
\usepackage{amssymb}
\usepackage{amsmath}
\usepackage{euscript}
\usepackage{setspace}
\usepackage{fancyhdr}
\usepackage[usenames,dvipsnames]{color}
\usepackage{afterpage,rotating,url}
\usepackage{times}

\newcommand{\fmu}{\mbox{\,$f_\mu$}} 
\newcommand{\Msolar}{\mbox{\,$\rm M_{\odot}$}}        
\newcommand{\Lsolar}{\mbox{\,$\rm L_{\odot}$}}        
\newcommand{\Ldust}{\mbox{\,$\rm L_{dust}$}}        
\newcommand{\Mdust}{\mbox{\,$\rm M_{dust}$}}

\newcommand{\mdld}{\ensuremath{M_{\mathrm{dust}}}\slash\ensuremath{L_{\mathrm{dust}}}}
\newcommand{\mdms}{\ensuremath{M_{\mathrm{dust}}}\slash\ensuremath{M_{\mathrm{star}}}}

\newcommand{\tcold}{\ensuremath{T_{\mathrm{cold}}}}

\newcommand{\tgrey}{\ensuremath{T_{\mathrm{grey}}}}
\newcommand{\ldust}{\ensuremath{L_{\mathrm{dust}}}}
\newcommand{\ldustnopacs}{\ensuremath{L_{\mathrm{dust}}^{\mathrm{no PACS}}}}
\newcommand{\mdust}{\ensuremath{M_{\mathrm{dust}}}}
\newcommand{\mdustnopacs}{\ensuremath{M_{\mathrm{dust}}^{\mathrm{no PACS}}}}
\newcommand{\mstars}{\ensuremath{M_{\mathrm{stars}}}}
\newcommand{\mstar}{\ensuremath{M_{\mathrm{stars}}}}
\newcommand{\mstarsnopacs}{\ensuremath{M_{\mathrm{stars}}^{\mathrm{no PACS}}}}

\newcommand{\hatlas}{{\it H}-ATLAS}
\newcommand{\iras}{{\it IRAS}}
\newcommand{\wise}{{\it WISE}}

\newcommand{\gtsim}{\mbox{{\raisebox{-0.4ex}{$\stackrel{>}{{\scriptstyle\sim}}$}
}}}

\title[{\it Herschel} ATLAS: SEDs]{{\it Herschel}-ATLAS\thanks{{\it
      Herschel} is an ESA space observatory with science instruments
    provided by European-led Principal Investigator consortia and with
    important participation from NASA}: Multi-wavelength SEDs and
  physical properties of 250\,$\mu$m-selected galaxies at $z < 0.5$}

\author[D.J.B.~Smith et al.]{D.J.B. Smith$^{1,2}$\thanks{E-mail:
    daniel.j.b.smith@gmail.com(DS)}, L. Dunne$^{1,26}$, E. da
  Cunha$^{3}$, K. Rowlands$^{1}$, S.J. Maddox$^{1,26}$, \newauthor
  H.L. Gomez$^{4}$, D.G. Bonfield$^2$, S. Charlot$^{5}$,
  S.P. Driver$^{6,7}$, C.C. Popescu$^8$, R.J. Tuffs$^9$, \newauthor
  J.S. Dunlop$^{10}$, M.J. Jarvis$^{2}$, N. Seymour$^{11,12}$,
  M. Symeonidis$^{12}$, M. Baes$^{13}$, N. Bourne$^{1}$,  \newauthor
  D.L. Clements$^{14}$,   A. Cooray$^{15}$,  G. De Zotti$^{16,17}$, 
  S. Dye$^{1}$, S. Eales$^4$, D. Scott$^{18}$, A. Verma$^{19}$,\newauthor
  P. van der Werf$^{20}$, E. Andrae$^{9}$, R. Auld$^{4}$, S. Buttiglione$^{16}$,
  A. Cava$^{21}$, A. Dariush$^{22,4}$,  \newauthor
  J.Fritz$^{13}$, R. Hopwood$^{22}$, E. Ibar$^{23}$,
  R.J. Ivison$^{23,10}$, L. Kelvin$^{6,7}$, B.F. Madore$^{24}$,
  \newauthor M. Pohlen$^{4}$, E.E. Rigby$^{1}$,
  A.Robotham$^{6,7}$, M. Seibert$^{24}$, P. Temi$^{25}$ \\
  $^{1}$School of Physics and Astronomy, University of Nottingham, University Park, Nottingham NG7 2RD, UK \\
  $^2$Centre for Astrophysics, Science \& Technology Research Institute, University of Hertfordshire, Hatfield, Herts, AL10 9AB, UK \\
     $^{3}$Max-Planck Institute for Astronomy, K\"onigstuhl 17, 60115 Heidelberg, Germany \\
     $^{4}$School of Physics and Astronomy, Cardiff University, The Parade, Cardiff, CF24 3AA, UK \\
     $^5$Institut d'Astrophysique de Paris, CNRS, Universit\'e Pierre \&\ Marie Curie, UMR 7095, 98bis bd Arago, 75014 Paris, France \\
     $^6$SUPA, School of Physics and Astronomy, University of St. Andrews,  North Haugh, St. Andrews, KY16 9SS, UK \\
     $^7$International Centre for Radio Astronomy Research, University of Western Australia, 7 Fairway, Crawley, Perth, Western Australia, WA6009, Australia \\
     $^8$ Jeremiah Horrocks Institute, University of Central Lancashire, Preston, PR1 2HE, UK \\
     $^9$Max Planck Institut f\"ur Kernphysik (MPIK), Saupfercheckweg, 69117 Heidelberg, Germany \\ 
  $^{10}$Institute for Astronomy, University of Edinburgh, Royal Observatory, Blackford Hill, Edinbugh, EH9 3HJ, UK\\
  $^{11}$CSIRO Astronomy \& Space Science, PO Box 76, Epping, NSW 1710, Australia\\
  $^{12}$University College London, Department of Space \&\ Climate Physics, Mullard Space Science Laboratory, Holmbury St. Mary, Dorking, Surrey, RH5 6NT, UK\\
  $^{13}$Sterrenkundig Observatorium, Universiteit Gent, Krijgslaan 281 Sg, B-9000 Gent, Belgium \\
  $^{14}$Astrophysics Group, Imperial College London, Blackett Laboratory, Prince Consort Road, London SW7 2AZ, UK\\
  $^{15}$University of California, Irvine, Department of Physics \&\ Astronomy, 4186 Frederick Reines Hall, Irvine, CA 92697-4575, USA\\
  $^{16}$INAF - Osservatorio Astronomico Di Padova, Vicolo Osservatorio 5, I-35122, Padova, Italy \\ 
  $^{17}$SISSA, Via Bonomea 265, I-34136 Trieste, Italy\\
  $^{18}$Department of Physics and Astronomy, 6224 Agricultural Road, University of British Columbia, Vancouver, BC, V6T 1Z1, Canada\\
  $^{19}$Astrophysics, Department of Physics, Denys Wilkinson Building, Keble Road, Oxford, OX1 3RH, UK\\
  $^{20}$Leiden Observatory, Leiden University, PO Box 9513, NL - 2300 RA Leiden, The Netherlands \\
  $^{21}$Departamento de Astrof\'{\i}sica, Facultad de CC. F\'{\i}sicas, Universidad Complutense de Madrid, E-28040 Madrid, Spain \\
  $^{22}$Department of Physics, Imperial College, South Kensington Campus, London, SW7 2AZ, UK\\
  $^{23}$UK Astronomy Technology Centre, Royal Observatory, Edinburgh, EH9 3HJ, UK\\
  $^{24}$Observatories of the Carnegie Institution, 813 Santa Barbera St., Pasadena, CA 91101, USA\\
  $^{25}$Astrophysics Branch, NASA Ames Research Center, Mail Stop 2456, Moffett Field, CA 94035, USA\\
  $^{26}$Department of Physics and Astronomy, University of Canterbury, Private Bag 4800, Christchurch, 8140, New Zealand\\
}
    
\begin{document}

\date{\today}

\pagerange{\pageref{firstpage}--\pageref{lastpage}} \pubyear{2012}

\maketitle

\label{firstpage}
\clearpage

\begin{abstract}
We present a pan-chromatic analysis of an unprecedented sample of 1402
250\,$\mu$m-selected galaxies at $z < 0.5$ ($\bar z = 0.24$) from the
{\it Herschel}-ATLAS survey. We complement our {\em Herschel\/}
100--500\,$\mu$m data with UV--K-band photometry from the Galaxy And
Mass Assembly (GAMA) survey and apply the {\sc magphys} energy-balance
technique to produce pan-chromatic SEDs for a representative sample of
250\,$\mu$m selected galaxies spanning the most recent 5 Gyr of cosmic
history. We derive estimates of physical parameters, including star
formation rates, stellar masses, dust masses and infrared
luminosities. The typical H-ATLAS galaxy at $z<0.5$ has a far-infrared
luminosity in the range $10^{10} - 10^{12}~L_\odot$ (SFR: 1--50
$M_{\odot}~yr^{-1}$) thus is broadly representative of normal star
forming galaxies over this redshift range. We show that 250\,$\mu$m
selected galaxies contain a larger mass of dust at a given infra-red
luminosity or star-formation rate than previous samples selected at
60\,$\mu$m from {\em IRAS\/}. We derive typical SEDs for
\hatlas\ galaxies, and show that the emergent SED shape is most
sensitive to specific star-formation rate. The optical-UV SEDs also
become more reddened due to dust at higher redshifts. Our template
SEDs are significantly cooler than existing infra-red templates. They
may therefore be most appropriate for inferring total IR luminosities
from moderate redshift submillimetre selected samples and for
inclusion in models of the lower redshift submillimetre galaxy
populations.
\end{abstract}


\section{Introduction}

In the past couple of decades, our understanding of the Universe has
flourished as a result of our new-found ability to observe in almost
all regions of the electromagnetic spectrum. This advance is in no
small part due to our ability to associate observations at different
wavelengths with particular astrophysical phenomena and link them
together, making modern astronomy truly pan-chromatic. By observing an
astronomical source at multiple wavelengths, we may piece together its
spectral energy distribution (SED), and by comparing the observed SED
to models, we may infer the physical properties of the source (or
sample of sources) that we are studying.

At ultraviolet, optical and near-infrared wavelengths the SED of the
average galaxy is dominated by emission from stars; there are many
tens of different models to which we may compare our observations, in
the hope of understanding the stellar components of astrophysical
sources (e.g. Bruzual \& Charlot, 2003, Fioc \& Rocca-Volmerange,
1997, V\'azquez \& Leitherer, 2005, Anders \& Fritze-von Alvensleben,
2003, Jimenez et al. 1995, 2004, Maraston, 2005, Pietrinferni et
al. 2004). Such SED model analysis may be used to determine basic
properties of a galaxy's stellar components, such as its age,
metallicity or stellar mass (see e.g. Carter et al., 2009, Smith \&
Jarvis, 2007, Collins et al., 2009, Pacifici et al., 2012, Pforr,
Maraston \&\ Tonini, 2012).

While the ultraviolet to near-infrared emission tells us about the
stellar content of a normal galaxy (subject to correcting for
attenuation by dust of the different stellar components; Charlot
\&\ Fall, 2000, Tuffs et al. 2004, Pierini et al. 2004), the
far-infrared and sub-millimetre wavelengths probe its cool dust
content, which is itself crucial to our understanding of star
formation, since approximately half of the energy ever radiated by
stars has been absorbed by dust and re-radiated at these wavelengths
(e.g. Puget et al., 1996, Fixsen et al., 1998). The sub-millimetre
region has been a difficult part of the electromagnetic spectrum in
which to conduct large galaxy surveys (e.g. Smail et al. 1997, Hughes
et al. 1998, Eales et al. 1999). Previous ground-based sub-millimetre
surveys had to be either pointed at pre-selected targets, or limited
to relatively small regions of sky covering areas of $<1$ deg$^2$
(Coppin et al., 2006, Weiss et al., 2009). The combined effects of the
large negative $k$-correction at these wavelengths, sensitivity and
the steep number counts, have meant that the average 850\,$\mu$m
selected sub-millimetre galaxy is extremely luminous
($10^{12}-10^{13}~L_\odot$) and resides at high redshift ($z \sim 2$,
e.g. Chapman et al. 2005). Few relatively local galaxies have been
found in blind sub-mm surveys, due to the small local volumes probed
in these surveys coupled with the observing wavelength targetting the
faint Rayleigh-Jeans tail of the dust SED at low redshift. Our
undertstanding of the local Universe at sub-mm wavelengths has come so
far from targetted surveys such as the SCUBA Local Universe Galaxy
Survey (SLUGS, Dunne et al. 2000), which observed a sample of 184
\iras- and optically-selected galaxies (Vlahakis, Dunne \& Eales,
2005). Pre-selected galaxies in this way can lead to biases if there
are classes of sub-mm emitting galaxies which are not bright at either
optical or 60\,$mu$m wavelengths. The SLUGS surveys were also limited
to very nearby galaxies and so could not address the question of
evolution of sub-mm properties in the relatively recent past.

With the advent of the PACS (Poglitsch et al. 2010) and SPIRE (Griffin
et al. 2010) instruments aboard the ESA {\it Herschel Space
  Observatory} (Pilbratt et al. 2010), we now have our first
opportunity to survey a large area of sky at sub-mm wavelengths. The
angular resolution and sensitivity of {\it Herschel} allow us to
robustly determine the counterparts to thousands of local
sub-millimetre {\it selected} galaxies across the whole
electromagnetic spectrum, thus gaining invaluable insight into their
physical processes. This paper uses a sample from the {\em Herschel}
Astrophysical TeraHertz Large Area Survey ({\it H}-ATLAS: Eales et
al. 2010) and presents fits to their UV--sub-mm SEDs.  This is the
first relatively local ($z<0.5$) sub-mm selected sample for which such
complete SED modelling has been performed. This work is based on only
3 percent of the final data-set, but is still large enough to provide
a statistical study of the optical and IR properties of $>1000$
250\,$\mu$m selected galaxies, and templates for SEDs which can be
applied more widely.

Studies of the multi-wavelength properties of the relatively small
number of galaxies detected in sub-millimetre surveys have been
extensive (e.g. Swinbank et al. 2009). At high redshifts, galaxy star
formation rates have been frequently estimated based on a single
sub-millimetre flux measurement (e.g. at 850\,$\mu$m), and a local
template SED belonging to e.g. M82 or Arp\,220 (e.g. Silva et al.,
1998), chosen not because they are known to be representative of the
average sub-millimetre galaxy, but rather because they are
comparatively well-studied.

Another commonly-used method of describing far-infrared galaxy SEDs is
to assume one or more components with modified black-body (the
so-called ``grey-body'') profiles. In these simple parametrisations,
the observed flux densities depend only on the temperatures ($T$) and
dust emissivity index ($\beta$), which may be either assumed or
derived, depending on the available observations. Such simple
grey-body profiles have been widely shown to broadly reproduce the
sparsely-sampled far-IR SEDs of galaxies at all redshifts (e.g. Dunne
et al. 2000, Blain et al. 2002, Blain, Barnard \&\ Chapman 2003, Pope
et al. 2006, K\'ovacs et al. 2006, Dye et al., 2010), although when
the SED is sampled from 60$\mu$m to the sub-mm additional greybody
components may be required to reproduce the observations (e.g. Dunne
\&\ Eales, 2001, Galametz et al., 2011, Smith et al. 2010, Dale et
al., 2012).

Empirical templates have been created for use with sparsely-sampled
far-infrared data based on observations of small samples of local
galaxies with good coverage from mid- to far-IR wavelengths
(e.g. Chary \&\ Elbaz, 2001, Dale \&\ Helou, 2002, Rieke et
al. 2009). Selecting galaxies at shorter FIR wavelengths tends to
favour those with substantial warm dust components, which may not be
representative of populations selected at longer wavelengths with {\em
  Herschel\/} and ground based sub-mm instruments. Several studies
have found that sub-mm selected galaxies (so far mostly at higher
redshifts) may have colder dust than their local equivalents at
similar far infrared luminosities (e.g. Pope et al. 2006, Coppin et
al. 2008, Hwang et al., 2010, Magnelli et al., 2012).

In this paper, we use a model that relies on energy balance - the idea
that the energy absorbed by dust at ultra-violet and optical
wavelengths must be re-radiated in the far-infrared - combined with a
statistical fitting approach, to consistently model each galaxy's full
SED, and gain robust constraints on the star formation activity,
stellar and dust content of 250\,$\mu$m selected galaxies from
\hatlas.

In Section \ref{sec:data} we discuss the {\it Herschel}-ATLAS survey
and the multi-wavelength data used in generating the catalogue, while
in Section \ref{sec:method} we discuss the SED-fitting method used in
the analyses which we present in Section \ref{sec:Results}. In Section
\ref{comparisons} we compare our median SEDs with other templates
available, and in Section \ref{conclusions} we present some
conclusions based on our results for the population of sub-millimetre
galaxies in general. Throughout this paper, we use a standard
cosmology with $H_0 = 71$\,km\,s$^{-1}$\,Mpc\,$^{-1}$, $\Omega_M =
0.27$ and $\Omega_\Lambda = 0.73$, and an Initial Mass Function (IMF)
from Chabrier (2003).

\section{Catalogue Construction}
\label{sec:data}

The {\em Herschel\/} ATLAS (Eales et al. 2010) is the widest area
survey being conducted with {\em Herschel\/} covering 570 square
degrees of sky in five FIR--sub-mm bands from 100--500\,$\mu$m. One
primary aim of {\it H}-ATLAS is to provide a census of dust and
obscured star formation in the local Universe, with galaxies selected
on the basis of their dust mass for the first time. Our current study
is based on the {\it Herschel}-ATLAS Science Demonstration Phase data
(SDP) covering $\sim 14$~deg$^2$ centred on the 9hr GAMA field (Driver
et al. 2011). The SPIRE and PACS map-making procedures are described
in Pascale et al. (2011) and Ibar et al. (2010). From these maps a
catalogue of sources which are $\geq 5\sigma$ in any of the three
SPIRE bands was produced using the {\sc MADX} algorithm (Maddox et
al. in prep) and described in detail in Rigby et al., (2011). PACS
sources were added to the catalogue based on the flux in apertures
placed at the locations of SPIRE 250\,$\mu$m sources. The catalogue we
used for this sample is 250\,$\mu$m selected and contains 6621 sources
at $>5\sigma$ (though not all of these are detected in all other bands
-- see below and Table \ref{tab:coverage}). The $5\sigma$ point source
flux limits are 132, 126, 32, 36 \&\ 45 mJy in the 100\,$\mu$m,
160\,$\mu$m, 250\,$\mu$m, 350\,$\mu$m \&\ 500\,$\mu$m bands,
respectively (and including confusion), with beam sizes of
approximately 9, 13, 18, 25 and 35 arcsec {\sc FWHM} in the same five
bands.

\begin{table*}
\caption{Table showing the coverage of the 250\,$\mu$m sources in our
  catalogue, detailing which sources are detected in which
  far-infrared bands, and detailing the number of sources with
  photometry from {\it GALEX}. The sensitivity limits indicate the
  properties of the input catalogues, note that in order for a source
  to have aperture photometry, we require a 5$\sigma$ 250\,$\mu$m
  detection with an $R \ge 0.8$ counterpart from the upper row.}
\centering
  \begin{tabular}{lcccccccccc}
    \hline
    \hline
    \multicolumn{1}{r}{Instrument:} & \multicolumn{3}{c}{{\it Herschel}-SPIRE} & \multicolumn{3}{c}{{\it Herschel}-PACS} & {\it IRAS} &\multicolumn{3}{c}{{\it GALEX}} \\
    \multicolumn{1}{r}{Band:} & 250\,$\mu$m & 350\,$\mu$m & 500\,$\mu$m & 160\,$\mu$m & 100\,$\mu$m & Both & 60\,$\mu$m & FUV & NUV & Both\\ 
    \multicolumn{1}{r}{Catalogue Sensitivity:} & $\ge 5\sigma$ & $>0$ & $>0$ & $\ge 5\sigma$ & $\ge 5\sigma$ & $\ge 5\sigma (\times 2)$ & $\ge 5\sigma$ & $\ge 5\sigma$ & $\ge 5\sigma$ & $\ge 5\sigma (\times 2)$ \\
    \hline
    N(detections) & {\bf 6621} & 5346 & 1717 & 304 & 151 & 117 & 34 & & & \\
    N(Galaxies, $R>0.8$) & {\bf 2417} & 1636 & 344 & 245 & 142 & 111 & 34 & & & \\
    \hline
    Aperture Photometry &  {\bf 1402} & 902 & 170 & 197 & 116 & 93 & 24 & 529 & 726 & 522 \\
    $z_{\mathrm{spec}}$ & 1095 & 710 & 139 & 199 & 118 & 98 & 33 & & & \\
    Aperture Photometry \&\ $z_{\mathrm{spec}}$ & 1052 & 682 & 128 & 186 & 108 & 89 & 24 & 520 & 700 & 513 \\
    \label{tab:coverage}
  \end{tabular}
\end{table*}

A likelihood-ratio analysis (LR -- Sutherland \& Saunders, 1992,
Ciliegi et al., 2005) was performed to identify robust optical
counterparts to the sub-millimetre selected sources, using the SPIRE
250\,$\mu$m channel and SDSS $r$-band positions down to a limiting
magnitude of SDSS $r$ modelmag = 22.4. The method used is described in
detail in Smith et al. (2011), but to summarize; the LR method uses
both positional and photometric information of both individual sources
and of the population in general to quantify the reliability, $R$, (or
equivalently, the probability) of an association between two
sources. Star--galaxy separation was performed following a method
similar to that in Baldry et al. (2010), and the LR calculations were
applied to each population separately, as the 250\,$\mu$m properties
of stars and galaxies are quite different (see Smith et al.,
2011). For this study we have chosen a reliability limit of $R \ge
0.8$ which gives 2417 250\,$\mu$m sources with reliable galaxy
counterparts, and a contamination rate of less than five percent. We
have also removed the five gravitationally lensed SMGs identified in
Negrello et al. (2010) from the subsequent analysis.

These data were combined with the Galaxy And Mass Assembly (GAMA --
Driver et al. 2011) catalogue over the same field (Hill et al. 2011),
which contains thousands of spectroscopic redshifts, in addition to
$r$-band defined aperture-matched photometry for 1402 of the reliable
galaxy counterparts. Since we require well-sampled multi-wavelength
SEDs to constrain the physical properties of these
250\,$\mu$m-selected galaxies, we base our analysis on this sub-sample
of 1402 galaxies, with optical\slash near-infrared aperture-matched
photometry. The $r$-defined aperture-matched photometry is based on
pixel- and seeing-matched images derived from the Sloan Digital Sky
Survey (SDSS - York et al., 2000) and the UK Infrared Deep Sky Survey
(UKIDSS) Large Area Survey (LAS - see e.g. Lawrence et al. 2007) in
the $ugrizYJHK$ bands. Additional photometry in the Galaxy Evolution
Explorer ({\it GALEX}) far- and near-ultraviolet channels has been
included from the {\it GALEX}-GAMA survey (Seibert et al., in prep.),
as have spectroscopic redshifts from the GAMA, SDSS and 6dFGS
surveys. There are 1052 sources with spectroscopic redshifts and
$r$-defined aperture photometry. For the 350 sources with no
spectroscopic redshifts available, we adopt ANN$z$--derived (Collister
\&\ Lahav, 2004) neural network photometric redshifts from Smith et
al. (2011).

All available photometry has been brought on to the AB magnitude
system. Simulations have shown that the photometric errors estimated
by {\sc SExtractor} (Bertin \& Arnouts, 1996) are underestimated by a
factor of four in the GAMA resampled images (Hill et al. 2011); this
factor was applied to the catalogue values. A further 0.1 magnitude
error was added in quadrature to all optical and NIR photometry to
account for the global uncertainties in the total flux measurements
and calibrations between the various surveys.

Additional photometry was compiled for known detections from the {\it
  InfraRed Astronomical Satellite (IRAS)} Faint Source Catalogue
(Moshir et al., 1992, Wang \& Rowan-Robinson, 2009), using a two
arcsecond maximum match radius between cross-identified counterpart
positions, including those sources with updated optical counterparts
based on the higher-quality SPIRE and PACS images discussed in Smith
et al. (2011). As noted in Table \ref{tab:coverage}, there are 34
sources with detections in the \iras\ 60\,$\mu$m band. The errors on
the \iras\ fluxes were assumed to be 20 per cent (including
calibration error) and we included upper limits in the \iras\ 12, 25
and 60$\mu$m channels based on the $5\sigma$ sensitivity limits given
in Wang \& Rowan-Robinson (2009).

To reflect the uncertainty in the SPIRE and PACS photometric
calibration, the SPIRE errors had a factor of 15 per cent added in
quadrature to the catalogue values, and the PACS errors had 10 per
cent and 20 per cent added in quadrature to the errors in the 100 and
160\,$\mu$m bands, respectively (e.g. Poglitsch et al. 2010, Griffin
et al. 2010). Although we only require 250\,$\mu$m fluxes for the
far-infrared selection, we include SPIRE fluxes in our input catalogue
for each SPIRE band, irrespective of the signal to noise ratio in the
350 and 500\,$\mu$m bands, provided that their measured flux is
positive. For PACS, we only include those sources detected at $\ge
5\sigma$ significance in each band. This is due to residual $1 \slash
f$ noise in the current version of the PACS maps which limits the
level at which we can extract reliable photometry at this current
time. We plan to alleviate these problems in future releases.

The number of sources with coverage in each of the far-infrared
photometric bands, as well as the number of sources with
aperture-matched photometry and spectroscopic redshifts are listed in
Table \ref{tab:coverage}. To summarize, our analysis will focus on a
sub-sample of 1402 (i.e. 21 per cent) of the $5\sigma$ 250\,$\mu$m
detections for which well-matched multi-wavelength data are available.

\subsection{The impact of selection effects}
\label{sec:selection}

Since the sample presented in this paper is not selected purely at
250\,$\mu$m, and relies on the identification of an optical
counterpart brighter than 22.4\,mag in the SDSS $r$-band data, it is
necessary to consider the impact that this additional selection
criterion might have on the results of this study.

\begin{figure}
  \centering
  \includegraphics[height=0.90\columnwidth, angle=0]{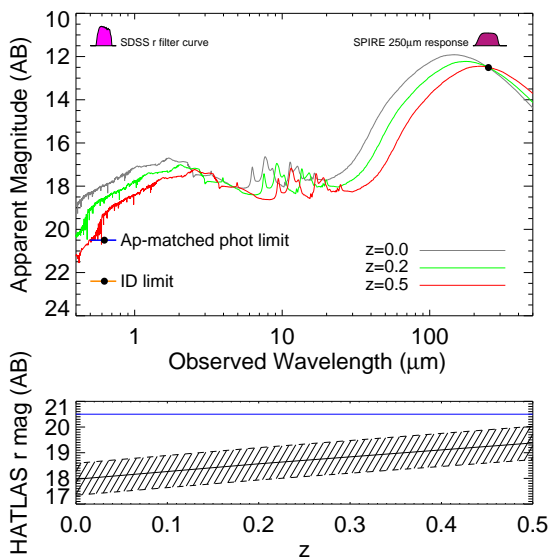}
  \caption{Top: expected magnitude of a median 250\,$\mu$m selected
    galaxy template (discussed in Section \ref{trends}) normalised to
    the \hatlas\ 250\,$\mu$m flux limit at redshifts between $0.0 < z
    < 0.5$. The $r$-band limits for the optical cross-identification
    from Smith et al. (2011) and the GAMA aperture-matched photometry
    (Hill et al., 2011) are shown by the labelled blue and orange
    lines on the left hand side. The response curves of SPIRE at
    250\,$\mu$m and the SDSS r-band data are shown in the top
    corners. Bottom: the variation of the expected brightness of a
    median 250\,$\mu$m selected galaxy template in the SDSS r band as
    a function of redshift. The aperture-matched photometry limit of
    $r_{\mathrm{petro}} = 20.5$ is shown by the blue horizontal line
    (i.e. the same colour as in the upper panel); the median
    250\,$\mu$m selected galaxy is at least a magnitude brighter than
    the $r$-band aperture-matched photometry limit even out to $z \sim
    0.5$. The shaded area corresponds to the range of SEDs in the
    \hatlas\ template SED.}
  \label{sensitivity}
\end{figure}

In the top panel of Figure \ref{sensitivity}, we show the variation in
$r$-band magnitude for a representative template spectral energy
distribution based on the galaxies in our sample (see Section
\ref{trends} for more details of how this SED template was
derived). At each redshift being considered, we fix the 250\,$\mu$m
flux of the template to the 5\,$\sigma$ limit of our survey data
(which corresponds to $\sim$\,12.5\,mag in the AB system), and
convolve the template with the SDSS $r$-band filter curve to determine
the expected $r$-band magnitude that would be observed. We are then
able to compare these values to the limits for the
cross-identification ($r_{\mathrm{model}} = 22.4$\,mag) and for the
GAMA aperture-matched photometry ($r_{\mathrm{petro}} = 20.5$\,mag),
both of which are shown as the black labelled points and coloured
lines in the upper panel. In the bottom panel of Figure
\ref{sensitivity}, we compare the predicted observed $r$-band
magnitude as a function of redshift (solid line, with the surrounding
shaded region corresponding to the uncertainty in the {\it H}-ATLAS
template SED). The GAMA aperture-matched photometry limit is shown as
a horizontal blue line. This shows that typical 250\,$\mu$m-selected
galaxies should still be more than a magnitude brighter than the
aperture-matched photometry limit even out at z = 0.5.

\begin{figure}
  \centering
  \includegraphics[height=0.90\columnwidth, angle=0]{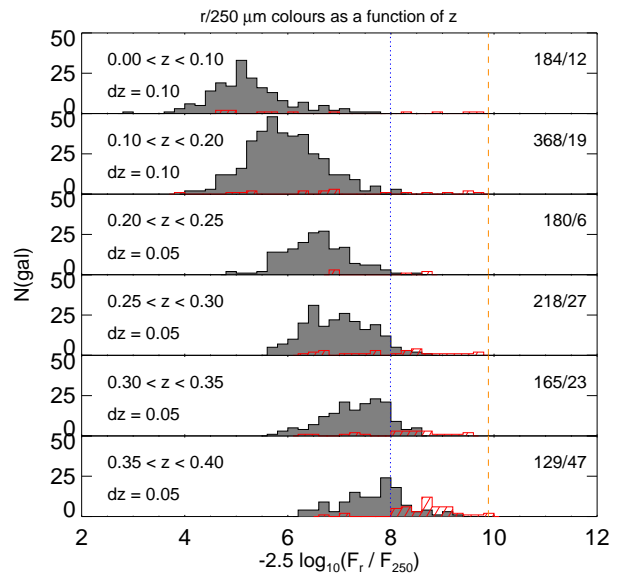}
  \caption{The $(r - $250\,$\mu$m$)$ colours (in AB magnitudes) of the
    galaxies in our sample, for different bins of redshift from $0.0 <
    z < 0.10$ (top), to $0.35 < z < 0.40$ (bottom). The grey
    histograms represent those galaxies for which we have
    aperture-matched photometry from the $u$--$K$ bands from the GAMA
    survey, while the dashed red histograms show the colours of those
    galaxies for which this aperture-matched photometry is not
    available (these are fainter than $r_{\mathrm{petro}} = 20.5$ or
    below $\delta = -1.0^\circ$ declination due to the GAMA magnitude
    limit, and to the peculiarities of the GAMA\slash {\it H}-ATLAS
    survey overlap, respectively). The number of sources falling into
    each of these categories is shown in the top right hand corner of
    each panel. The vertical dashed line indicates the $(r -
    $250\,$\mu$m$)$ colour corresponding to the sensitivity limits of
    our $r$-band and 250\,$\mu$m catalogues, while the vertical dotted
    line indicates the colour defined by the magnitude limit for the
    GAMA aperture-matched photometry relative to the 250\,$\mu$m flux
    limit. It is clear that we sample a representative range of
    250\,$\mu$m selected galaxy colours out to $z \approx 0.35$, but
    beyond this we are only sensitive to the less obscured part of the
    population.}
  \label{selection_colour}
\end{figure}

We may still lose galaxies of certain types from the sample at higher
redshifts due to the optical limit, such as those with higher dust
obscurations (i.e. higher 250\,$\mu$m to $r$-band flux ratios). To
investigate this further, in Figure \ref{selection_colour} we show
histograms of the $(r - $250\,$\mu$m$)$ colour (in AB magnitudes) of
those galaxies with reliable counterparts in our sample, for different
redshifts spanning $0.0 < z < 0.5$. Counterparts with aperture-matched
photometry in the GAMA catalogue (i.e. those galaxies to which we
apply our SED fitting method) are shown as the filled grey histograms,
while galaxies without are hatched red. The orange vertical dashed
line indicates the $(r - $250\,$\mu$m$)$ colour of sources at the
sensitivity limits in both the $r$ band and 250\,$\mu$m data, while
the blue vertical dotted line shows the same colour for sources at the
250\,$\mu$m flux limit and the nominal completeness limit for the GAMA
photometry (20.5\,mag in Petrosian magnitudes). It is clear from
Figure \ref{selection_colour} that our $r$-band selection criterion
does not prevent us from obtaining a representative sample of
250\,$\mu$m sources with a full spread of $(r - $250\,$\mu$m$)$
colours, at least out to $z < 0.35$.  At redshifts higher than $z \sim
0.35$ however, the $r$-band selection criterion does suggest that we
are biased towards the lower obscuration sources (see also Dunne et
al., 2011).

\section{Method}
\label{sec:method}

\subsection{Spectral energy distribution modelling}
\label{modelling}

We use the model of da Cunha, Charlot \& Elbaz (2008 -- hereafter
DCE08\footnote{The da Cunha, Charlot \& Elbaz (2008) models are
  publicly available as a user-friendly model package {\sc magphys}
  at: www.iap.fr/magphys.}) to interpret the panchromatic SEDs of the
galaxies in our {\it Herschel}-ATLAS\slash {\it IRAS}\slash GAMA\slash
{\it GALEX} data set in terms of physical properties related to their
star formation activity and dust content. This physically-motivated
model relies on an energy balance technique to interpret the
(attenuated) stellar emission at ultraviolet, optical and
near-infrared wavelengths consistently with the dust emission at
mid\slash far-infrared and sub-millimetre wavelengths. Therefore this
model is ideal to interpret the multi-wavelength observations
available for this sample of galaxies. Here we briefly summarize the
main features of this model; for more details we refer to the
exhaustive description provided in DCE08.

The dust-free ultraviolet to near-infrared emission from stellar
populations in galaxies is computed using the latest version of the
Bruzual \& Charlot (2003) stellar population synthesis models (Bruzual
2007; Charlot \& Bruzual, in prep.).  The attenuation of starlight by
dust is described by the two-component prescription of Charlot \& Fall
(2000), which also provides the total energy absorbed by dust in the
birth clouds (i.e. molecular clouds where stars form) and in the
ambient (i.e. diffuse) interstellar medium (ISM). The spectral
distribution of the energy re-radiated by dust at infrared and
sub-millimetre wavelengths is then computed by assuming that the
energy re-radiated by dust in the birth clouds and diffuse ISM is
equal to the energy absorbed, and that starlight is the only
significant source of heating (i.e. that there is no AGN
contribution). In stellar birth clouds, the dust emission is described
as a sum of three components: polycyclic aromatic hydrocarbons (PAHs),
hot mid-infrared continuum and warm dust in thermal equilibrium with
temperature in the range 30 -- 60~K. In the ambient ISM, the emission
by dust is described using these three components (whose relative
proportions are fixed for simplicity), plus a component of cold dust
in thermal equilibrium with temperature in the range 15 -- 25~K. The
prior distribution of both the warm and cold dust temperatures is
flat, such that all temperatures between the bounds of the prior have
the same probability.\footnote{Our choice of prior distribution for
  the cold dust temperature is discussed in more detail in Appendix
  \ref{tcold_prior_appendix}} The hot and cold dust components of the
spectrum are assumed to be optically thin, and are described in the
same way as in DCE08 using modified greybody template spectra,
$\propto \kappa_\lambda B_\lambda(T)$, with emissivity index $\beta =
1.5$ and $2.0$ for the warm and cold components, respectively, and
dust mass absorption coefficient approximated as a power law, such
that

\begin{equation}
\kappa_\lambda \propto \lambda^{-\beta},
\end{equation}

\noindent with the normalisation defined such that $\kappa_{850\mu m}
= 0.077\,{\rm kg}^{-1}\,{\rm m}^2$ as in Dunne et al., 2000.

The simplicity and versatility of the DCE08 model make it ideal to
interpret our rich multi-wavelength data set, as it allows us to
derive statistical constraints (including probability density
functions, hereafter PDFs) for several physical properties of the
galaxies (such as star formation rate, stellar mass, dust attenuation,
dust luminosity measured between 3--1000\,$\mu$m, dust temperature and
dust mass), from the consistent modelling of their observed
ultraviolet to sub-millimetre spectral energy distributions. To do so,
we adopt the Bayesian approach used in DCE08 (see also da Cunha et
al. 2010, hereafter dC10).

We use two stochastic libraries of models, as described in DCE08; the
first contains 25,000 stellar population models, including a wide
range of star formation histories, metallicities and dust
attenuations, while the second consists of 50,000 dust emission models
including a large range of dust temperatures and fractional
contributions of PAHs, hot mid-infrared continuum, warm dust and cold
dust to the total infrared luminosity. These two libraries are
combined by associating models with similar values of
$\fmu=L_{\mathrm{dust}}^{\,\mathrm{ISM}}/L_{\mathrm{dust}}^{\,\mathrm{tot}}$
(the fraction of total dust luminosity contributed by the diffuse
ISM), which are scaled to the same total dust luminosity
$L_{\mathrm{dust}}^{\,\mathrm{tot}}$. For each combined model
spectrum, we compute a library of synthetic photometry making use of
the filter transmission curves for the same photometric bands as our
observations, at intervals of $dz = 0.01$. We do not consider redshift
to be a free parameter in our SED fits, and use either the
spectroscopic redshift or the best-fit photometric redshift from Smith
et al. (2011) for these purposes. For ease of reference, some of the
output parameters to which we refer in this analysis are
summarized in Table \ref{tab:parameters}. 

\begin{table*}
\caption{Summary of the SED model parameters used in this paper. For a
  more detailed description of each parameter, see da Cunha, Charlot
  \&\ Elbaz (2008).} \centering
  \begin{tabular}{|l|l}
    \hline
    \hline
    Parameter & Definition \\ 
    \hline
    $f_\mu$ & Fraction of total dust luminosity contributed by the diffuse interstellar medium \\
    sSFR & Specific star formation rate is defined as the star formation rate per unit stellar mass, averaged over the last 0.1\,Gyr, units of yr$^{-1}$ \\
    SFR & Star formation rate averaged over the last 0.1\,Gyr in \Msolar\,yr$^{-1}$ \\
    \Mdust & Dust mass in solar units, M$_\odot$ \\
    \Ldust & Dust luminosity integrated between 3 \&\ 1000\,$\mu$m,  in units of L$_\odot$ \\
    \mstars & Galaxy stellar mass in units of M$_\odot$ \\
    \tcold\ & Temperature of the cold dust in thermal equilibrium in the diffuse ISM
    \label{tab:parameters}
  \end{tabular}
\end{table*}

\subsection{Spectral fits}
\label{fits}

We use the {\sc magphys} model to fit the observed SEDs for the 1402
Herschel-ATLAS sources with robust (i.e. Reliability $> 0.8$)
counterparts and matched-aperture photometry in the GAMA catalogue.
For each galaxy we compared the observed ultraviolet to sub-mm fluxes
to the predicted fluxes of every pair of models in the stochastic
libraries satisfying the energy balance criterion, by computing the
$\chi^2$ goodness-of-fit parameter for each model. This allowed us to
build the PDF of any given physical parameter for the observed galaxy
by weighting the value of that parameter in each model by the
probability $\exp(-\chi^2/2)$. We then determine the median value of
each PDF, corresponding to our best estimate for each parameter.  We
also determine an associated uncertainty, which corresponds to the
16th \&\ 84th percentiles of the PDF. We note that the PDFs generated
in this analysis are marginalised; this is particularly important
since by definition they include parameter uncertainties due to
e.g. the co-variances between parameters in the model. In what
follows, the values of the physical properties of the galaxies
mentioned refer to the median values of the PDF unless explicitly
stated otherwise. We also create 'stacked PDFs' when discussing the
parameter properties of samples of sources, this procedure is
described fully in Appendix A1 but is designed to give our best
estimate of the distribution of parameter values for sources in a
sample convolved with our ability to constrain them. In addition to
the PDFs for each model parameter, we also obtained the best-fit model
SED for each galaxy, which is the model that minimizes $\chi^2$.

It is important to determine whether or not the best-fit model
reasonably reproduces our observed data, which are not uniform across
the sample. Since neighbouring photometric bands are not independent
of one another, we conducted several sets of simulations, designed to
empirically estimate the variation of the number of degrees of freedom
in our spectral fits, as a function of the number of bands of input
photometry available for a particular galaxy. The details of these
simulations are presented in detail in Appendix
\ref{chi2_sims_appendix}. To summarize, we determined a 99 percent
confidence interval on $\chi^2$, which depends solely on the number of
photometric detections for a particular source, such that if the
derived value of $\chi^2$ is higher than the upper bound of the
interval, there is only a $< 1$\,per cent chance that the galaxy is
well described by our model, and is then removed from our sample.

We find that $\sim 92$ per cent of our sample are well-described by
our model. The galaxies with larger $\chi^2$ values than our limit may
have problems with photometry, contributions from AGN components, be
lensed systems, or have catastrophic photometric redshift errors
etc. Inspection of the bad fits ($\sim$8 per cent) reveals that the
vast majority are due to serious problems with the aperture-matched
photometry (e.g. catastrophic failures in multiple bands), while at
least one is a QSO, and there are two possible lensed objects with
far-IR colours not consistent with the redshifts of their $R \ge 0.80$
counterparts, similar to those discussed in Negrello et al. (2010).

We note that there are 320 galaxies in our sample which rely on
photometric redshift estimates, of which 126 lie at $z<0.35$, where
our sample is thought to be representative of the broader 250\,$\mu$m
selected population. Nine per cent of those galaxies relying on
photometric redshifts have $\chi^2$ values outside the range of
acceptable values, as compared with seven per cent of those galaxies
with spectroscopic redshifts, suggesting that sources with
catastrophic photometric redshift errors do not make up the majority
of unreliable SED fits.

Finally, we note that dC10 conducted a series of tests of this
SED-fitting model using a local galaxy sample detected with SDSS and
{\it IRAS}, and demonstrated the general robustness of the technique
to the effects of inclination. They showed that while weak inclination
effects may be present (using the ratio of the apparent major and
minor axes of each particular galaxy as a proxy for inclination), they
do not dominate the dispersion in estimates of galaxy properties.

\section{Results}
\label{sec:Results}

\begin{figure*}
  \centering
  \includegraphics[width=0.98\textwidth]{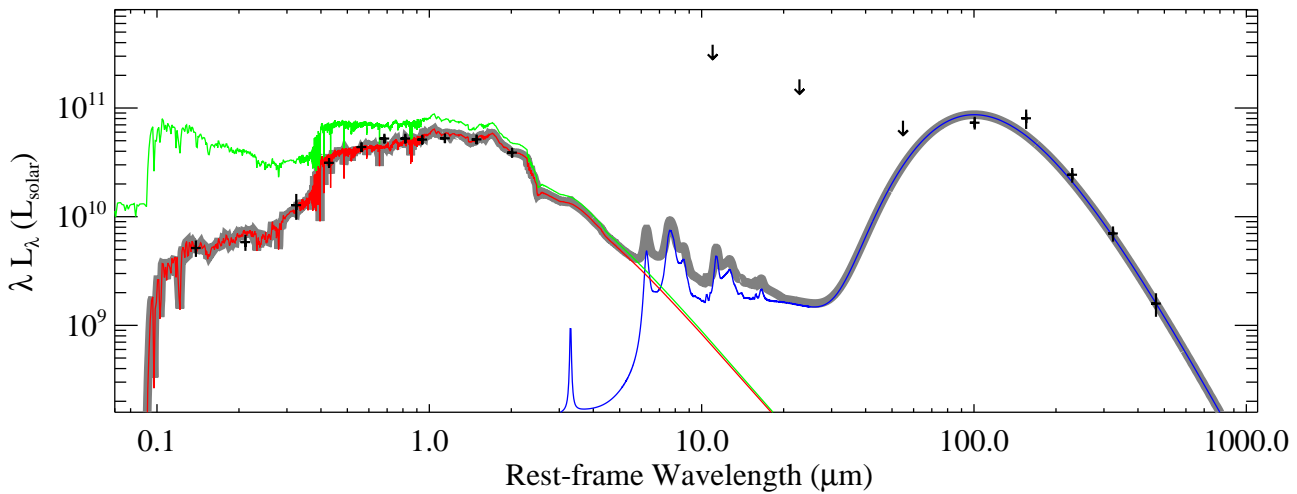}
  \caption{An example best fit SED for one of the galaxies in our
    sample (HATLAS J090713.1-000322). This SED is derived based on our
    compiled data from SPIRE and PACS, plus upper limits at 12, 25
    \&\ 60\,$\mu$m from {\it IRAS}, and aperture-matched photometry
    from $u$-$K$ bands as well as photometry in the {\it GALEX} FUV
    and NUV bands. The data points are shown as the black crosses,
    with the error bars as discussed in Section \ref{sec:data}. The
    upper limits are displayed as down-pointing arrows. The best fit
    total SED is plotted in grey (thick), with the stellar component
    in red and the corresponding dust model in blue. The unattemuated
    (i.e. dust-free) model stellar SED is shown in green. The most
    uncertain area of the SED is the mid-infrared, which is
    constrained only weakly by the {\it IRAS} upper limits, with
    additional constraint coming from the energy balance
    criterion. The energy balance technique of da Cunha et al. (2010)
    enables us to only combine optical\slash near infrared SEDs that
    are physically consistent with the sub-millimetre SED.} 
  \label{SEDresult}
\end{figure*}

\begin{figure*}
  \centering 
  \includegraphics[width=0.9\textwidth]{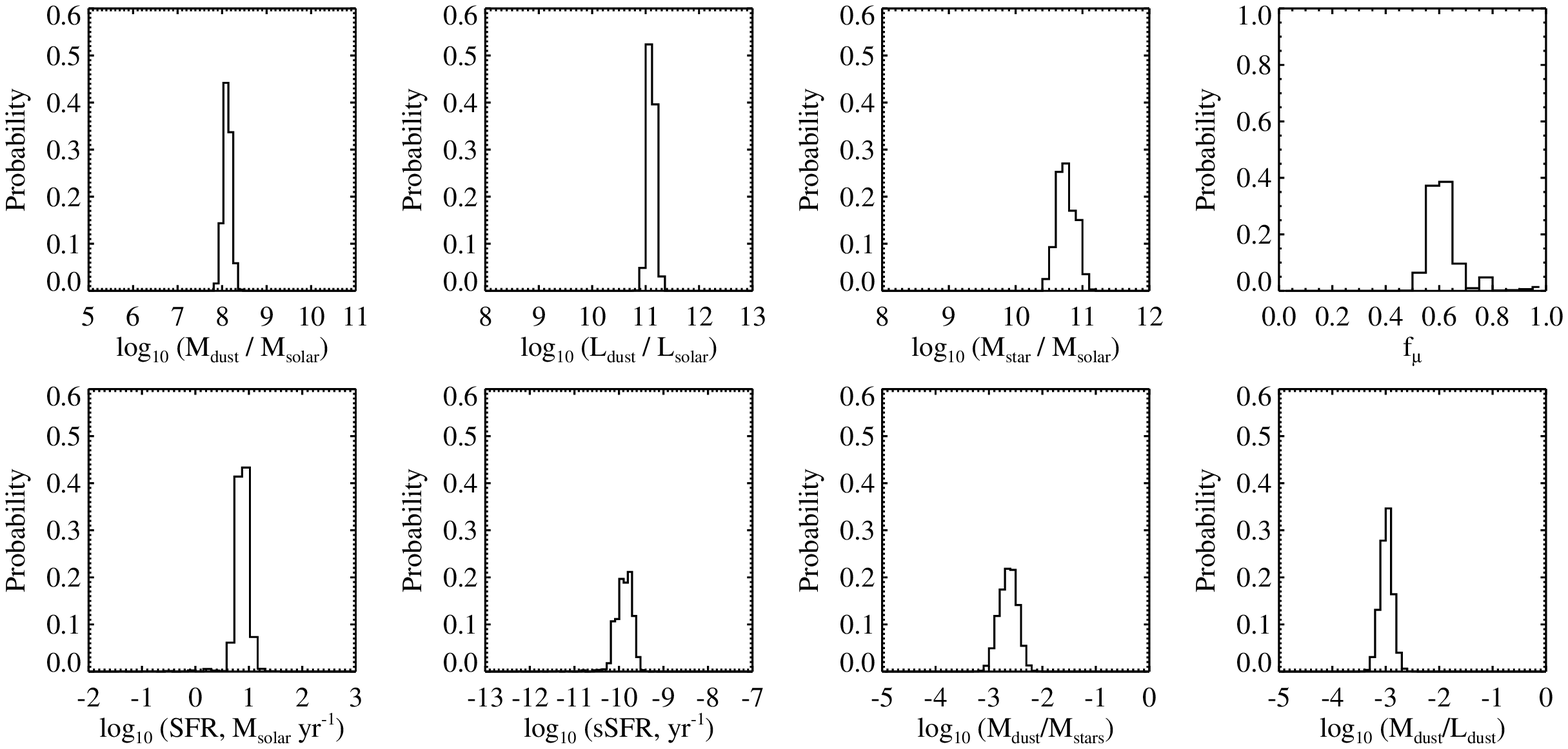}
  \caption{Probability density functions (PDFs) for the galaxy shown
    in figure \ref{SEDresult} (HATLAS J090713.1-000322); the dust
    mass, dust luminosity, stellar mass, \fmu (the fraction of the
    total dust luminosity contributed by the diffuse ISM), the
    [specific] star formation rate, dust to stellar mass ratio, and
    the ratio of dust mass to dust luminosity. Our results suggest
    that this galaxy is a massive luminous infrared galaxy
    (LIRG). Using these PDFs we can estimate not only the median
    values for each parameter, but also the 16th \&\ 84th percentiles,
    corresponding to the confidence range of our estimates.}
  \label{SEDparams}
\end{figure*}

Here we present the results of our SED-fitting analysis. For each
galaxy we not only determine best-fit SEDs (Figure \ref{SEDresult}),
but also the PDFs for each parameter (Figure \ref{SEDparams}). The
parameters that we focus on are those shown in Figure \ref{SEDparams}
(namely \mdust, \ldust, \mstar, $f_\mu$, SFR, sSFR,
\mdust\slash\mstars\ and \mdust\slash\ldust), since we are interested
in investigating the star formation activity and dust mass of normal,
star-forming galaxies. Our sample represents almost an order of
magnitude's improvement upon the largest sub-mm selected samples until
now, even though the {\it H}-ATLAS SDP data comprise only $\sim$3 per
cent of the total eventual {\it H}-ATLAS data set.

\subsection{Importance of the available infrared observations}
\label{biastest}

Given the inhomogeneous set of data which we have compiled for the
{\it H}-ATLAS sample, it is important to have some understanding of
the sensitivity of the derived physical parameters to the
absence\slash presence of data at certain wavelengths. Some tests of
this nature were performed by DCE08 and dC10, and demonstrated that
the results derived were generally robust, although certain parameters
(e.g. sSFR) are better constrained when UV data are included, for
example. Our concerns for this study relate to determining the
reliability of the estimates of dust luminosity, SFR and mass
trends. Given that not all galaxies in the sample have data spanning
the peak of their rest-frame far-infrared SED (from {\it IRAS} or
PACS) we need to assess the impact of this heterogeneity on our
results. We have performed three tests. The first test was to
determine the reliability of our method when only {\it IRAS} data
between 60--100\,$\mu$m are used to constrain the far-infrared SED;
secondly, we attempted to determine the influence of incomplete PACS
data on our results, and finally, we attempted to probe the
reliability of our results in the absence of mid-IR data, which are
not available over the {\it H}-ATLAS SDP field at the time of writing.

\subsubsection{Comparison with {\it IRAS}-selected samples}
\label{IRAStest}

In Section \ref{properties}, we compare the star formation activity
and dust content of \hatlas\ 250\,$\mu$m-selected galaxies with those
of a previous sample of local, star-forming galaxies selected at
60\,$\mu$m with {\it IRAS} (da Cunha et al., 2010). To compare these
two samples, we need to understand possible differences\slash biases
in the derived physical parameters that may arise from the different
selection of the samples. Therefore, in this section, we investigate
the effects of including SPIRE data in the SED fitting for a
sub-sample 250\,$\mu$m selected galaxies for which we also have {\it
  IRAS} data. This allows us to assess whether any difference between
our results and those of dC10 are due solely to the lack of SPIRE data
for that study, or if they are rather due to effects of selection.

\begin{figure*}
  \centering
  \includegraphics[width=0.9\textwidth]{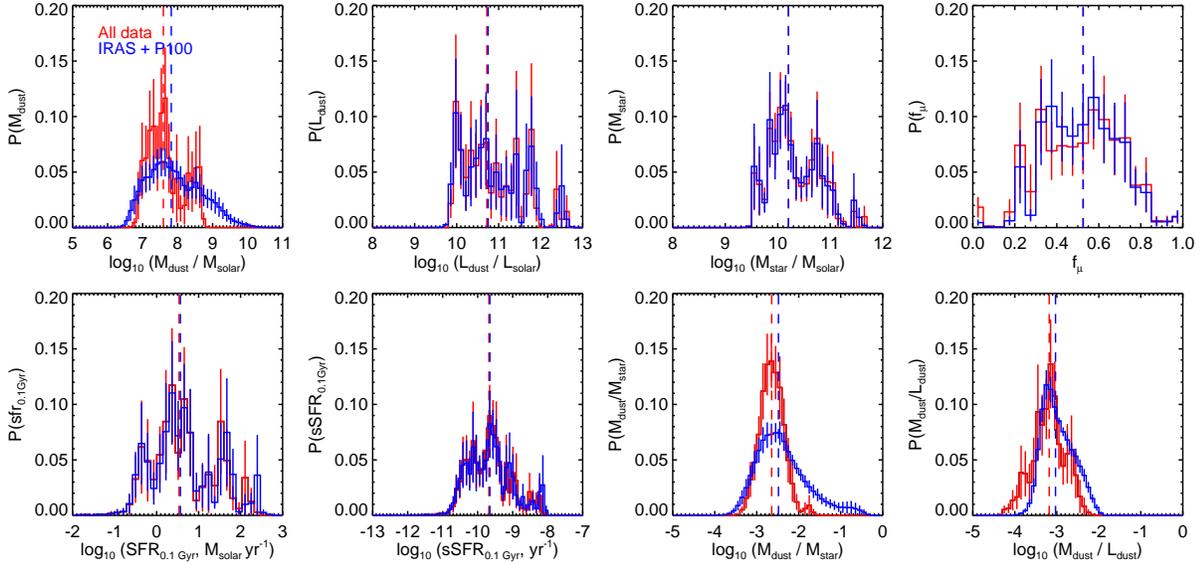}
  \caption{Stacked probability density functions and errors for
    sources in our sample detected by {\it IRAS} at 60\,$\mu$m,
    derived using the full {\it IRAS}, PACS, \&\ SPIRE data set (red
    histograms), and using only the {\it IRAS} 60\,$\mu$m and PACS
    100\,$\mu$m data (blue histograms). The median values of each
    stacked PDF are indicated by the vertical dashed lines. The
    increased uncertainty in dust mass when the $\lambda >100$\,$\mu$m
    data are not included in the fits is apparent in the wider range
    of values in the blue (i.e. no {\it Herschel})
    \mdust\ PDF. Stellar masses, \fmu, dust luminosity and star
    formation rates are much less strongly influenced by the inclusion
    of longer wavelength data i.e. they are well constrained without
    data beyond 100\,$\mu$m. }
  \label{best_checks}
\end{figure*}

We take a sub-sample of the \hatlas\ 250\,$\mu$m-selected galaxies
which are also detected by {\it IRAS} at 60\,$\mu$m, and applied our
fitting procedure twice; once including all available data, and a
second time omitting all data longward of the {\it PACS} 100\,$\mu$m
band. The PACS 100\,$\mu$m data were included for both sets of fitting
to ensure that our results are as comparable as possible with dC10,
since their galaxy sample was bright enough to be well-detected with
{\it IRAS} at 100\,$\mu$m. There are 18 {\it IRAS}-detected galaxies
with aperture--matched $u$- to $K$-band photometry which are
well-described by our models. For these galaxies fitted both with and
without the 160--500\,$\mu$m data, we compare the medians of the
stacked PDFs for a given parameter in each run. We estimate the
uncertainty associated with each bin in a stacked PDF according to the
16th and 84th percentiles of the cumulative frequency distribution of
values in each PDF bin (this method is discussed in greater detail in
Appendix \ref{tcold_prior_appendix}). These stacked PDF comparisons
and their associated uncertainties are displayed in Figure
\ref{best_checks}, with the full data set PDFs shown in red, and the
    {\it IRAS}-only results shown in blue.

In general, removing the $\lambda\geq 160$\,$\mu$m data from the SED
fitting for those sources detected by {\it IRAS} results in small
variations of the median values of the stacked PDFs for each
population (dotted vertical lines in Figure \ref{best_checks}), but
the changes are always less than the confidence interval derived
according to the 16th and 84th percentiles of the stacked PDFs.  The
most noticeable effect of removing the $\lambda \geq 160$\,$\mu$m data
is the considerably larger uncertainty in \mdust; this is not
surprising since the cold dust component dominates the total dust
mass, and constraints on this component come primarily from the longer
wavelength SPIRE data. There is a tendency for the model to
overestimate the dust mass when using only {\it IRAS} data (see the
tail to high \Mdust\ values in Figure \ref{best_checks}), as the lack
of SPIRE data allows the model to add in more cold dust without any
strong constraint from the energy balance (since the contribution to
the overall \Ldust\ varies $\propto T^{4+\beta}$, warmer dust far
outweighs colder dust in its effect on \Ldust). The quantities
\mdust\slash\ldust, and \mdust\slash\mstars\ are much better
constrained when including the longer wavelength data, for the same
reasons discussed for the dust mass estimates above. The other
parameters, \ldust, \mstars, SFR, sSFR and \fmu\ are all comparable in
both samples.

\subsubsection{The effect of lacking PACS observations}
\label{pacscomp}

\begin{figure} 
  \centering
  \includegraphics[width=0.95\columnwidth]{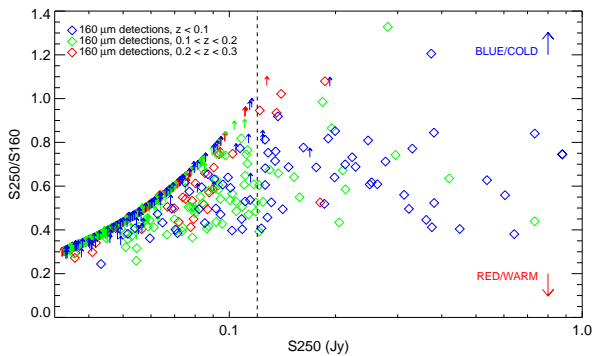}
  \caption{$S_{250}/S_{160}$ colour as a function of 250\,$\mu$m flux
    density, binned in redshift according to the colours as shown in
    the legend. At the fainter 250\,$\mu$m fluxes our PACS data detect
    only the warmer galaxies, as shown by the large number of lower
    limits on the $S_{250}/S_{160}$ colour at these fluxes. At
    $S_{250} > 120$ mJy we sample the full range of colours (and
    therefore temperatures) in our 250\,$\mu$m selected sample, so it
    is these galaxies that we use to study the impact of missing PACS
    data for the majority of our sources on the derived
    parameters. The vertical dashed line shows S250 = 120\,mJy, the
    250\,$\mu$m flux limit for the PACS-complete sub-sample.}
  \label{pacs_completeness}
\end{figure}

Whilst there are 1289 $5\sigma$ SPIRE sources with reliable optical
counterparts in the {\it H}-ATLAS survey and good SED fits, we have
5$\sigma$ PACS detections for only 207 and must rely on upper limits
for the remaining sources. It is important to understand the effect of
missing PACS data on our results, and so we now investigate the impact
on our SED fits when we omit the PACS data for a sub-sample of
PACS-detected galaxies drawn from our main sample. To determine a
representative sample, we show in Figure \ref{pacs_completeness} the
$S_{250}/S_{160}$ colour as a function of 250\,$\mu$m flux, binned in
redshift. The PACS data are not deep enough to probe colder colours at
fainter fluxes or higher redshifts, but do sample the full range of
colours above $S_{250} \approx 120$\,mJy, at least for $z < 0.2$. To
assess the impact of missing PACS data on the results of our SED
fitting, we use these 59 sources with 160\,$\mu$m PACS detections and
$S_{250} \gtsim 120$\,mJy at $z < 0.2$, to check how the SED
parameters change when the PACS data are removed from the
fitting\footnote{the equivalent sample for the PACS 100\,$\mu$m
  channel, which requires $S_{250} \gtsim 200$\,mJy, contains no
  additional sources above those selected here}. We will refer to this
sample as the ``PACS-complete'' sample.

\begin{figure*}
  \centering 
  \includegraphics[width=0.9\textwidth]{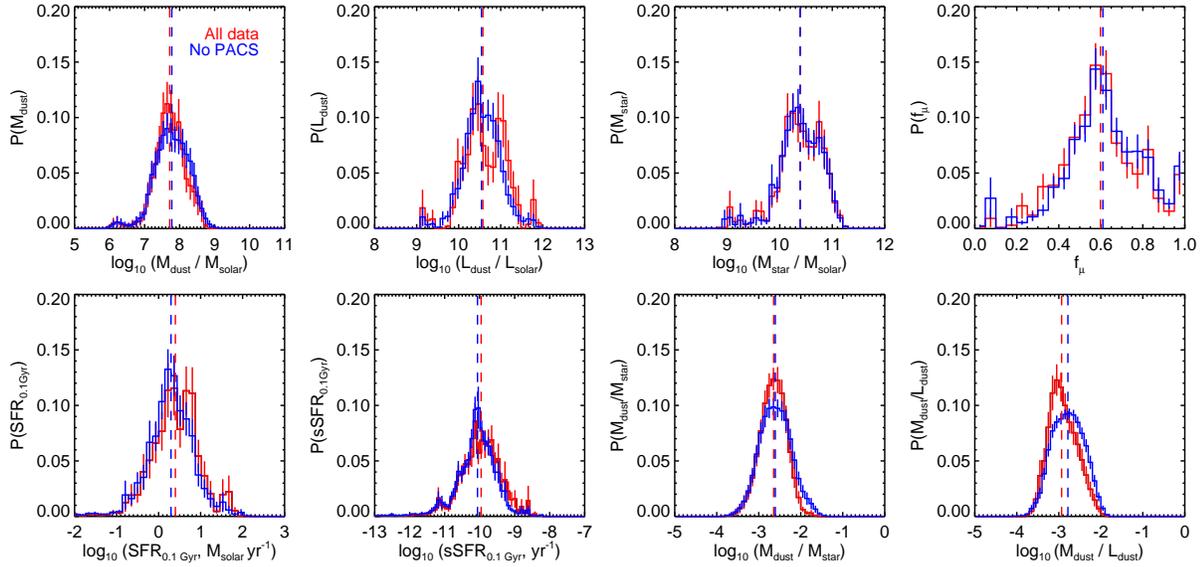}
  \caption{Stacked PDFs comparing the dust temperatures, dust mass and
    luminosity, stellar mass, $f_\mu$, specific star formation rate,
    dust mass to luminosity ratio and dust to stellar mass ratio for
    the representative PACS-complete subsample. The results including
    all available data are shown by the red histograms, while those
    results derived neglecting the PACS data are shown in blue. The
    vertical dotted lines indicate the median values of the stacked
    PDFs. We find that fits lacking PACS data have SSFR underestimated
    by $\sim 0.1$ dex, and \mdust overestimated by $\sim
    0.06$\,dex. Estimates of the stellar mass and $f_\mu$ are largely
    unaffected by omitting the PACS data, while the effects on the
    \mdust\slash\ldust and \mdust\slash\mstar are consistent wth the
    changes to their input parameters.}
  \label{nopacs_figure}
\end{figure*}

In Figure \ref{nopacs_figure}, we show a comparison of the stacked
PDFs when the PACS data are included and excluded from the fitting for
the ``PACS-complete'' sample. This is not quite the same effect which
will apply to the full sample, as there we do have some PACS
information (in the form of upper limits), while in this test we have
no information when the PACS data are removed. This makes this
comparison conservative, in the sense that the effects on the full
sample are going to be no larger than the worst-case scenario we study
here.


Removing the PACS information from the SED fitting causes the average
estimate of the specific star formation rate to decrease by $\sim 0.1$
dex. Estimates of the stellar mass are, unsurprisingly, barely
altered, while $f_\mu$ and \ldust\ are also robust in a sample
average. Dust masses of the population are overestimated by
approximately $\sim 0.06$\,dex. The median offset for each parameter
in the absence of PACS data is tabulated in Table
\ref{tab:nopacs_changes}.

Looking at the changes in globally averaged parameters in this way is
reassuring, however, it is important to check that this is not masking
a potential correlation of a bias in one parameter as a function of
another. For example, we may overestimate \mdust\ at low values of
\ldust\ and underestimate at high values -- i.e. our estimates may be
skewed -- but still have an average offset consistent with zero
bias. As we will next investigate trends of one parameter against
others, and later bin SEDs by parameter, we must consider these
effects now. We focus on these issues in detail in Appendix
\ref{bias_tests}, though to summarise, we find that the lack of PACS
data does not introduce bias in any parameters as a function of
redshift, \mstar, or \fmu. The same is generally true of our estimated
\ldust\ and \mdust, though these parameters possibly show weak
correlations (i.e. they may be skewed). In the absence of PACS data,
\ldust\ may be under-estimated at large \mdust, and over-estimated
toward lower dust masses, though the offset averaged over the whole
population is small. \mdust\ shows weak bias with \mdustnopacs; in
that our PACS-free estimates of \mdust\ are slightly high for large
\mdust, though the correlation shows considerable scatter and the
overall offset across all values is small. Our estimates of SFR and
sSFR show larger scatter than the other parameters, reflected in the
larger error bars on $\Delta (s)SFR$, though it is difficult to
discern any skewed bias in the derived values. We will discuss the
impact of possible bias with these parameters later in this paper.

It is worth noting that because of the good multi-wavelength coverage
and lack of temperature\slash colour bias in the $S_{250} > 120$\,mJy
PACS-complete sample (Figure \ref{pacs_completeness}), we can use it
to determine our best estimate of the dust temperature in low-redshift
250\,$\mu$m-selected galaxies. We determine a median likelihood
estimate of $\tcold = 20.4 \pm 0.4$\,K from the DCE08 model, and an
isothermal value from simple grey-body fitting to the FIR data of
$\tgrey = 26.1 \pm 3.5$\,K (assuming $\beta=1.5$ for comparison with
literature values). These estimates are colder than values in the
literature pre-dating {\it Herschel}; for example, SLUGS found $\tgrey
= 36 \pm 5$\,K, $\beta = 1.3$, for the {\it IRAS}-selected sample
(Dunne et al., 2000) and $\tgrey = 31.6 \pm 0.6$\,K, for the
optically-selected sample (Vlahakis, Dunne \&\ Eales, 2005).  Our
temperatures are comparable to those observed in galaxy samples
selected at longer wavelengths but including data from 24-160\,$\mu$m
using e.g. BLAST ($26 \pm 5$\,K, $\beta = 1.5$, Dye et al., 2009) or
selected in the K-band and observed with {\it Herschel} (e.g. $\tgrey
\approx 20$\,K, $\beta = 2.0$, Boselli et al., 2010, $\tgrey \approx
25.8\,$K for spirals in Skibba et al. 2011, or $\approx 23$\,K, $\beta
\approx 1.5$, Dale et al., 2012). Multiple-component modified
black-body SED fits have long noted the presence of substantial cold
dust components with lower temperatures, consistent with our findings
(e.g. Dunne \& Eales 2001; Contursi et al., 2001; Vlahakis, Dunne
\&\ Eales 2005; Clements, Dunne \&\ Eales, 2011, Rowan-Robinson et
al., 2011; Galametz et al., 2011, Planck Collaboration, 2011).

\begin{table}
\caption{The effects of omitting PACS data from the SED fitting
  results for those sources in the ``PACS-complete'' sample. These
  offsets are also shown in Figure \ref{nopacs_figure}. Note that all
  values are consistent within the uncertainties derived from the
  16$^{\mathrm{th}}$ and 84$^{\mathrm{th}}$ percentiles of the stacked
  PDFs. Furthermore, and as can be seen in figure
  \ref{further_bias_tests} of appendix \ref{bias_tests}, the offsets
  for individual galaxies are almost always consistent with zero once
  the errors are taken in to account.}  \centering
  \begin{tabular}{|l|c}
    \hline
    \hline
    \multirow{2}{*}{Parameter} &  Offset  \\
    & (Best $-$ no PACS) \\
    \hline
    $\log_{10} L_{\mathrm{dust}}$ & 0.03\,dex \\
    $\log_{10} M_{\mathrm{dust}}$ & -0.06\,dex \\
    $\log_{10} M_{\mathrm{stars}}$ & 0.00\,dex \\
    $f_\mu$ & -0.01 \\
    sSFR & -0.10\,dex \\
    SFR & -0.10\,dex \\
    $M_{\mathrm{dust}}$\slash$L_{\mathrm{dust}}$ & -0.15\,dex \\
    $M_{\mathrm{dust}}$\slash$M_{\mathrm{stars}}$ & -0.04\,dex 
    \label{tab:nopacs_changes}
  \end{tabular}
\end{table}

\subsubsection{The effect of missing mid-IR observations}
\label{midirtest}

At the time of writing, mid-infrared data over the {\it H}-ATLAS
fields are unavailable (the first public data release of the
Wide-Field Infrared Survey Explorer -- {\it WISE}, Wright et al., 2010
-- survey does not include the \hatlas\ SDP field), we investigate the
effects of not having such data on the PDFs by applying the same
fitting procedure to a sample of galaxies selected at 250\,$\mu$m in
data from the Balloon-borne Large Aperture Sub-millimetre Telescope
(BLAST -- Devlin et al., 2009). We derive multiwavelength SEDs and
PDFs for a sample of 14 BLAST galaxies in the Extended Chandra Deep
Field South (ECDFS, Lehmer et al., 2005) which have $\geq$5 $\sigma$
detections in all three BLAST bands (250, 300, 500\,$\mu$m, Dye et
al., 2009), and spectroscopic redshifts from Eales et al. (2009), with
additional photometry in the {\it GALEX} (FUV and NUV -- Morrissey et
al., 2007), optical ($ugriz$), 2MASS $J$ and $K_s$ (Skrutskie et al.,
2006), MIR (3.6-8.0, 24, 70\,$\mu$m), as well as far-infrared
(160\,$\mu$m) bands from the \emph{Spitzer Space Telescope} (Lonsdale
et al., 2003).

\begin{figure*}
  \centering
  \includegraphics[width=0.9\textwidth]{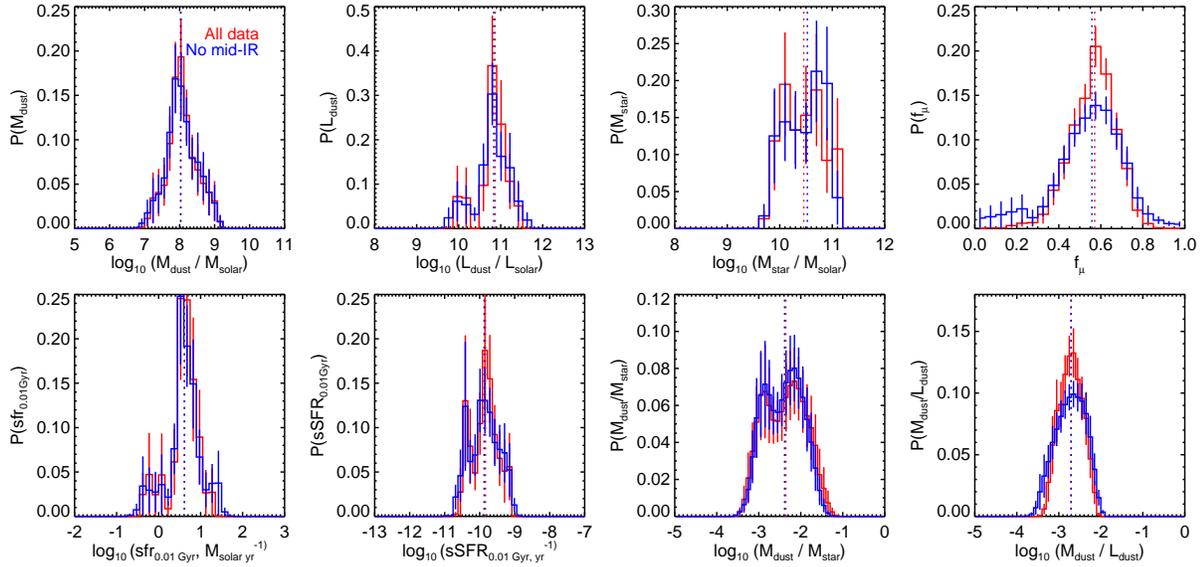}
  \caption{A comparison between the stacked PDFs derived by fitting 14
    galaxies selected from the {\it BLAST} survey of the E-CDFS. In
    red we show histograms of the results obtained when the
    mid-infrared data (i.e. those at 3.6-8.0, 24 and 70\,$\mu$m) are
    included in the fitting, whilst the blue histograms show the
    results in the absence of these data. The similarity of these two
    sets of PDFs indicates that our estimates of these parameters are
    not biased by the absence of mid-infrared coverage in our data
    set, and consequently that the absence of such data does not limit
    the validity of our results. Once more, the median values of the
    PDFs are indicated by the vertical dotted lines.}
  \label{midir_test_figure}
\end{figure*}

In Figure \ref{midir_test_figure} we show the stacked probability
density functions for the same parameters as in figure
\ref{nopacs_figure} derived for these {\it BLAST} galaxies. The red
histograms show the PDFs determined when we include the complete
data-set, while the blue histograms show the PDFs derived in the
absence of the mid-infrared data. The similarity between these two
sets of histograms, and the absence of bias between them, suggests
that our estimates of the dust mass, luminosity and SFR are robust to
the absence of mid-infrared data in our wider {\it H}-ATLAS data set,
although these tests have necessarily only been applied for a small
number of sources. The mid-IR accounts for only a small fraction of
the total infrared emission, and the similarity of the PDFs highlights
the power of the energy balance criterion in constraining the dust
luminosity even in the absence of mid-infrared data. The detailed
shape of the SED in the mid-IR is clearly not well defined for our
sample and constraints on this can only come from comparison with
mid-IR data, e.g. from WISE.

\subsection{The properties of Sub-millimetre selected galaxies in {\it H}-ATLAS}
\label{properties}

\subsubsection{First results, and comparison with previous studies}

\begin{figure}
  \centering
  \includegraphics[height=0.70\columnwidth, angle=0]{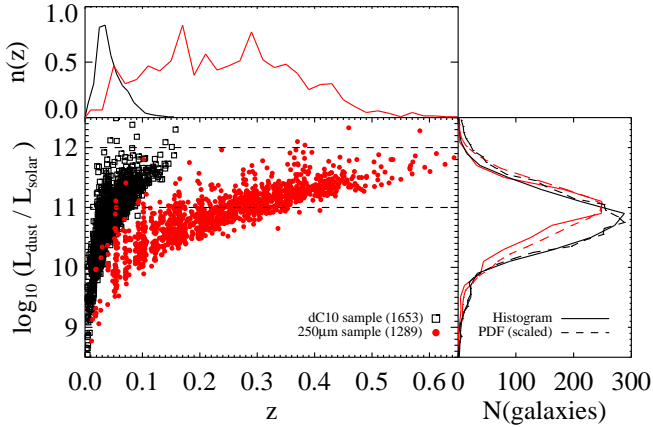}
  \caption{The dust luminosity of {\it Herschel}-ATLAS-selected
    galaxies with reliable counterparts, and reliable SED fits to our
    compiled photometry, as a function of redshift, is shown in
    red. The median redshift is $\bar z = 0.239$ and the median dust
    luminosity of our full sample is $5.6 \times 10^{10}
    $L$_\odot$. We also include data points from dC10 for comparison,
    in black, highlighting the different redshift and luminosity
    properties of each sample. The horizontal dashed lines delineate
    those galaxies with luminosities in the LIRG and ULIRG categories,
    at $10^{11}$ and $10^{12}\,$L$_\odot$, respectively. We also show
    histograms detailing the redshift distribution of each sample
    (top), in which the relative heights of the two histograms are
    arbitrary to bring out the contrasting properties of the two
    samples. On the right we show that the dust luminosity
    distributions for the two samples are comparable.}
  \label{z_ldust}
\end{figure}

By stacking the PDFs for galaxies well described by our model, we
determine a median dust luminosity for our whole sample of $5.6 \times
10^{10} L_\odot$, placing the average \hatlas\ galaxy's luminosity
just below what would traditionally have been considered a luminous
infrared galaxy (LIRG). Figure~\ref{z_ldust} shows the dust luminosity
as a function of redshift for \hatlas\ (red) and it is important to
note that \hatlas\ traces typical star forming spirals (with log
L$_{\rm{dust}} < 11.0$ L$_\odot$) out to much higher redshifts ($z\sim
0.35$) than was possible with {\it IRAS} ($z<0.05$). Our SED-fitting
results indicate that the star formation rate of the average
low-redshift \hatlas\ galaxy is $\sim 3.4$ M$_\odot$ yr$^{-1}$, and
that the median dust to stellar mass ratio is $\sim 0.4\%$.

\begin{figure*}
  \centering
  \subfigure{\includegraphics[width=0.9\textwidth, angle=0]{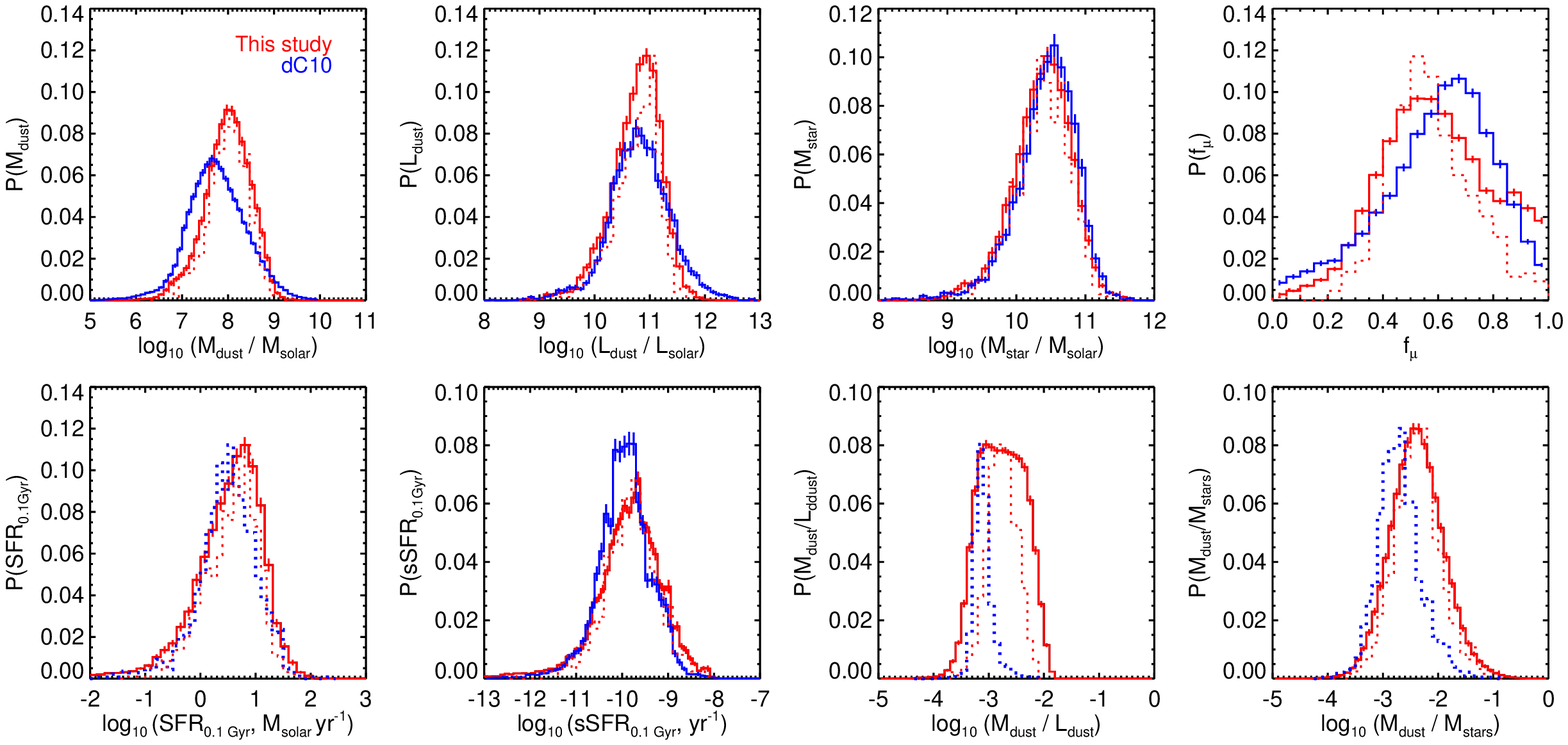}}
  \caption{Comparisons of the stacked PDFs from \hatlas\ SED fitting
    (red histograms) with those of dC10 (blue histograms). The red
    dotted histograms indicate the distribution of the median
    likelihood values for \hatlas\ (i.e. the 50th percentiles of the
    individual PDFs, arbitrarily renormalised for ease of comparison),
    which are narrower than the stacked PDFs since they do not reflect
    the uncertainty in the derived values. The histograms compare
    parameter distributions for the $z<0.35$ sample which are well
    described by our model, with the results of dC10.  We note that
    the dC10 results do not include PDFs for SFR,
    $M_{\mathrm{dust}}$\slash$L_{\mathrm{dust}}$ or
    $M_{\mathrm{dust}}$\slash$M_{\mathrm{stars}}$, and so we include
    the quotient of the median values of the relevant PDFs for the
    dC10 sample shown as thick blue dotted histograms. The stacked
    results suggest that galaxies selected in \hatlas\ contain on
    average a factor $\sim 2$ more dust than those galaxies in the
    dC10 sample, although the luminosities are broadly comparable.}
  \label{distributions}
\end{figure*}

In Figure \ref{distributions}, we present histograms of the results of
our SED fitting, and those from dC10, who applied the same method to a
{\it GALEX}-SDSS-2MASS-{\it IRAS} data set and included photometry
only up to 100\,$\mu$m. The different selection criteria of the dC10
sample produces a different redshift distribution from the
\hatlas\ sample, however the stellar mass and dust luminosity
distributions are similar, and comparison of specific parameters can
still be instructive. In Table \ref{tab:sampleprop} we provide results
for the dC10 sample ($z<0.16$), and for our sample of 250\,$\mu$m
selected galaxies (limited to $z < 0.35$).

On average, the galaxies selected at 250\,$\mu$m have slightly higher
specific star formation rates than those in dC10, by $\log_{10}$\,sSFR
$\approx 0.1$ dex. Though the dust luminosities and stellar masses are
roughly the same, the dust masses in the two samples differ by
0.3\,dex, with the \hatlas\ galaxies being more dusty. This translates
into higher ``specific'' dust masses for \hatlas\ galaxies (higher
\mdust\slash\ldust\ and \mdust\slash\mstar). Part of this difference
is due to the evolution in dust masses with redshift (Dunne et
al. 2011), as the \hatlas\ sample probes a higher redshift
range. However, this is not the whole story as we will discuss in the
next section.

In figure \ref{distributions}, we compare the stacked PDFs of several
physical parameters for the {\it IRAS}-selected sample of dC10 (blue)
and the \hatlas\ 250\,$\mu$m-selected sample (red). The dC10
distribution of \mdld\ values is considerably narrower than that based
on the results of this study.  This may be because Herschel's
selection at 250\,$\mu$m is intrinsically more sensitive to a range of
\mdld\ values, as it is not solely sensitive to the warmer dust but
also picks up the cold, dusty galaxies which have large \mdld\ but are
not necessarily warm enough for \iras\ to detect.

\begin{table}
\caption{Comparison between the median properties of the dC10 {\it
    IRAS}-selected sample and the \hatlas\ 250\,$\mu$m-selected
  sample.} 
  \centering
  \begin{tabular}{|l|c|c}
    \hline
    & \multicolumn{2}{c}{Median values} \\
    & dC10 & 250\,$\mu$m   \\
    Parameter & $z < 0.16$ & $z < 0.35$ \\
    \hline
    SFR $($M$_\odot$ yr$^{-1})$ & 3.25$^\dagger$ & 4.17$^\dagger$\\
    $\log$\,sSFR (yr$^{-1}$) & -9.94 & -9.80\\
    $\log$\,\mstars\ $($M$_\odot)$ & 10.48 & 10.40\\
    $\log$\,M$_{\mathrm{dust}}$ $($M$_\odot)$ & 7.74 & 8.01\\
    $\log$\,L$_{\mathrm{dust}}$ $($L$_\odot)$ & 10.81 & 10.81\\
    $\log$\,(M$_{\mathrm{dust}}$\slash L$_{\mathrm{dust}}$) & -3.12$^{\dagger}$ & -2.77$^{\dagger}$ \\
    $\log$\,(M$_{\mathrm{dust}}$\slash M$_{\mathrm{stars}}$) & -2.72$^{\dagger}$ & -2.38$^{\dagger}$ \\
    \fmu & 0.62 & 0.57 \\
    \hline 
    N(galaxies) & 1653 & 1032
  \end{tabular}
  \label{tab:sampleprop}
\end{table}

\subsubsection{Star formation and dust in 250\,$\mu$m-selected galaxies}

\begin{figure*}
  \centering 
  \includegraphics[width=0.90\textwidth,angle=0]{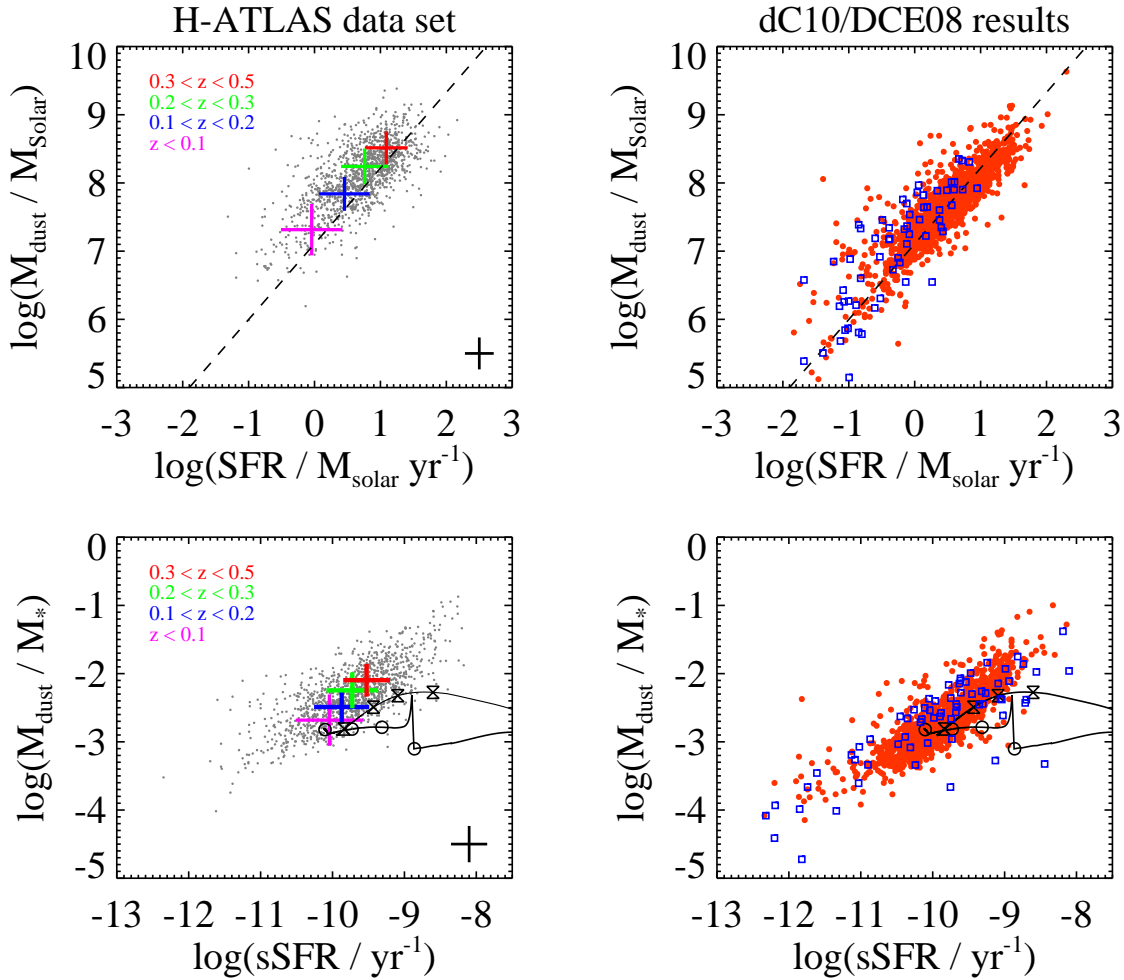}
  \caption{Comparisons between \hatlas\ 250$\mu$m-selected galaxies
    and other local galaxy surveys. Top Row: the variation of the
    total dust mass as a function of the star formation rate for the
    three samples of galaxies; on the left, we show the
    \hatlas\ 250$\mu$m-selected sample with at least least two far-IR
    detections in our data set are shown in grey. The coloured error
    bars indicate the mean positions of those galaxies in the redshift
    range shown by the colours, with size equal to the standard
    deviation of the values in each bin. On the right we present the
    same results for the {\it GALEX}--SDSS--2MASS--{\it IRAS} sample
    from dC10 in orange circles, with the SINGS galaxy sample from
    dC08 in blue squares. Overplotted in each frame is the best fit
    relationship from da Cunha et al. 2010, which is offset with
    respect to the results of our best fit SEDs. Bottom row: the ratio
    of dust mass to stellar mass as a function of specific star
    formation rate, with the same colour scheme as above. The
    ``solar-neighourhood'' and ``dwarf-irregular galaxy'' theoretical
    models of Calura et al. (2008) are overplotted as the open circles
    and the open bow-ties, respectively, with the symbols indicating
    the locations of the models at ages of 1, 3, 6, and 12\,Gyr. The
    models are described in considerable detail in the main text, and
    the majority of \hatlas\ sources do not overlap with the locus of
    theoretical points.}
  \label{correlations}
\end{figure*}

We now explore the star formation activity and dust properties of
\hatlas\ galaxies. In Figure \ref{correlations}, we plot relationships
for three different galaxy samples; the results of this study (left),
and the dC10\slash DCE08 samples (right, orange and blue squares,
respectively). Each individual galaxy is shown in grey, and we split
our sample in bins of redshift, with colours as shown in the
legend. The positions of the error bars correspond to the mean of the
galaxies in that redshift bin, while the size of the error bar
represents the standard deviation of the derived values within that
redshift bin. The best fit relation between SFR and dust mass from
dC10 (dashed line) appears to trace the low dust-mass edge of the
250\,$\mu$m selected population. The slope of the relationship for
\hatlas\ appears to be the same, but the \hatlas\ sources are offset
such that they have larger dust masses for a given SFR compared to
{\it IRAS} selected galaxies.

It is unlikely that this offset is a result of biases in the fitting
given the difference far-infrared coverage for the two samples. As we
showed in Section \ref{IRAStest} for the \iras\ sample, we do not
expect \mdust\ to be under-estimated (if anything, the converse
applies) due to the lack of data at wavelengths longer than
100\,$\mu$m. For \hatlas\ sources without PACS data, there was a small
tendency to overestimate the dust mass at the highest values of SFR,
but not at a level which could explain this offset which is present at
all SFR values. This effect is also not due to redshift differences
between the two samples, as the error bars show, increasing redshift
drives a given galaxy along the plotted slope and not away from
it. Comparing only sources within the same redshift range (i.e. if we
limit the \hatlas\ results to $z < 0.16$), we still find an increase
in dust mass per unit star formation rate for the \hatlas\ sample,
relative to the DCE08 and dC10 samples. At low values of dust mass,
the \hatlas\ sample includes few sources with $\log_{10}
($M$_{\rm{dust}}\slash \Msolar ) < 6.5$. This is due to the flux limit
in \hatlas\ combined with the small survey area in SDP. The ``missing
\hatlas\ sources'' following the {\it IRAS} and SINGS points at the
lower left of the plot are simply below the detection threshold of the
\hatlas\ SDP sample.

The lack of {\it IRAS} sources with high \mdust/SFR may reflect the
lack of sensitivity of {\it IRAS} to cold dust; galaxies with low SFR
and larger masses of dust would tend to have colder dust temperatures
and therefore be absent from {\it IRAS} selected samples. This
comparison suggests that {\it IRAS} preferentially selects those
galaxies with the highest SFR per unit mass of dust, since more star
formation for a given mass of dust will result in stronger dust
heating and more emission at the {\it IRAS} selection wavelength of
60\,$\mu$m. For the DCE08 SINGS sample (squares), which are not
selected from a flux limited FIR survey, the trend is similar to that
seen in \hatlas\ but extended to lower values of dust mass.

In the bottom panel of Figure \ref{correlations}, we also plot the
ratio of the dust to stellar mass as a function of specific star
formation rate. This relationship was first noted by dC10, and their
data are plotted in the right panel of Figure
\ref{correlations}. Compared to the sample in dC10, the {\it H}-ATLAS
galaxies appear to have higher specific dust content (relative to
stellar mass) for a given specific star formation rate. The reason for
this difference is likely to be the same as that in the upper
relationship between \mdust\ and SFR since the stellar mass
distributions of the two samples are very similar.

We overlay the predictions of dust evolution tracks from the chemical
evolution models of Calura et al. (2008), which are based on the model
in Dwek (1998) and follow the build-up of heavy elements and dust
formed in low-and-intermediate mass stars (LIMS) during their AGB
phase and in both Type Ia and Type II supernovae. The upper black
curve shows the evolution of dust from a dwarf-irregular galaxy with
continuous star formation. The solar-neighbourhood model, which
reproduces the properties of the Galactic disc and centre (see Calura
et al. for more detail), is indicated by the lower curve.  As
discussed in dC10, the observed trend between specific dust mass and
specific star formation rate can be explained as follows: dust is
produced through stellar sources, its production rate closely linked
to the star formation rate and rises steadily as the galaxy starts to
build up stellar mass. As gas is consumed, the star formation rate
declines and so less dust is formed. Given the destruction of dust
via astration, outflows and supernova shocks, at this stage, galaxies
can no longer replenish their dust content through star formation and
the dust mass decreases. The chemical evolution models trace the
evolutionary history of the galaxy; increasing the star formation
rate, the gas mass available to form stars and/or the amount of dust
from supernovae will drive the models towards the upper right in
Figure \ref{correlations}, i.e. towards those \hatlas\ galaxies with
the highest specific dust masses.

The dust mass evolution as traced by the Calura\slash Dwek chemical
evolution models are likely to be a best case scenario since their
assumed condensation of dust required from the heavy elements ejected
by LIMS during their stellar wind phases and/or massive star
supernovae are rather optimistic compared to the dust masses observed
for stellar sources (e.g. Morgan \& Edmunds 2003). Indeed, Dunne et
al.\ (2011) find it extremely difficult to explain those galaxies with
the highest dust masses in the {\it H}-ATLAS sample without grain
growth in the ISM as the main contributor to the interstellar dust
budget, or a top-heavy initial mass function (see also Gomez et al.,
in prep).

\begin{figure*}
  \centering
  \includegraphics[width=0.90\textwidth]{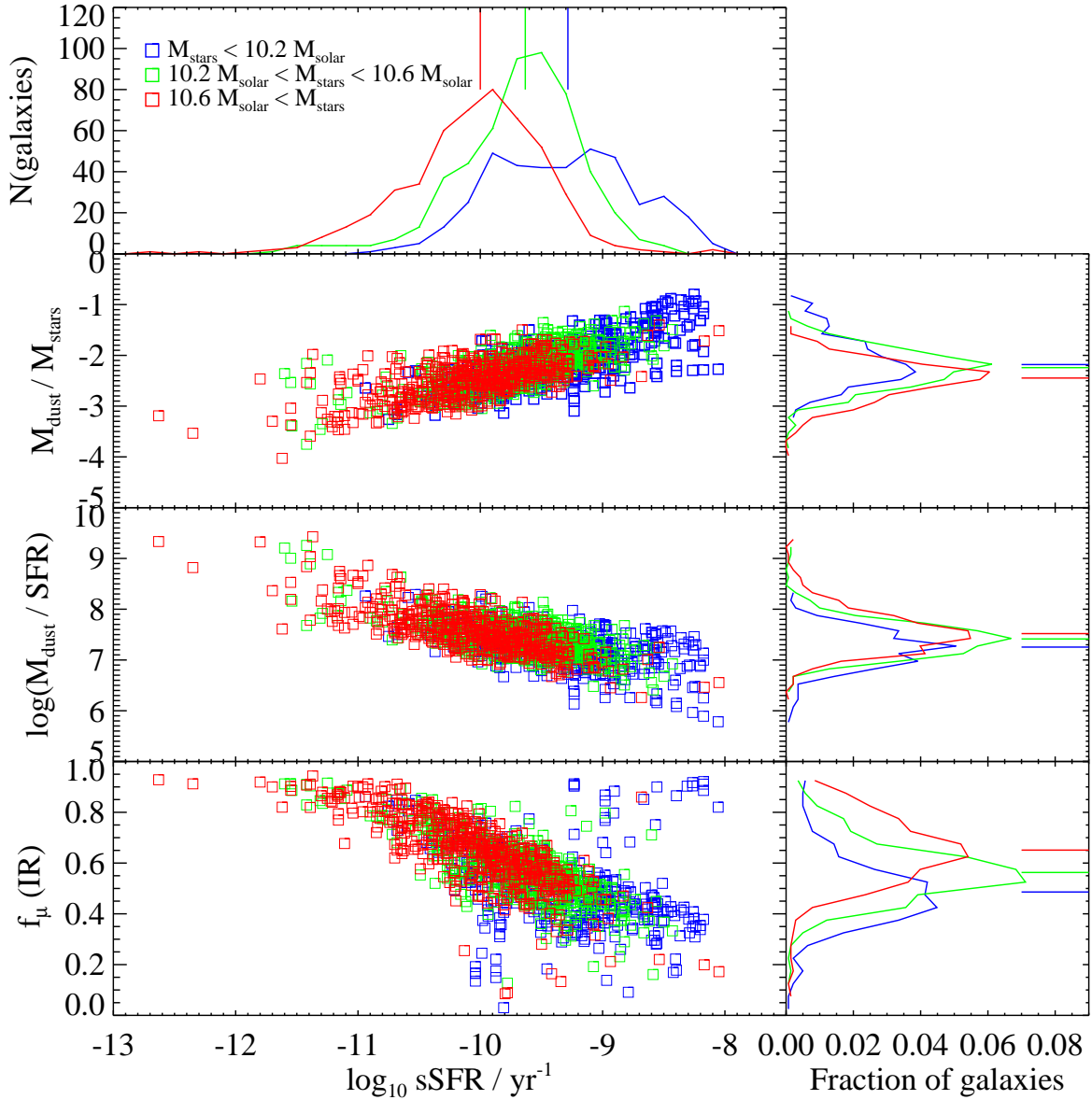}
  \caption{The variation of the dust to stellar mass ratio, dust mass
    per unit star formation rate and the $f_\mu$ parameter as a
    function of specific star formation rate derived from our SED
    fitting. The galaxies are additionally binned into three
    populations based on their stellar mass, with the low-,
    intermediate- and high-mass galaxies (with limits of $\log_{10} (M
    \slash \Msolar) < 10.2$, $10.2 - 10.6$, and $> 10.6$), shown by
    the blue, green and red squares, respectively. To the right of
    each panel we also show histograms corresponding to that mass
    bin's distribution relative to the vertical axis, and in the very
    top panel we show histograms of the specific star formation rate
    in each mass bin.}
  \label{Properties_vs_SSFR}
\end{figure*}

We now investigate whether the dust and star formation properties of
our sample vary as a function of stellar mass.  In Figure
\ref{Properties_vs_SSFR} we show the variation of $M_{\mathrm{dust}}
\slash M_{\mathrm{stars}}$, \mdust/SFR, and $f_\mu$ as a function of
specific star formation rate binned by stellar mass, such that each
bin has approximately equal numbers of galaxies. In the upper and
right panels, we plot histograms for each of the bins in stellar mass,
with the median values overlaid as vertical\slash horizontal
lines. The lower stellar mass galaxies have higher specific dust mass,
a similar trend is also observed in stacking analyses of
optically-selected galaxies in the {\it H}-ATLAS SPIRE data (Bourne et
al., 2012) and also in volume limited samples of very local galaxies
(e.g. Cortese et al. 2012), suggesting that this effect is not the
result of selection bias in our data.

It is also clear that the lower-mass galaxies are considerably more
actively star-forming than the high-mass galaxies, consistent with the
idea that the massive galaxies have consumed more of their available
baryonic fuel through either the process of star formation or the
aftermath of AGN feedback than their less massive neighbours
(e.g. Cowie et al. 1996, Bundy et al. 2006, 2009, Hopkins et al.,
2007, Pozzetti et al. 2010).

The lower panel of Figure \ref{Properties_vs_SSFR} suggests that lower
mass galaxies have smaller contributions to their total infrared
luminosity from dust in the ambient ISM, than their more massive
counterparts (lower $f_{\mu}$). This, once more, is due to the fact
that the less massive galaxies are undergoing proportionally more star
formation and so the stellar birth clouds make a larger contribution
to the total far infrared energy output.

As in dC10, we observe a small fraction of low-mass galaxies with high
specific star-formation rates and high $f_\mu$. The value of \fmu\ in
the model enters both in the optical part of the star formation
history libraries (based on the age of the stellar population and
opacity of the birth clouds) and from the combination of the dust
components. Clearly, there are degeneracies in the IR part of the SED;
for example a high \fmu\ with a warm temperature for the ISM component
could produce a very similar looking far-infrared SED to a lower
\fmu\ with a cool temperature for the birth cloud component. However,
the optical colours for these two scenarios may look different as a
result of the different stellar ages and attenuations in the model and
so the IR data are not the only (or even the strongest) constraint on
this parameter.

The high \fmu\ values indicate that dust in the diffuse ISM heated by
stars older than 10\,Myr dominates the far-infrared emission in these
sources, but the high sSFR averaged over the last 0.1\,Gyr suggests
that these sources have very recent star formation activity. If we
instead use the sSFR values averaged over the last 10\,Myr, these
galaxies are assigned more modest values of sSFR and shift to the left
suggesting that the models which best fit the data are ones in which
the star formation was recently truncated. These sources are all
strong emission line objects, not displaying classical
'post-starburst' spectra and so the high \fmu\ and sensitivity of
model SSFR to the timescale of integration could indicate that they
are in a short-lived phase transitioning from their obscured birth
clouds to the more diffuse ISM. If global SED fitting can potentially
isolate sources in specific stages of evolution, this could be a
powerful technique, however, full testing on a larger sample including
a detailed analysis of the optical spectra is required in order to
confirm this.

\subsection{Variation of SEDs within the 250\,$\mu$m galaxy population}
\label{trends}

We now analyse the shape of the averaged population SEDs as a function
of model parameter, in order to understand the main physical
properties driving the shape of galaxy SEDs. We obtain median SED
templates by stacking the SEDs of our galaxies according to their
best-fit parameters, following the method described in detail in
Appendix \ref{stack_calc}.

In Figure \ref{ssfr_comparisons}, we show the median SEDs in stacks
binned on (a) specific star formation rate, (b) dust luminosity, (c)
stellar mass, and (d) redshift. We include only galaxies at $z < 0.35$
in this analysis, i.e. we only consider the range in redshifts where
we believe that our sample is representative. We also show the
unattenuated starlight SED in the same bins in Figure
\ref{unat_comparisons} to illustrate changes in the underlying stellar
populations fitted by the models.

As a further check of the influence of having PACS detections for only
a fraction of our sources, we have performed the stacking in each bin
twice; once including all available data in the fitting (solid lines)
and once omitting the PACS data for all sources (dotted lines). The
number of sources in each stack is shown in the legend to each
sub-figure, as is the fraction of these with $5\sigma$ PACS
detections. The broad similarity between the solid and dotted stacked
SEDs -- i.e. those including and neglecting the PACS data -- is
generally reassuring; we now discuss each set of stacked SEDs in
detail.

\subsection*{Specific star formation rate}

The most striking trends are with specific star formation rate, shown
in Figure \ref{ssfr_comparisons}\,(a) and Figure
\ref{unat_comparisons}\,(a); those galaxies with the highest specific
star formation rates not only have the youngest stellar populations,
but also the hottest effective dust temperatures manifest by the bluer
optical colours, and shorter peak wavelengths of the FIR bump
respectively. Similar results were also found by DCE08 for the much
smaller sample of SINGS galaxies.

\subsection*{Total dust luminosity} 

In Figure \ref{ssfr_comparisons}\,(b) we show the relatively weak
variation of the SED properties in our sample as a function of dust
luminosity. The UV-optical part of the transmitted SED is similar in
all but the highest \ldust\ bin, which as shown in figure
\ref{unat_comparisons} is dominated by a younger intrinsic SED with
greater reddening than the less luminous bins. The template PAH
luminosity increases markedly with \ldust\, though this area of the
SED is only indirectly constrained by the model priors and energy
balance (due to the absence of mid-infrared observations in this
study) and so we cannot determine how significant this is. The
FIR/optical ratio increases with \ldust, indicating that galaxies with
higher \ldust\ are also more obscured with a greater fraction of their
bolometric luminosity being re-radiated by dust. The shape of the FIR
peak seems to be largely uncorrelated with \ldust\ until the highest
bin, when it shifts to the blue; thus the most luminous dust sources
have warmer temperatures, but it does not appear to be a monotonic
trend across the range of \ldust\ probed. This change in peak
wavelength is however consistent with the differing intrinsic
(i.e. unattenuated) starlight SEDs shown in figure
\ref{unat_comparisons}; it is clear that the more dominant young
stellar population in the most luminous bin of \ldust\ is associated
with the apparently hotter dust template in the same bin.

At this point, we need to proceed with caution in our conclusions
because of the potential bias in \ldust\ when PACS data are missing
(as discussed in Section \ref{pacscomp} and Appendix
\ref{bias_tests}). The legend in figure \ref{ssfr_comparisons} shows
the fraction of PACS-detected sources in each bin, and also shows in
dotted lines the same stacked template SEDs compiled when the PACS
data are neglected from the fitting (i.e. even for those galaxies
which are detected by PACS). As figure \ref{ssfr_comparisons}\,(b)
shows, the fraction of PACS-detected sources in each template bin of
\ldust\ is approximately constant, and the results do not change when
the PACS data are neglected altogether, suggest that they are robust
to the presence\slash absence of PACS data. The number of sources in a
stack in \ldust\ is almost always in the minority, however, which
could also be the root cause of the similarity between the PACS and
no-PACS stacks. To check that this is not the case, we applied our
stacking analysis to the PACS-complete sample, dividing it into two
luminosity bins containing approximately equal numbers of galaxies and
stacking the best-fit SEDs derived in the two bins of \ldust, both
including and excluding the PACS data. Figure
\ref{fig:ldust_bins_bias_check} shows the results for the most/least
luminous sources in the top/bottom panel, including the PACS data (in
red) and neglecting them (in blue), with each stack normalised at
250\,$\mu$m. At lower luminosities, the lack of PACS data has no
effect on the stacked FIR SED shape and so for bins below $\log_{10}
(\ldust\slash\Lsolar) < 10.5$ we can be confident that the lack of
trend of SED morphology with \ldust is robust. For the bin with
$\log_{10} (\ldust\slash\Lsolar) > 10.5$, there is more luminosity in
the mid-IR and PAH component when PACS data are included compared to
when they are not, but this is an area of the SED which we cannot
confidenctly discuss with the present data set. The FIR peak is
broadened when the PACS data are included, but not shifted
significantly, while the optical\slash far-infrared ratios are
consistent, and the median optical templates are almost identical.  We
do observe an increase in the range of values (i.e. the bounds of the
16$^{\mathrm{th}}$ and 84$^{\mathrm{th}}$ percentiles) seen at optical
wavelengths derived in the absence of PACS data, though they are
broadly consistent. Indeed, these stacked SEDs are compiled from
smaller numbers of input galaxies than any of the individual stacked
SEDs in figure \ref{ssfr_comparisons}, and so the stacked median SEDs
in figure \ref{fig:ldust_bins_bias_check} (and their percentiles,
dotted) in particular are more susceptible to the influence of small
numbers at all wavelengths than those in our wider study.

\subsection*{Stellar Mass}

In Figure \ref{ssfr_comparisons}\,(c) we show the variation of the
galaxies in our sample as a function of their stellar
mass. Unsurprisingly, the most massive sub-sample has a considerably
more dominant old stellar population in the optical wavelengths than
the least massive subset, consistent with the unattenuated stellar
SEDs stacked in the same way in figure
\ref{unat_comparisons}\,(c). The lowest mass galaxies have broader FIR
peaks, suggesting they have a larger warm dust component than the
larger stellar mass sources. They also have stronger PAH emission in
the templates, but again we caution that this is not constrained by
data for this sample. This shift in the FIR SED shape with mass is
likely due to the lower mass galaxies having the highest sSFR
(i.e. ``downsizing'', as previously discussed) and consequently their
average SED shows more of a warm component heated by the ongoing star
formation.

\subsection*{Redshift}

\begin{figure*}
  \centering
  \subfigure[Galaxies binned by specific star formation rate]{\includegraphics[width=0.49\textwidth]{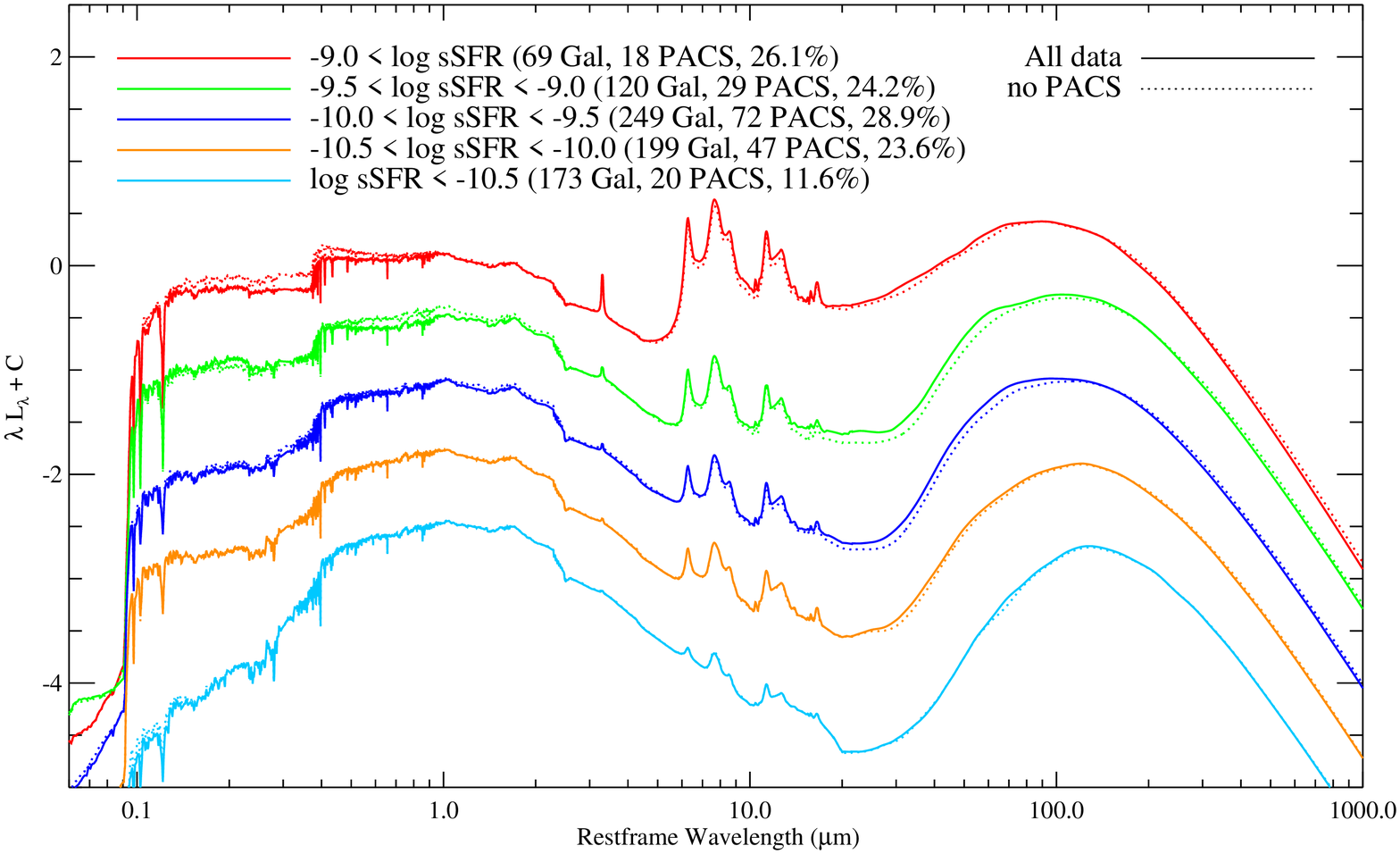}}
  \subfigure[Binned by dust luminosity]{\includegraphics[width=0.49\textwidth]{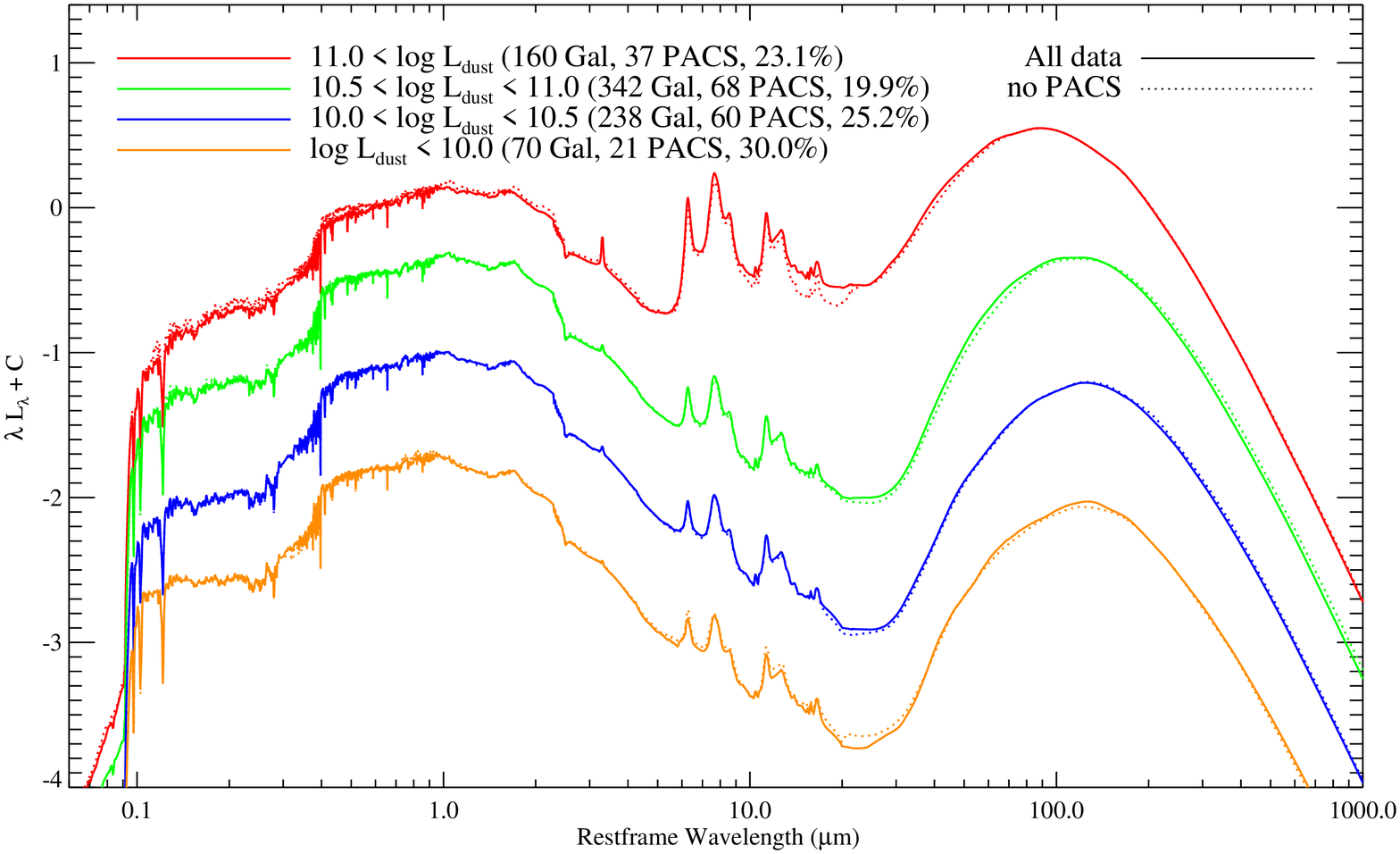}}
  \subfigure[Binned by stellar mass]{\includegraphics[width=0.49\textwidth]{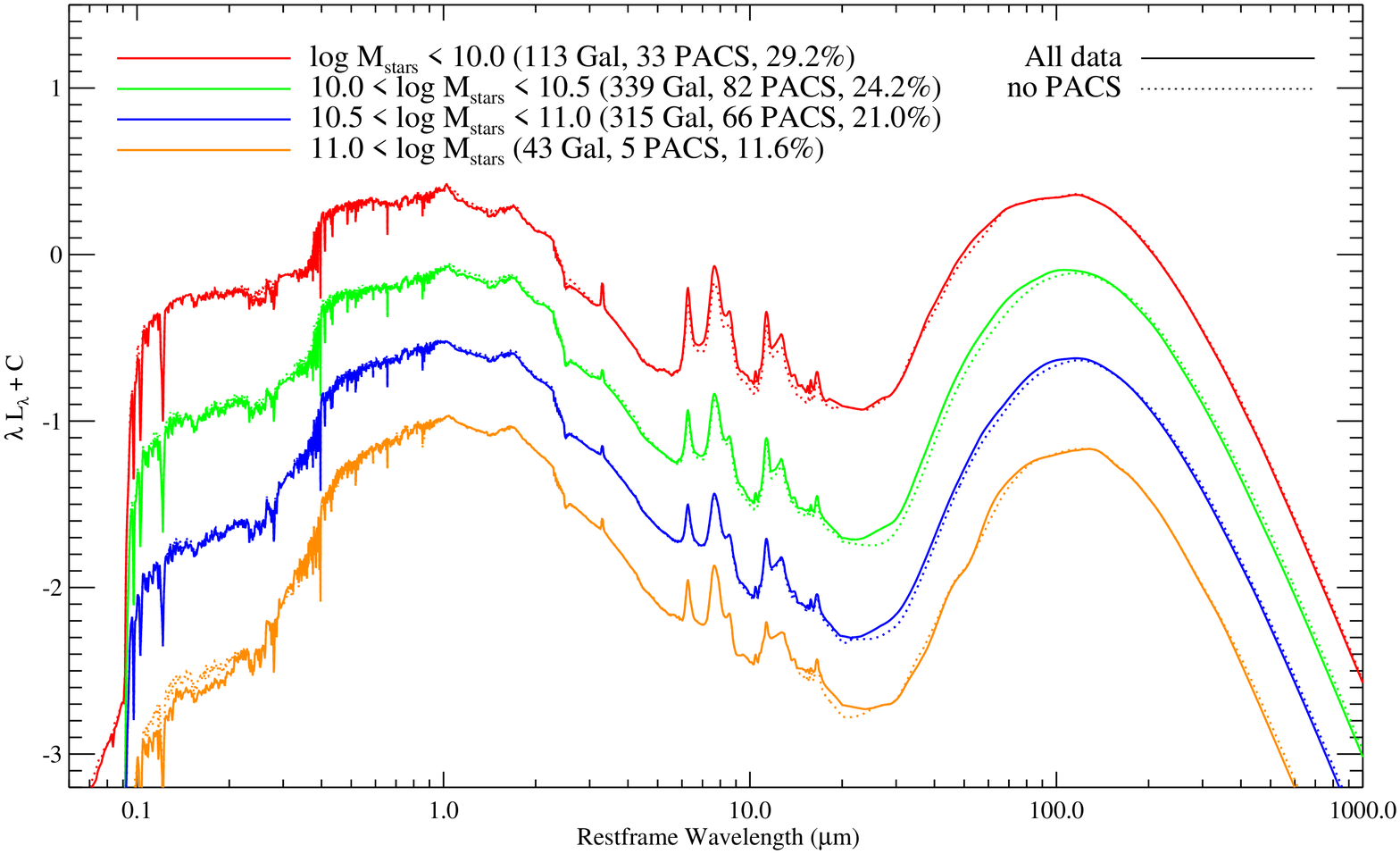}}
  \subfigure[Binned by redshift]{\includegraphics[width=0.49\textwidth]{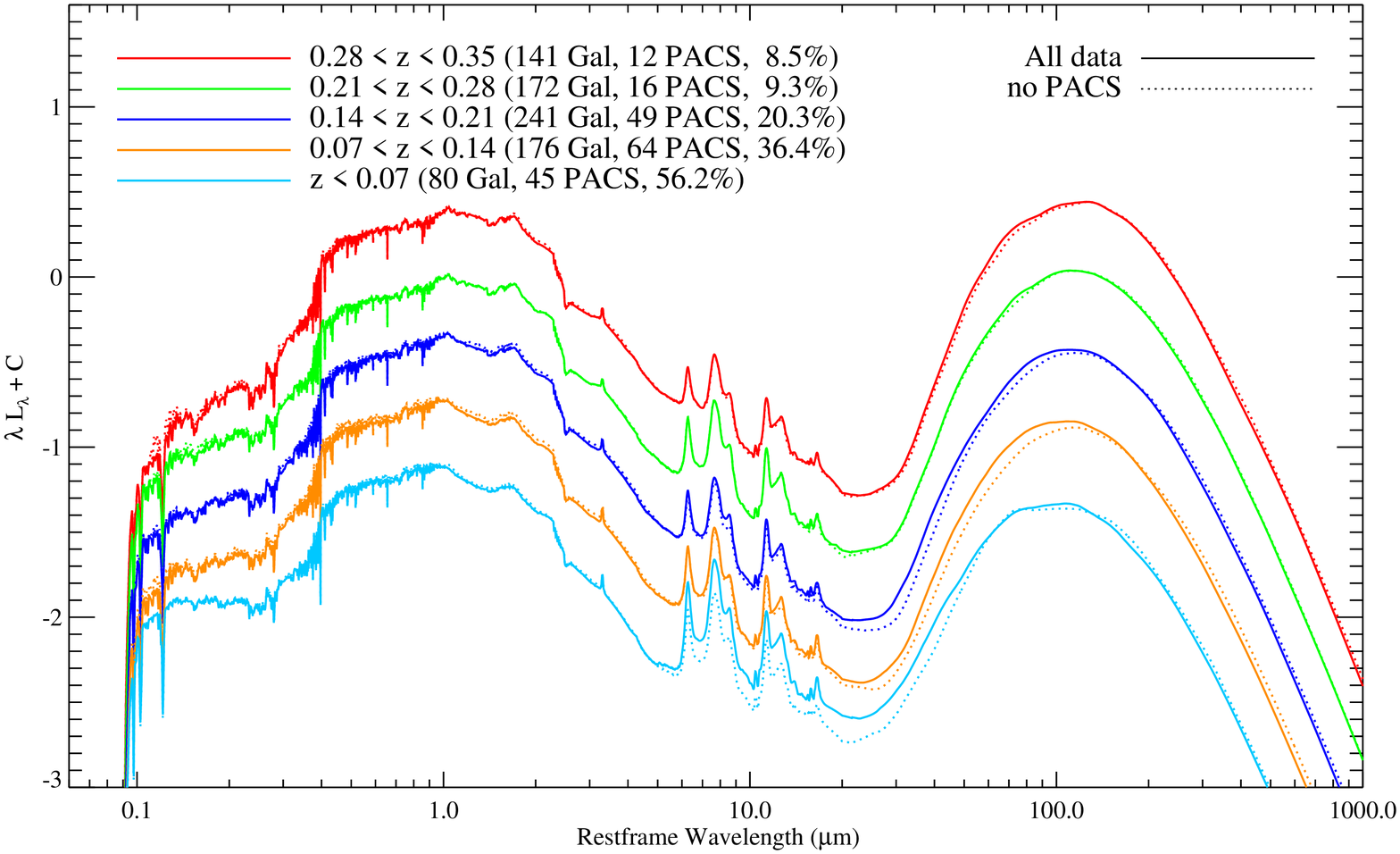}}
  \caption{The variation in the SEDs of 250\,$\mu$m selected galaxies
    in bins of their (a) specific star formation rates, (b) dust
    luminosity (c) stellar mass and (d) redshift. The range of values
    for each bin, and the number of galaxies in each bin is as
    described in the legend for that particular plot. These median
    galaxy SEDs have been normalised to their mean flux between 0.2
    and 500\,$\mu$m, and offset from one another so as to reveal the
    different properties of each median SED. Whilst the vertical
    position of each galaxy is arbitrary, the morphlogy of each median
    SED is not. The differing specific star formation rates (with
    increasing specific star formation rates from the bottom of the
    plot upwards) are apparent in the dust properties of these median
    spectra in the far-infrared wavelengths; the most vigorously
    star--forming galaxies have warmer dust temperatures due to the
    extra source of heating in the form of a recent yield of OB
    stars. The specific star formation rate appears to be the key
    driver of the properties of a 250\,$\mu$m selected galaxy's SED,
    whereas the variation between different bins of other properties
    is weaker. The stacked SEDs constructed omitting the PACS data are
    shown in the dotted lines, whilst those derived using all
    available data are shown as the solid lines. }
  \label{ssfr_comparisons}
\end{figure*}

\begin{figure*}
  \centering
  \subfigure[Galaxies binned by sSFR]{\includegraphics[width=0.49\textwidth]{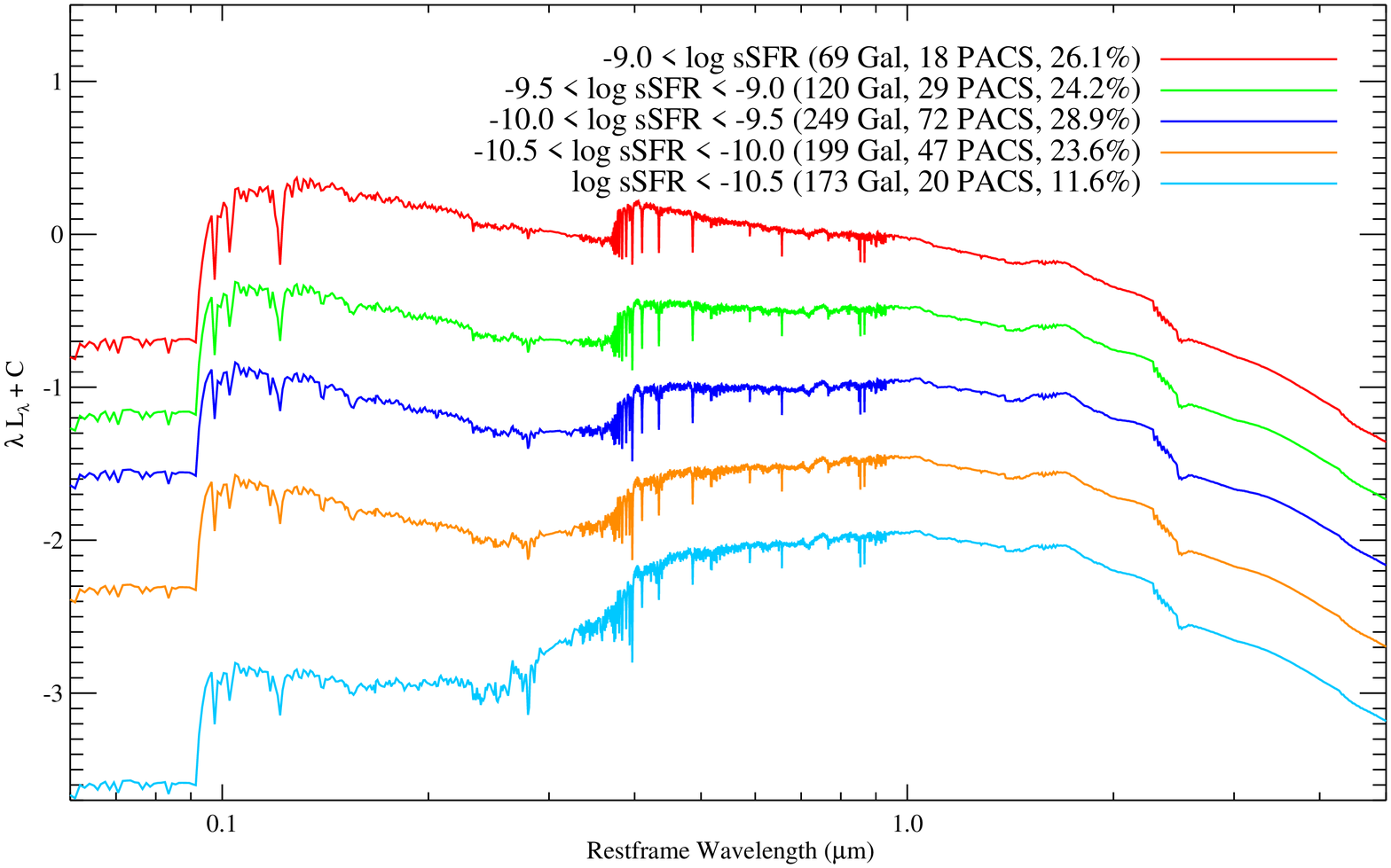}}
  \subfigure[Galaxies binned by \ldust]{\includegraphics[width=0.49\textwidth]{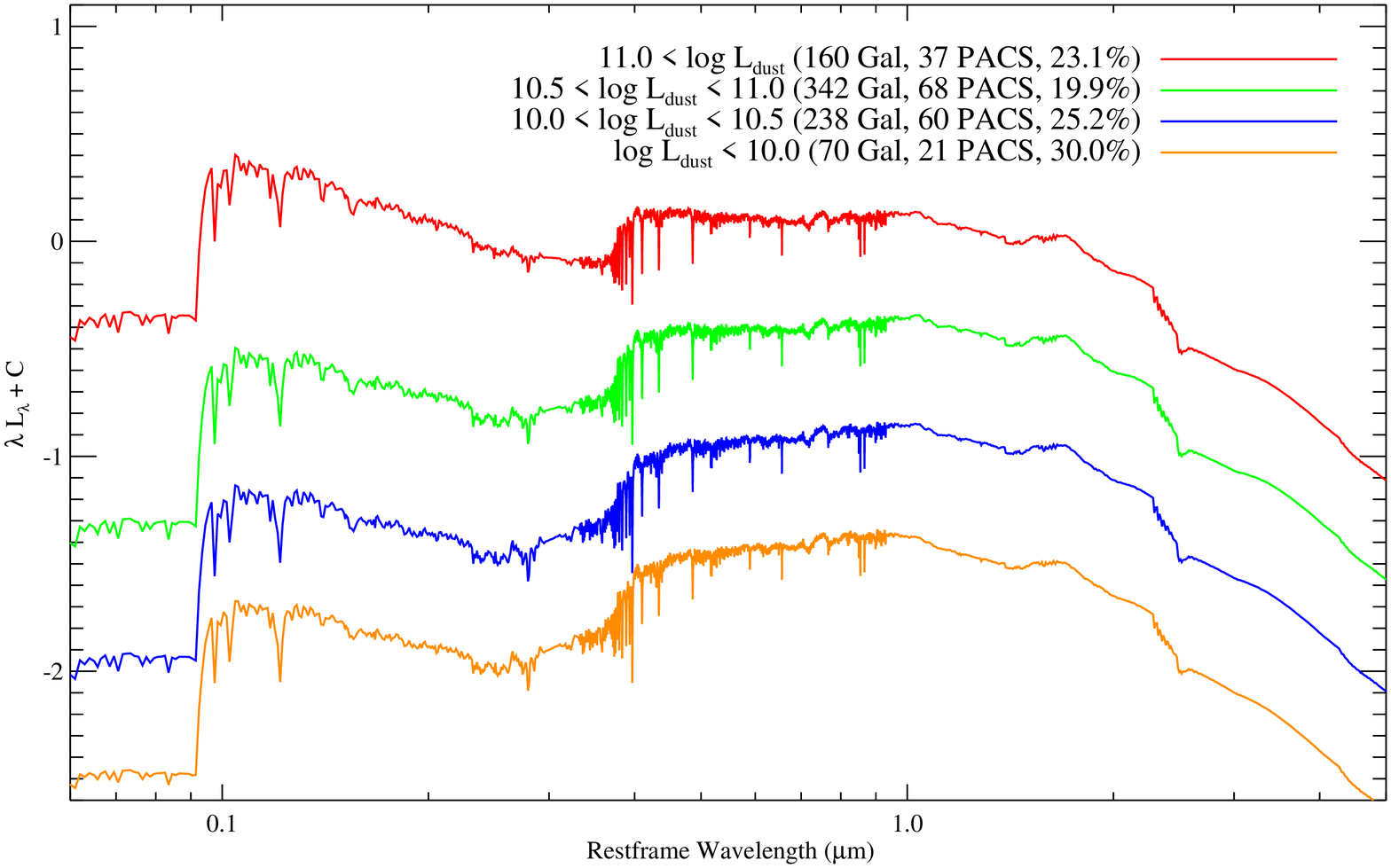}}
  \subfigure[Galaxies binned by stellar mass]{\includegraphics[width=0.49\textwidth]{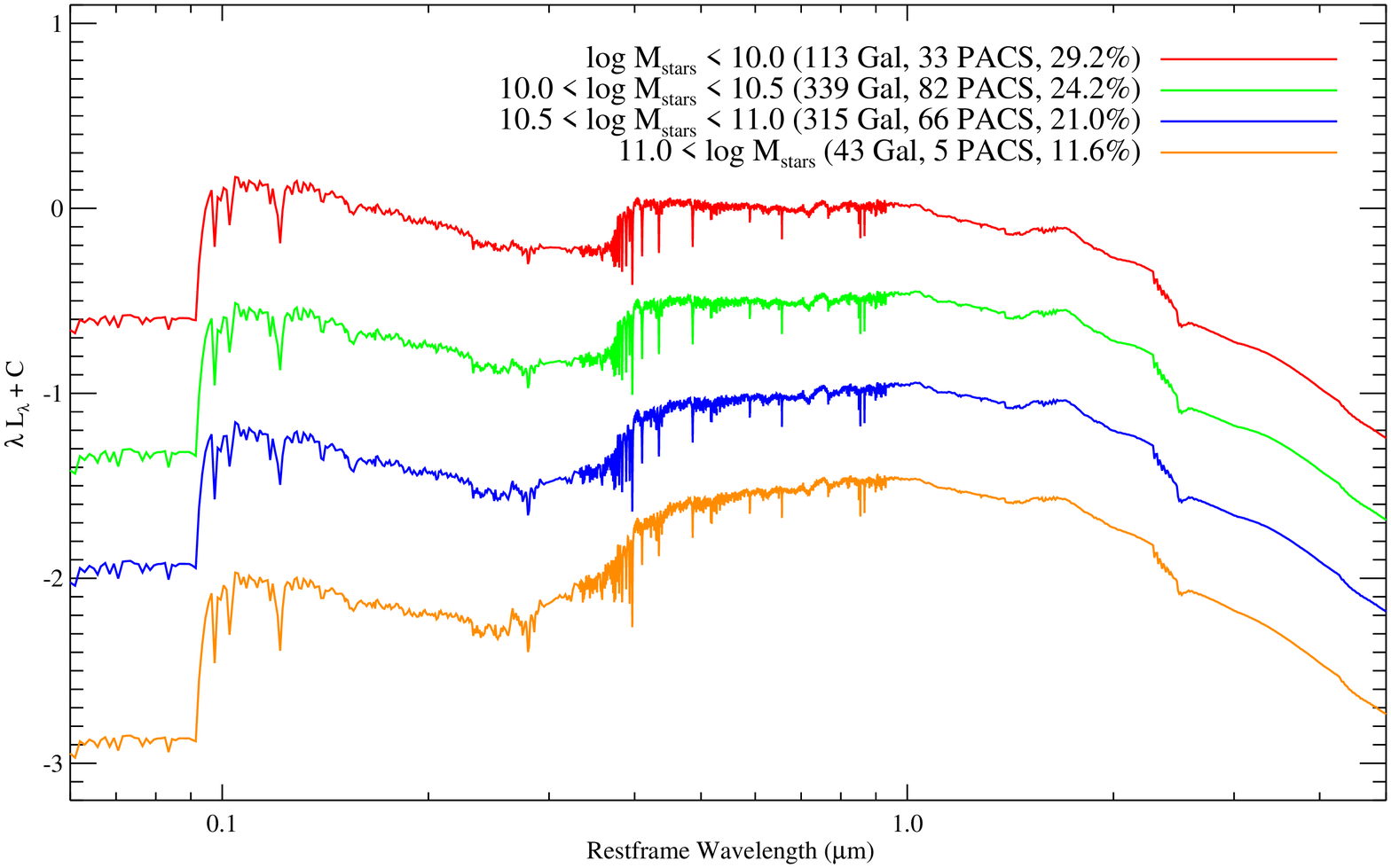}}
  \subfigure[Galaxies binned by redshift]{\includegraphics[width=0.49\textwidth]{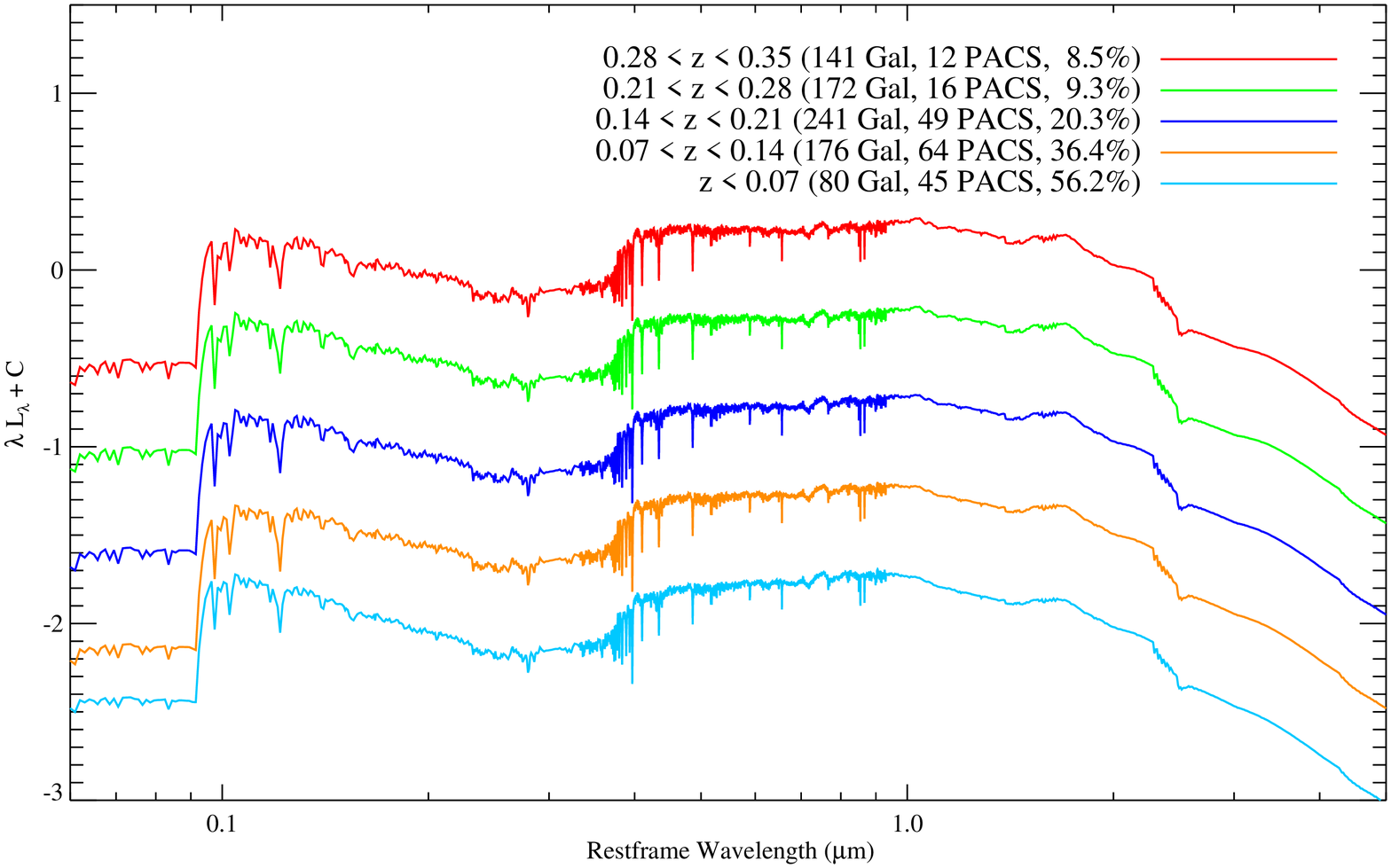}}
  \caption{Stacked best-fit unattenuated (i.e. intrinsic) stellar
    SEDs, derived in the same $z < 0.35$ bins of (a) sSFR, (b)
    \ldust\ (c) \mstar, and (d) redshift as for the full panchromatic
    SEDs in figures \ref{ssfr_comparisons}. The colours are identical
    to figures \ref{ssfr_comparisons}\,(a)-(d), and each stack has
    been derived including all available data; once more, the
    normalisation of each SED is arbitrary, but the morphology is
    not.}
  \label{unat_comparisons}
\end{figure*}

Finally, we note the weak variation in the far-infrared SEDs of these
galaxies when binned by redshift, with the panchromatic stacks shown
in Figure \ref{ssfr_comparisons}\,(d) and the unattenuated stellar
componenent in figure \ref{unat_comparisons}\,(d). The far-infrared
SEDs of these populations all have similar temperatures, with the main
difference between them being among their optical colours, with the
higher redshift stacks appearing redder; figure
\ref{unat_comparisons}\,(d) shows that the intrinsic stellar
populations in each bin are very similar, suggesting that this is due
to increasing dust opacity in the UV-optical with increasing
redshift. This has also been noted by Dunne et al. (2011) for \hatlas\
galaxies, and we note that the fraction of energy emerging in the
optical/FIR is also changing with redshift, again the higher redshift
sources have more of their total bolometric output emerging in the FIR
compared to those at lower redshifts, consistent with the
aforementioned increased dust opacity at higher redshifts.

\begin{figure}
  \centering
  \includegraphics[width=0.98\columnwidth]{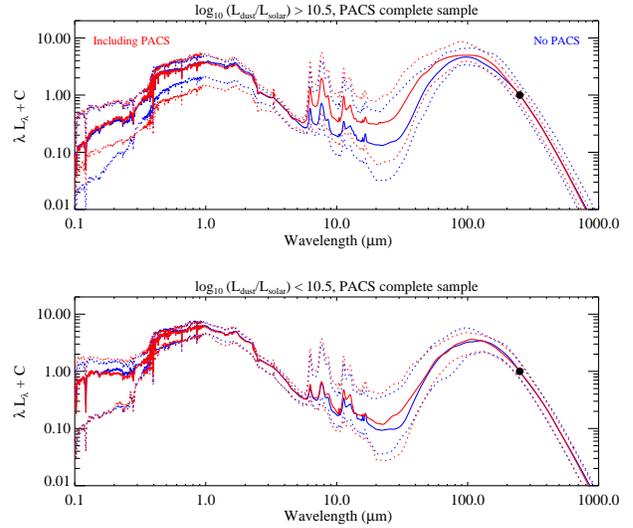}
  \caption{Stacked SEDs in two bins of \ldust\ derived using the
    PACS-complete sample. The top panel shows the stacks of those
    galaxies with high luminosity ($\log_{10} (\ldust\slash \Lsolar ) >
    10.5$, while the bottom panel shows the stack for the lower
    \ldust\ galaxies. The boundary values are chosen so as to have
    approximately equal numbers of galaxies in each bin, with the
    galaxies assigned to one bin or the other according to the
    \ldust\ estimate derived using the same set of data
    (i.e. including\slash neglecting the PACS data). In each panel we
    show the stacked SEDs (solid) and 1 $\sigma$ scatter within the bin
    (dotted) from the best fits derived including or omitting the PACS
    data (red and blue, respectively). The far-infrared luminosities,
    effective temperatures, and optical properties of the low luminosity
    stack in \ldust (i.e. including\slash omitting PACS) are almost
    identical, while for the higher \ldust bin there is more mid-IR and
    PAH luminosity when the PACS data are included, and the stacked SED
    shows a slightly broader, slightly warmer FIR peak.}
  \label{fig:ldust_bins_bias_check}
\end{figure}

\subsection*{The range of SEDs in each stack}

\begin{figure*}
  \centering
  \includegraphics[width=0.99\textwidth]{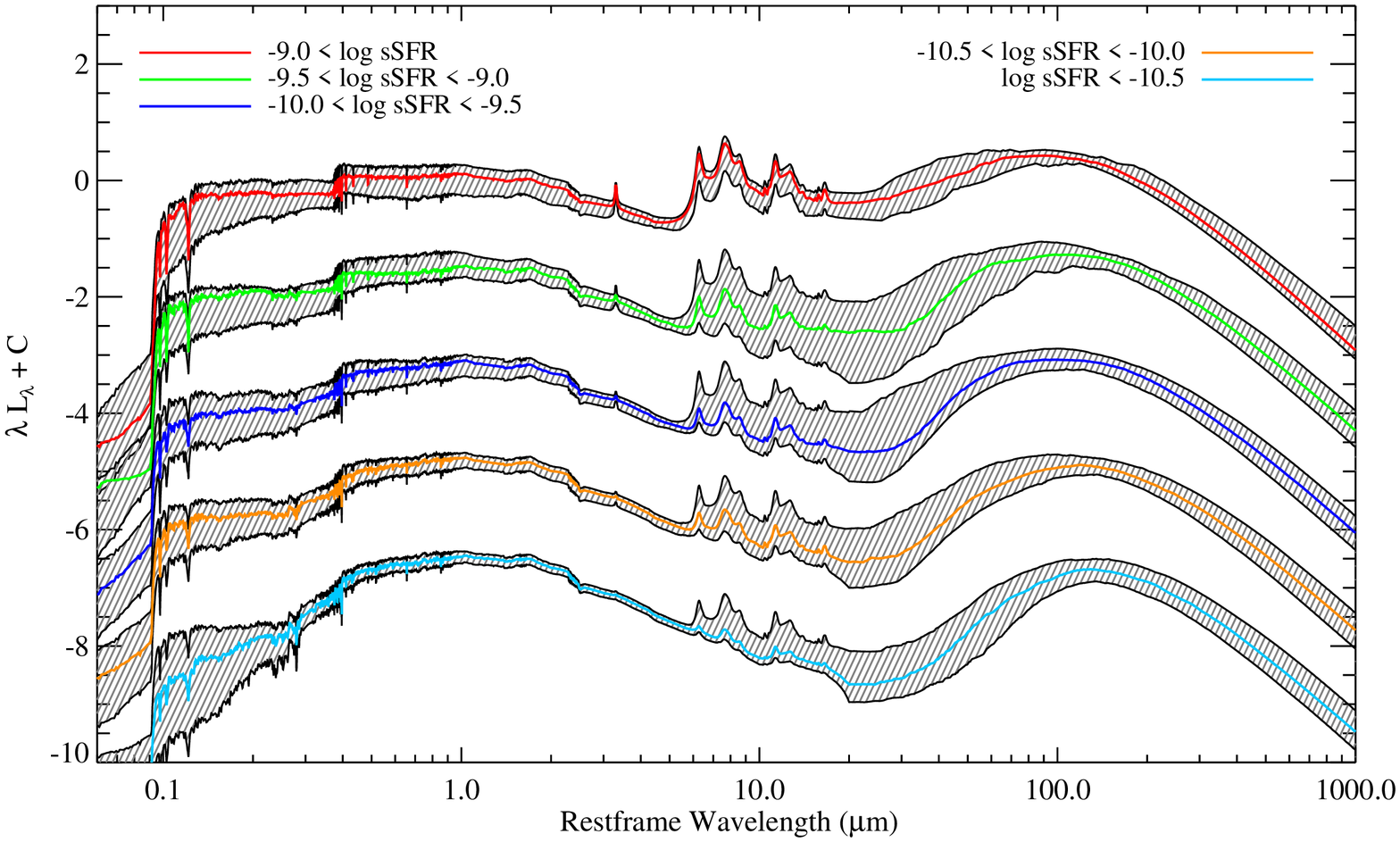}
  \caption{In addition to Figure \ref{ssfr_comparisons}, it is
    possible to show the variety of different SED models within a bin
    of e.g. specific star formation rate. The median best fit
    transmitted SEDs are shown by the coloured solid lines
    corresponding to the bins of sSFR, with the 1$\sigma$ spread of
    SEDs going into the stack shown by the shaded areas bounded by the
    black lines. In making these SEDs, each component SED in a
    particular bin of specific star formation rate has been normalised
    to its mean between 0.2 and 500\,$\mu$m and the resulting median
    SEDs (plotted) have been offset from one another to prevent them
    from overlapping. The large variety of best fit models at mid
    infrared wavelengths (between $\sim 6-60$\,$\mu$m) is due to the
    lack of observational constraints in this part of the spectrum.}
  \label{sSFR_variation_v2}
\end{figure*}

In addition to calculating the median stacked SED, we are also able to
quantify the spread of SEDs within each bin. In Figure
\ref{sSFR_variation_v2}, we demonstrate this for galaxies binned by
specific star formation rate. Once more, each bin has been normalised
to the mean of each SED between 0.2 and 500\,$\mu$m, and the offsets
between models are arbitrary for ease of comparison. As in Figure
\ref{ssfr_comparisons}, we present the transmitted galaxy templates as
the solid lines, with colours corresponding to the bins of sSFR. The
shaded grey regions show the region bounded by the 16$^{\mathrm{th}}$
and 84$^{\mathrm{th}}$ percentiles of the ensemble of normalised model
SEDs at a given wavelength, $\sigma(\lambda)$ (see Appendix
\ref{stack_calc} for more details); this is distinct from the much
smaller error on the median SED, which could be used for
e.g. population studies, or comparisons with other templates (section
\ref{comparisons}). However, this range in percentiles enables the
reader to see the range of SEDs which are included in each stack, and
is representative of how well any individual galaxy within a
particular bin may be expected to conform to the median template. The
large dispersion among the models at optical\slash ultraviolet
wavelengths is due to the varying degrees of dust attenuation
affecting the different intrinsic stellar populations, though as the
unattenuated SEDs in figure \ref{unat_comparisons} show, the effective
age of the stellar population also plays a role. In contrast, the
comparable dispersion in the sub-millimetre wavelength regime
($\lambda > 80$\,$\mu$m) is due to the varying dust properties
(e.g. temperatures, relative weights of components) of the best-fit
models.

The mid-infrared wavelength range (between $\sim 6$ and $60$\,$\mu$m),
shows the largest dispersion, and this is expected since we lack good
observational data at this time. Constraints in this region come from
the upper limits at 12, 25 and 60\,$\mu$m from \iras\ (though these
are often weak constraints), the priors fed into the stochastic far
infrared template library (which are based on observations with \iras,
ISO and Spitzer), and the energy balance criterion discussed in
Section \ref{modelling}. Mid-infrared data from the {\it WISE}
satellite will dramatically improve this situation, especially at $z <
0.2$ due to the large fraction of 250\,$\mu$m sources detected by {\it
  WISE} (Bond et al., 2012).

\section{Comparison with existing models}
\label{comparisons}

We now compare our binned SEDs to widely-used panchromatic SED
templates, such as the empirical templates described in Chary
\&\ Elbaz (2001, hereafter CE01), Dale \&\ Helou (2002, hereafter
DH02), or Rieke et al. (2009, hereafter R09). Each of these models has
a strong link between the dust luminosity, and the shape of the SED; a
link we have found to be weaker in our sample (Figure
\ref{ssfr_comparisons}).

The CE01 templates are derived as a function of their infrared
luminosity, and are designed to reproduce the SEDs of existing
\iras-selected galaxies. The data-sets used are sparsely-sampled and
heterogeneous, using up to one hundred local galaxies at any given
wavelength from 0.44\,$\mu$m through the {\it Infrared Space
  Observatory} ({\it ISO}) and \iras\ bands, and out to 850$\mu$m with
SCUBA, albeit with no coverage between 170 and 850\,$\mu$m.

The R09 templates are based on a variety of input imaging and
spectroscopy, including full optical photometry for 11 luminous and
ultra-luminous infrared galaxies (LIRGs and ULIRGs) from the NASA
extragalactic database (NED), 2MASS, \iras, {\it Spitzer} and $ISO$,
as well as the GALEXEV models from Bruzual \&\ Charlot (2003). The R09
models are binned in luminosity between $9.75 < \log_{10}
(\ldust\ \slash L_\odot) < 13.0$.

The DH02 models build upon the models of Dale et al. (2001), using the
sample of 69 ``normal'' galaxies, defined according to their optical
luminosities and Hubble types in Dale et al. (2000). The models have a
wavelength range from 3\,$\mu$m extending to radio wavelengths,
derived using data from the $ISO$ (between 52 and 170\,$\mu$m) and
SCUBA (at 450 and 850\,$\mu$m) to extend the observational constraints
from 3 to 850\,$\mu$m.  These models represent ``global'' spectra from
superpositions of local galaxy SEDs, assuming a power-law distribution
for dust mass over intensity of the interstellar radiation field
(ISRF), $U$, which is normalised such that $U = 1$ for the local ISRF
(with values spanning $0.3 \le U \le 10^5$). The templates are
constructed such that $dM_d(U) \propto U^{-\alpha} dU$, where $M_d(U)$
is the dust mass heated by an ISRF with intensity $U$, and the
exponent ($\alpha$) defines the relative contributions of the
individual local galaxy SEDs to each model spectrum. DH02 specify that
it is those models with $1 < \alpha < 2.5$ which describe the range of
normal galaxies.

\begin{figure}
  \centering
  \includegraphics[height=0.99\columnwidth]{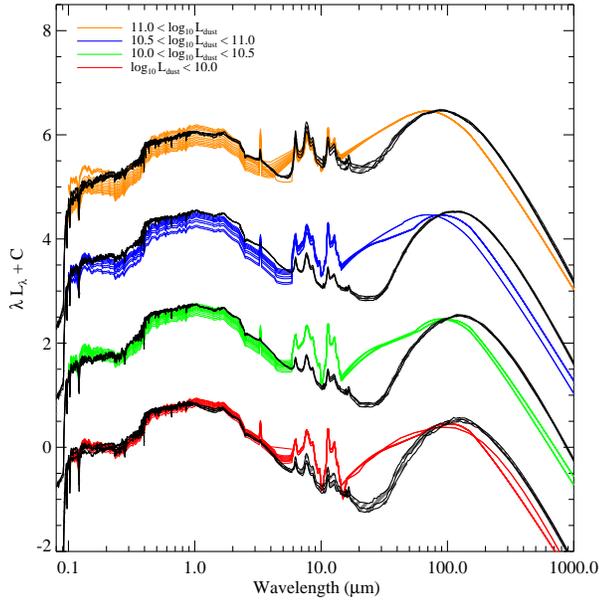}
  \caption{Comparison of our median, transmitted, and
    luminosity--binned templates (solid black lines, with the
    uncertainty on the median template SED in each stack -- defined
    using the median statistics method of Gott et al., 2001 --
    indicated by the shaded regions) with the templates of Chary \&
    Elbaz (2001 -- CE01, shown in colour and with the luminosity
    bounds indicated by the legend). Each set of transmitted models
    with the same range in dust luminosity (indicated by the colours)
    has been normalised to have the same mean between 6
    \&\ 500\,$\mu$m, and artificially offset in the vertical direction
    to prevent the models from overlapping.  }
  \label{modelcomp}
\end{figure}

In Figure \ref{modelcomp}, we overlay the CE01 models (colours) on the
median transmitted templates from our sample (black lines with grey
shaded regions to indicate the uncertainty on the median template SED
following the method of Gott et al. 2001 -- as opposed to the
variation across 16$^{\mathrm{th}}$-84$^{\mathrm{th}}$ percentiles of
the stacked ensemble of galaxy SEDs as a function of wavelength in
each bin as in figure \ref{sSFR_variation_v2}), with each overlapping
set of SEDs corresponding to the same range in far-infrared
luminosities. We normalise the SEDs with corresponding values of
luminosity to the mean of each SED between 6.0 and 500\,$\mu$m in
wavelength, and arbitrarily offset them from one another in the
vertical direction for ease of comparison. Whilst there is generally
good agreement between the models at optical wavelengths, the models
differ considerably in the far-infrared, where each CE01 model peaks
at shorter wavelengths (indicative of a hotter effective dust
temperature) than the corresponding {\it H}-ATLAS stacked SED. Whilst
there is some small bias in our stacked SED in the highest two
luminosity bins (see Section~\ref{trends}), this is not an issue for
the bins at $\log_{10} (\ldust \slash L_\odot) < 10.5$ and yet the
differences between the CE01 and \hatlas\ templates persist. The small
bias in our stacking at higher luminosities (figure
\ref{fig:ldust_bins_bias_check}) is also not large enough to account
for the differences in the $10.5< \log_{10} (\ldust \slash L_\odot)
<11.0$ bin. We also note that the sub-mm portion of the \hatlas\ SEDs
is well determined for all sources due to the high quality {\it
  Herschel} SPIRE data that form the foundation of this study; the
differences at these wavelengths relative to the CE01 templates
persist across the full luminosity range of our sample.

The largest disagreement between the two sets of models is at
mid-infrared wavelengths, however as lack observations in this region
of the SED the current mid-IR discrepancy with other templates is not
significant; a full analysis with WISE data will be required to see if
these differences persist.

\begin{figure}
  \centering
  \includegraphics[height=0.99\columnwidth]{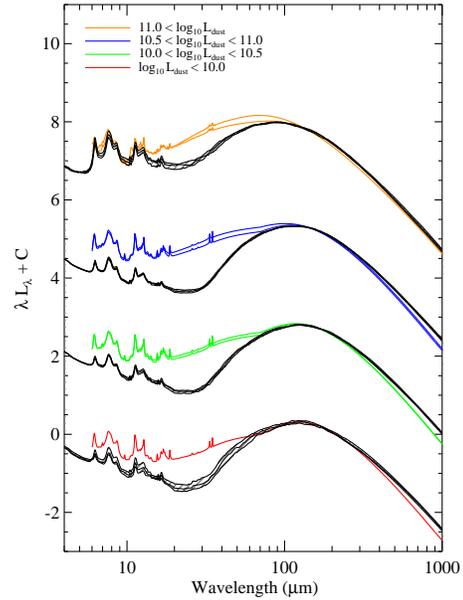}
  \caption{Comparison between the Rieke et al. (2009) and our template
    SEDs binned according to luminosity. As in Figure \ref{modelcomp},
    each set of SEDs is normalised between 100 \&\ 500\,$\mu$m (the
    Rieke et al models are not defined in the optical), and
    arbitrarily offset, with our templates and uncertainties indicated
    by the solid line and dashed areas, respectively. Once more the
    luminosities of the R09 templates are shown by the colour,
    tabulated according to the legend. Although the R09 templates are
    reasonably similar to the new {\it H}-ATLAS templates at the lower
    dust luminosities, at higher dust luminosities they become
    increasingly warmer than the \hatlas\ templates.}
  \label{rieke_modelcomp}
\end{figure}

In Figure \ref{rieke_modelcomp}, we compare our templates with those
of R09, this time normalised between 100 \&\ 500\,$\mu$m. The R09
templates are supplied only at $\lambda > 4$\,$\mu$m, and so the range
in wavelength values is smaller than for the other three sets of
models. We find that the R09 models have similar dust temperatures to
our models, as demonstrated by the similar peak wavelengths of the
far-IR SED, but that they all have considerably brighter mid-infrared
emission than our templates. Again, we cannot comment further on this
discrepancy until we have been able to consider the WISE data in the
fitting.

\begin{figure}
  \centering 
  \includegraphics[height=0.99\columnwidth]{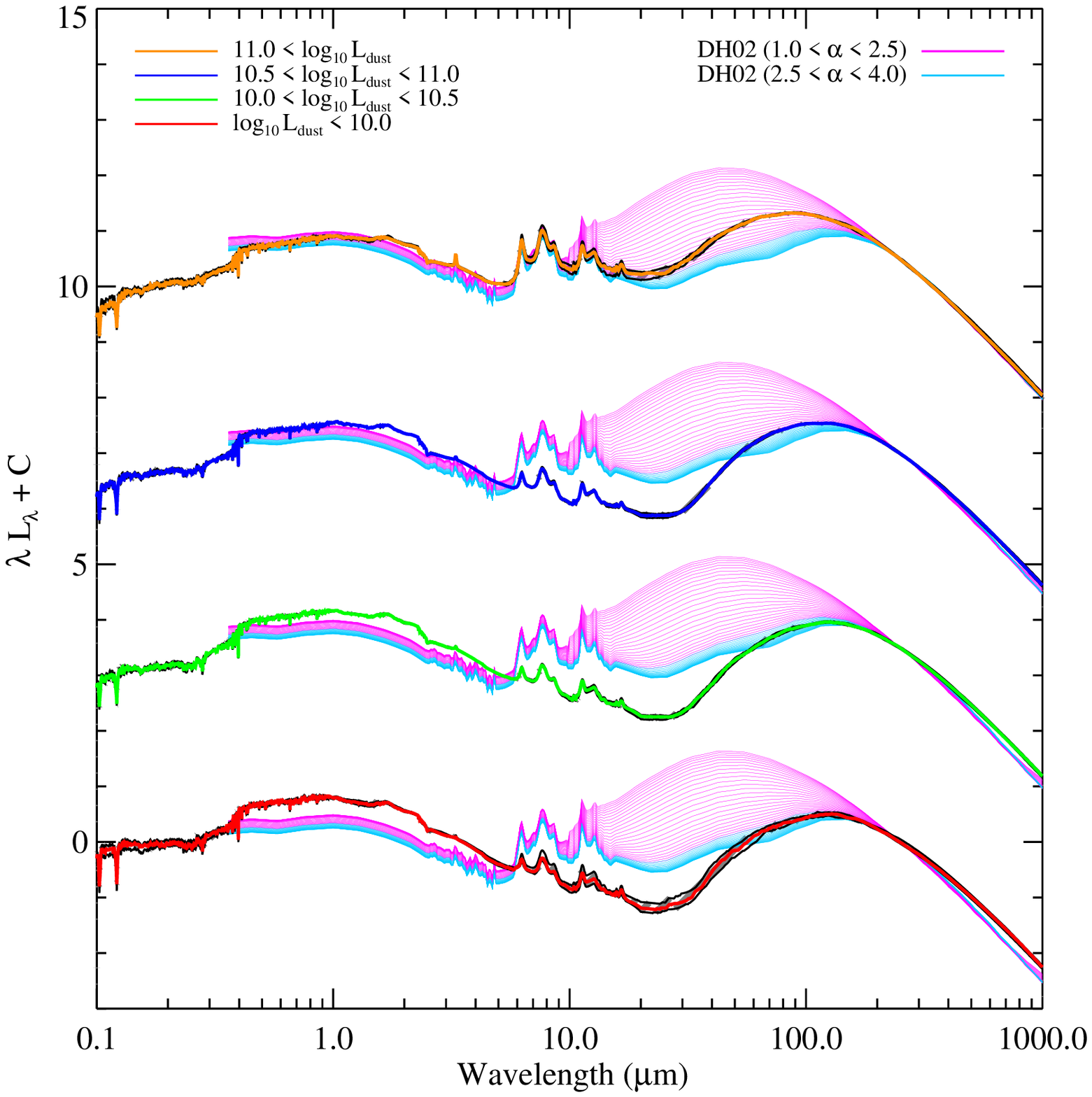}
  \caption{A comparison between our templates and those of Dale
    \&\ Helou (2002). The DH02 templates are shown in magenta for the
    range of $\alpha$ values thought by DH02 to describe normal
    galaxies, and in light blue for the most quiescent galaxies in the
    DH02 library (with $2.5 < \alpha < 4.0$). The luminosities of our
    templates are shown by the colour of the solid lines bounded by
    the shaded regions indicating the uncertainty associated with each
    template. For ease of comparison, each set of models being
    compared is normalised around 250\,$\mu$m, and arbitrarily offset
    in the vertical direction. It is clear that the DH02 templates
    contain considerably hotter dust than we observe in 250\,$\mu$m
    selected galaxies in {\it H}-ATLAS. }
  \label{dale_modelcomp}
\end{figure}

In Figure \ref{dale_modelcomp}, we compare our models with the DH02
templates, which are empirically constrained at $\lambda > 3$\,$\mu$m.
Rather than limit the comparison to the range of values of $\alpha$
which DH02 suggest span the range of normal galaxies in their input
sample ($1.0 < \alpha < 2.5$, magenta in Figure \ref{dale_modelcomp}),
we also compare our SEDs with the more quiescent range of DH02
templates spanning $2.5 < \alpha < 4.0$ (light blue in Figure
\ref{dale_modelcomp}). We normalise the models at 250\,$\mu$m for ease
of comparison. Whilst there are DH02 models which can match our
templates at the highest dust luminosities, at lower values the DH02
models suggest the presence of considerable hot dust components that
are not observed in our sample of 250\,$\mu$m-selected galaxies, and
the $\alpha > 2.5$ models are required to reproduce the far-IR
temperatures that we observe. Since the DH02 models are binned by
$\alpha$ rather than dust luminosity, we indicate the luminosity of
our models being compared using the colours as indicated in the
caption to Figure \ref{dale_modelcomp}, with the DH02 models overlaid.

These templates may be useful for studies of FIR selected samples in
the $z<0.5$ universe. They are also useful ingredients for any
evolutionary model which purports to explain the low redshift
FIR/sub-mm populations detected by {\it Herschel}. They do not,
however, appear to be representative of the high redshift ($z > 1$)
population detected in {\it H}-ATLAS (e.g. Lapi et al., 2011). The
template chosen for any SED fitting should always match the sample
under investigation as selection wavelength and redshift can have an
important impact on the SED types prevalent in a sample.

Irrespective of the templates to which we compare our model SEDs, the
comparative lack of hot dust in our stacks is striking. Though we plan
to investigate these details in future using {\it WISE} observations,
the contribution of the mid-infrared to the total dust energy budget
is not dominant; the impact of the {\it WISE} data near the peak of
the far-infrared SED is likely to be small.

We intend to make these new template SEDs, binned according to their
properties available to the community for further analysis, and
application to other data sets via the {\it H}-ATLAS
website\footnote{\url{http://www.h-atlas.org}} and the author's
webspace\footnote{\url{http://star.herts.ac.uk/~dsmith/}}.

\section{Conclusions}
\label{conclusions}

We have determined SEDs for a total of 1402 250$\mu$m-selected
galaxies from the Herschel-ATLAS science demonstation catalogue with
reliable counterparts and matched aperture photometry from the $u$- to
$K$-bands from the GAMA database. We also include far and near
ultraviolet data from the {\it GALEX}-GAMA survey, as well as the {\it
  H}-ATLAS data from PACS and SPIRE. Of these 1402 galaxies, 1289 are
well described by the model of DCE08, and we use these SEDs and the
model parameter probability density functions derived from the energy
balance SED-fitting, to determine the properties of these 250\,$\mu$m
selected galaxies out to $z=0.5$.
   \begin{enumerate}
   \item Studies of the colours of galaxies in our sample, and a suite
     of simulations, suggest that our sample is representative of the
     broader population of 250\,$\mu$m galaxies out to $z < 0.35$.
   \item The average \hatlas\ galaxy in our sample has a star
     formation rate of $\sim$\,4.0~M$_\odot $yr$^{-1}$, L$_{dust}
     \approx 6.4 \times 10^{10}$ ~L$_\odot$, and a dust to stellar
     mass ratio of $\sim$0.4 per cent, while the median redshift is
     $z=0.24$.
   \item Our results support the idea that \iras\ and
     \hatlas\ selected galaxies in the local Universe are different
     populations. Due to its lack of sensitiity and short selection
     wavelength, \iras\ preferentially selected galaxies with larger
     warm dust content, and consequently these galaxies are more
     luminous in the infrared for a given mass of dust. The
     \hatlas\ selection at 250\,$\mu$m is less biased towards strongly
     star forming objects over the same redshift range because of the
     longer selection wavelength and far superior sensitivity compared
     to \iras. \iras\ misses a population of massive dusty galaxies
     with colder dust temperatures, as was shown previously by
     Vlahakis, Dunne \& Eales (2005).
   \item The correlation between star formation rate and dust mass
     presented in da Cunha et al. (2010) is also present in this
     sample, although {\it Herschel} ATLAS-selected galaxies contain
     larger dust masses for a given star formation rate compared to
     the \iras\ selected sample of dC10. There is also a correlation
     between specific dust mass (\mdust\slash\mstar) and SSFR, which
     is not well reproduced by simple chemical and dust evolution
     models.
   \item The specific star formation rate of lower mass galaxies
     ($\log_{10} M_{\mathrm{star}}\slash \Msolar < 10.2$) is higher
     than that of the most massive galaxies in our sample (those with
     $\log_{10} M_{\mathrm{star}}\slash \Msolar > 10.6$) at all
     redshifts, supporting previous results that lower mass galaxies
     dominate the star formation rate density in the local universe.
   \item Stacks of SEDs show that sSFR is the strongest galaxy
     property driving the SED shape across both the UV/optical and
     FIR, as first noticed in the smaller sample of DCE08. Trends with
     \ldust\ are much weaker since smaller mass galaxies will have low
     \ldust\ and yet could have the highest values of sSFR. We see a
     signficant trend in this sample for galaxies to have more
     obscured optical/UV SEDs and higher reprocessed fractions with
     increasing redshift.
   \item Existing templates for panchromatic SEDs of galaxies show
     shorter FIR peaks and excess mid-IR emission compared to median
     stacked SEDs of galaxies in our \hatlas\ sample (binned by
     \ldust) although the mid-IR discrepancy is not significant at
     this time due to our lack of mid-IR data to constrain this part
     of the SED. Templates from Rieke et al. (2009) are the closest
     match to ours in terms of the FIR properties although they still
     predict a warmer FIR peak at the highest luminosities compared to
     our findings. We provide a new set of panchromatic SED templates
     from the UV--sub-mm, to enable more representative studies of
     dusty galaxies in the local Universe in the {\it Herschel} era.
   \item Data from the \wise\ satellite, which covers the wavelength
     range between 3 and 23\,$\mu$m will provide valuable constraints
     to the mid-IR and PAH features, as well as the hot dust component
     of these local galaxies. It will be interesting to see if the
     differences between templates in the mid-IR region persists when
     these data are included in the fitting.
   \end{enumerate}

\section*{Acknowledgments}

The authors wish to thank the anonymous referee for his\slash her
tireless work and insightful comments, which have substantially
improved this paper. The {\it Herschel}-ATLAS is a project with {\it
  Herschel}, which is an ESA space observatory with science
instruments provided by European-led Principal Investigator consortia
and with important participation from NASA. The {\it H}-ATLAS website
is \url{http://www.h-atlas.org/}. GAMA is a joint
European-Australasian project based around a spectroscopic campaign
using the Anglo-Australian Telescope. The GAMA input catalogue is
based on data taken from the Sloan Digital Sky Survey and the UKIRT
Infrared Deep Sky Survey. Complementary imaging of the GAMA regions is
being obtained by a number of independent survey programs including
{\it GALEX} MIS, VST KIDS, VISTA VIKING, \wise, GMRT and ASKAP
providing UV to radio coverage. GAMA is funded by the STFC (UK), the
ARC (Australia), the AAO, and the participating institutions. The GAMA
website is \url{http://www.gama-survey.org/}.  This work used data
from the UKIDSS DR5 and the SDSS DR7. The UKIDSS project is defined in
Lawrence et al. (2007) and uses the UKIRT Wide Field Camera (WFCAM;
Casali et al. 2007). Funding for the SDSS and SDSS-II has been
provided by the Alfred P. Sloan Foundation, the Participating
Institutions, The National Science Foundation, the U.S. Department of
Energy, the National Aeronautics and Space Administration, the
Japanese Monbukagakusho, the Max Planck Society and the Higher
Education Funding Council for England. The Italian group acknowledges
partial financial support from ASI\slash INAF agreement n. I/009/10/0.

\appendix

\section{Stacking PDFs and choice of cold temperature prior} 
\label{tcold_prior_appendix}

\subsection{Derivation and interpretation of stacked PDFs}

In this paper we make considerable use of stacked PDFs, which are our
best estimates of the distribution of the values of a given parameter
amongst the sources in a sample, convolved with our ability to
constrain them. To see why this latter point might be important,
imagine we have individual PDFs for a certain parameter which has very
weak constraints, all the PDFs will therefore appear flat. The
distribution of the medians of these PDFs (i.e. the median-likelihood
estimates for that particular parameter) will be centered near the
middle of the range (since the PDFs are all individually flat) and
will have a narrow scatter (since all medians are almost the same). If
we only considered the median-likelihood values, we might na\"ively
assume that we know the parameter distribution for the population
quite accurately, and that there was little scatter within the
population, even though in truth we merely had little ability to
constrain that parameter. This is an extreme example and does not
apply to the parameters we are exploring in this paper, but it
illustrates why we wish to present the stacked marginalised PDFs and
not simply the distributions of the median likelihood values.

To derive values of stacked PDFs, we start with the values in each bin
of the ensemble of PDFs that we wish to stack. For the value of the
stacked PDF in each bin we use the mean of the ensemble of values in
that bin. To estimate the error associated with the derived stacked
PDF in each bin, we simply use a symmetric value corresponding to the
mean of the 16-84th percentiles of the cumulative frequency
distribution of the values for each galaxy in that bin.

\subsection{Choice of \tcold\ prior distribution}

When choosing a prior distribution for a particular parameter in the
stochastic libraries of SEDs, we must ensure that our choice of prior
does not bias our results. This was of particular concern for the
prior on \tcold\, since the stacked PDFs do not always show a peak in
the range of the prior, but for some samples, increase towards the
lowest bounds of the prior. We wanted to determine whether or not this
was an indication that we should use a broader prior on \tcold. To
address this issue, we considered the properties of the galaxies in
our PACS-complete sample (see Section \ref{pacscomp}), which as we
showed in section \ref{pacscomp} and figure \ref{pacs_completeness},
are representative of the full range of colours in \hatlas\ sources,
and by virtue of their being detected in our PACS data, have our best
constraints on \tcold.

\begin{figure}
  \centering
  \includegraphics[width=0.99\columnwidth, angle=0]{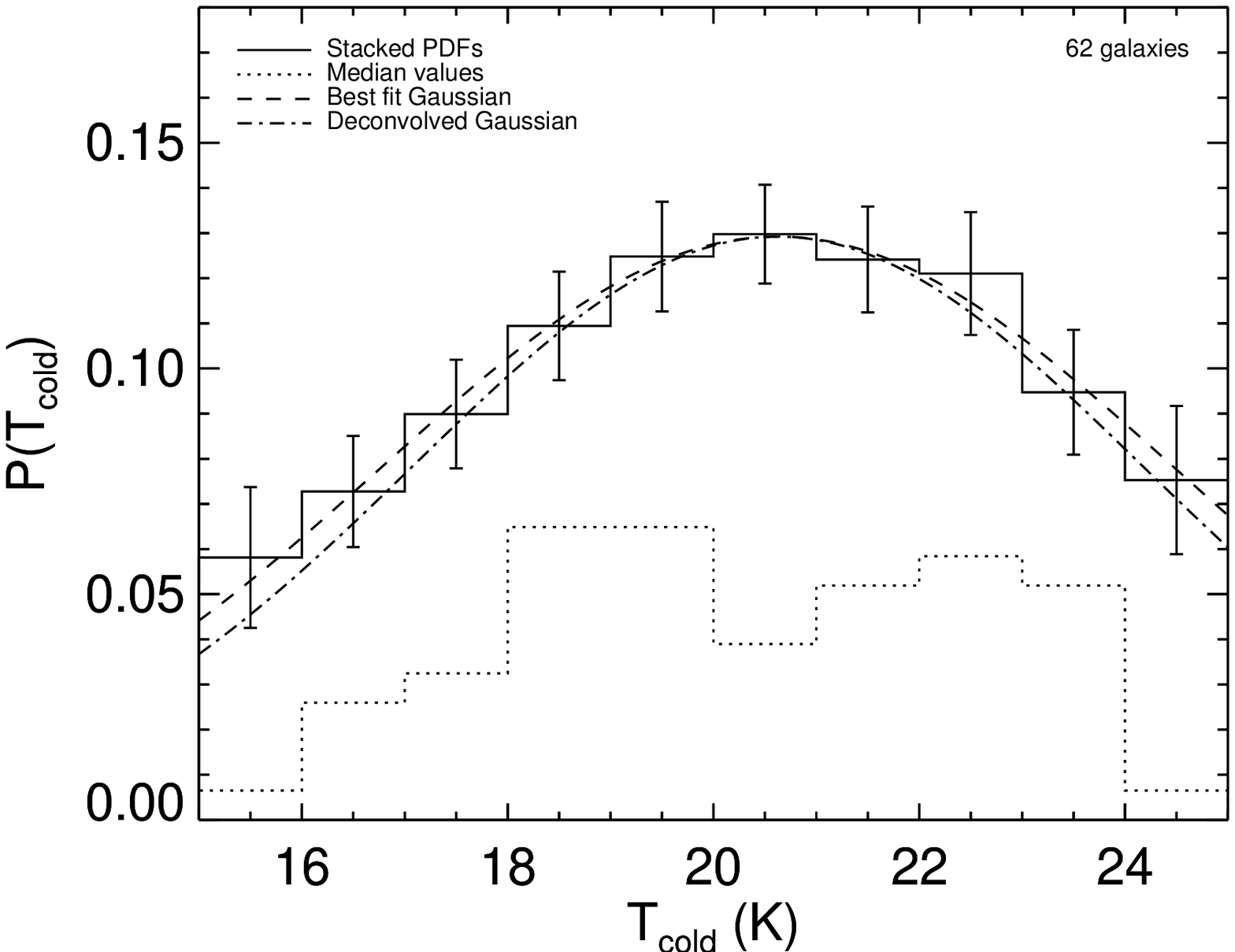}
  \caption{The stacked PDFs (solid line, with error bars) for those 59
    galaxies with PACS 160\,$\mu$m detections with $S_{250} >
    120$\,mJy at $z < 0.20$. We overplot the median values of the
    individual PDFs for these galaxies (dotted histogram), along with
    the best-fit Gaussian approximation of the PDF. The deconvolved
    version (described in the text) is shown in the dot-dashed line,
    and has a standard deviation of $3.54\pm0.41$\,K.}
  \label{pacs_complete_pdfstack}
\end{figure}

We stacked the PDFs for these galaxies, and the results are shown in
Figure \ref{pacs_complete_pdfstack}, in which the best fit Gaussian
approximation to the stacked PDF is shown as the dashed line, and a
histogram of the median values of each individual PDF that went in to
the stack in dotted lines. The best-fit Gaussian model of the stacked
PDF has a standard deviation of $3.84 \pm 0.41$\,K. In order to assess
the true range of \tcold\ that is present in our sample, we determined
the deconvolved best-fit Gaussian, by subtracting the mean of the
individual $1\sigma$ errors on \tcold\ for each galaxy (1.47\,K$ \pm
0.19$) in quadrature, leaving a $1\sigma$ uncertainty on the range of
\tcold\ in our sample as $3.54 \pm 0.41$\,K. The best fit deconvolved
Gaussian is shown as the dot-dashed line in Figure
\ref{pacs_complete_pdfstack}. We find that 85 percent of the true
\tcold\ PDF lies within the bounds of our temperature prior, with only
$\sim$six per cent of the PDF colder than 15\,K, and approximately 9
per cent warmer than 25\,K.

In that case, why not make a \tcold\ prior which is wider and
therefore encompasses the full range of temperatures possible in the
cold ISM? The reason not to do so is because of the strong non-linear
dependence of the \mdust\ parameter on \tcold\ when the value of
\tcold\ is below $\sim 15$\,K. At such cold temperatures, the SPIRE
bands no longer sample the Rayleigh-Jeans part of the SED (where
\mdust\ scales relatively linearly with \tcold\ ) but are nearer the
peak where the dependence on temperature is steeper.

The fitting becomes more prone to errors on the determination of
\tcold\ at low values, since the energy balance is not much affected
by the addition of very cold dust to the SED (which contributes little
to the total \ldust). Dust colder than 15\,K is essentially invisible
to our model (or any other for that matter), which combined with the
added sensitivity of mass to temperature at low \tcold\ results in an
asymmetry in the error on the dust mass; we have a larger
overestimation of the mass when \tcold\ is underestimated
(statistically likely to happen 50 per cent of the time) compared to
the size of our underestimate in \mdust\ when \tcold\ is
overestimated.

\begin{figure}
  \centering \includegraphics[width=0.99\columnwidth,
    angle=0]{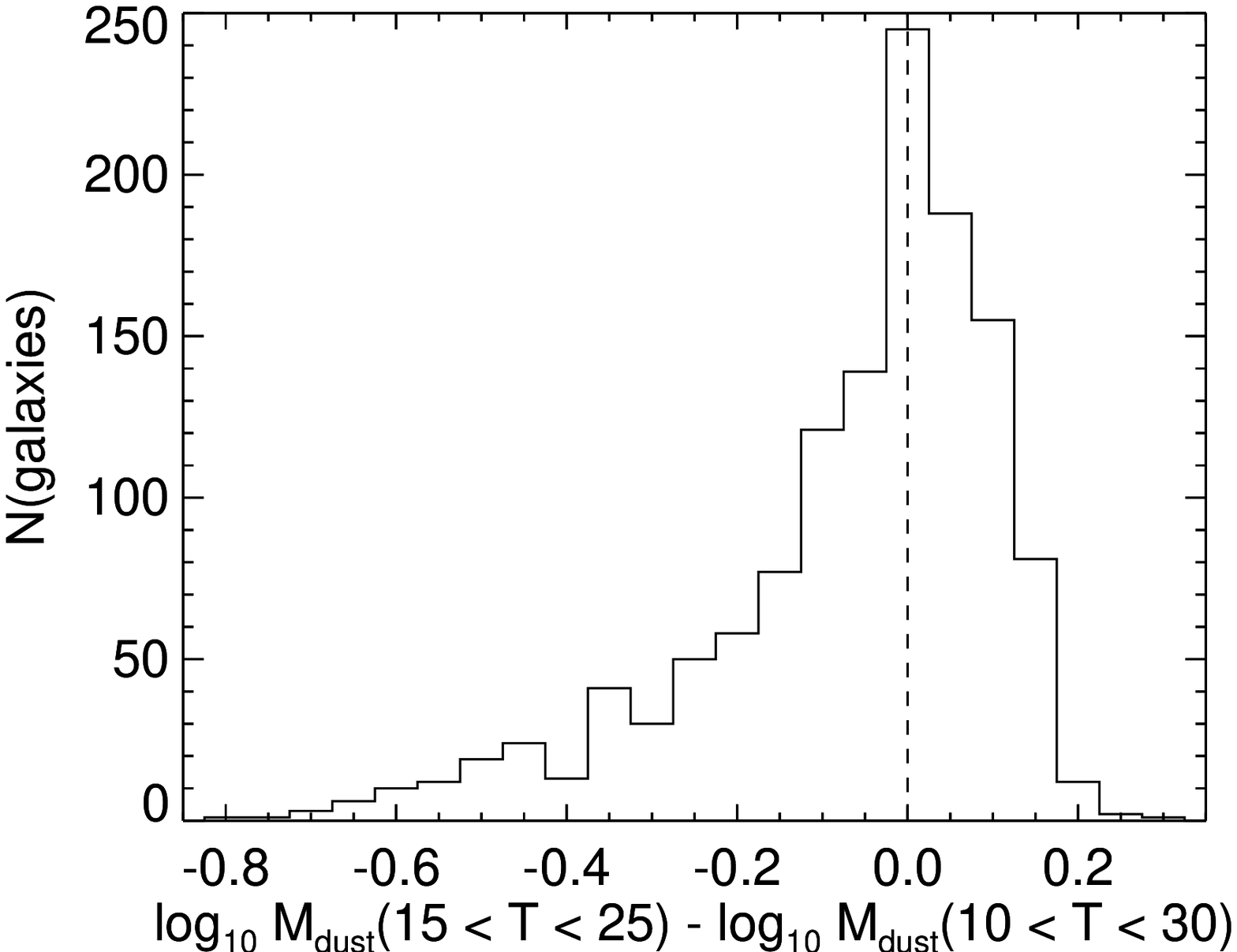}
  \caption{The difference in the dust mass estimates that we determine
    for the galaxies in our sample depending on whether we use the
    broader ($10 < \tcold < 30$) or narrower ($15 < \tcold < 25$)
    prior. The asymmetry in this histogram is caused by a greater
    sensitivity to temperature at cold temperatures.}
  \label{mdust_vs_priors}
\end{figure}

To demonstrate this, we created a hypothetical library of 1000 model
galaxies with Gaussian temperature and mass distributions, and simply
re-estimated the dust mass after adding on simulated measurement
errors to the true temperature distribution, using the relationship
between temperature and dust mass for a modified greybody emission
profile shown in Equation 3 of Dunne et al. (2011). In these simple
simulations, measurement errors introduced a systematic
anti-correlation between the estimated temperature and estimated dust
mass. In Figure \ref{mdust_vs_priors}, we show a histogram of the
difference in our median estimated dust masses, derived using the
broader and narrower \tcold\ prior distributions. The effect is
stronger at the coldest temperatures, which occur more often with the
broader \tcold\ prior, with some dust masses being overestimated by in
excess of 0.5\,dex. In order to limit the effects of this unphysical
bias towards large dust mass estimates in the colder galaxies in our
sample, we decided to use the narrower prior on \tcold.

We confirm the effects on \tcold\ by re-running the fitting using a
wider \tcold\ prior (10--30 K) on sub-samples of galaxies which
represent typical selections within the main analysis and compare
these results to those using the \tcold\ prior from DCE08 (15--25K).

We split our sample into five subsets, limited to $z < 0.2$ to limit
the possible influence of cosmic evolution on our results:

\begin{itemize}
\item galaxies in our PACS-complete sub-sample
\item galaxies detected at $>5\sigma$ in both PACS bands and
  $>3\sigma$ at 350 \&\ 500\,$\mu$m;
\item galaxies with at least one PACS $\ge 5\sigma$ detection;
\item galaxies detected at $\ge 3\sigma$ at 350\,$\mu$m;
\item all galaxies in our sample.
\end{itemize}

\noindent Of course, all galaxies discussed here are detected at $\ge
5\sigma$ at 250\,$\mu$m, since this is how our sample is defined.

In Table \ref{tab:prior_medians} we show the median-likelihood values
of \tcold\ derived from the PDFs for each sub-sample. None of the
stacked PDF median-likelihood \tcold\ values varies by more than
$\sim$0.6\,K when the broader prior is used rather than the narrower
prior; the impact on our global estimates of \tcold\ is therefore
minimal. When we use the broad prior on \tcold, we find that 13 per
cent of our sample have \tcold\ values of $< 15$\,K, approximately in
line with the 6 per cent that we expect from studying the
PACS-complete sample.

\begin{table*}
  \caption{The effects of our choice of prior distribution on our
    median-likelihood estimates of \tcold. The left column indicates
    the selection of each different sub-sample, in addition to the
    requirement that all sources have 5$\sigma$ detections in the
    SPIRE 250\,$\mu$m band. The middle two columns indicate the
    \tcold\ estimates determined for each sub-sample with the two
    prior distributions, while the right hand column shows the
    difference. Each sub-sample has been limited to $z < 0.2$, in
    order to limit the impact of evolution with redshift on these
    values.}
  \begin{tabular}{l|l|l|l}
    \hline
    \hline
    Detections & $10 < \tcold\ < 30$ & $15 < \tcold\ < 25$  & $\Delta(\tcold$, prior$)$ \\
    \hline
    PACS complete sub-sample & 20.66 & 20.61 & 0.05 \\                     
    All PACS $5\sigma$, S350 \&\ S500 $> 3\sigma$ & 21.59 & 21.49 & 0.10 \\
    P100 or P160 $> 5\sigma$ & 21.39 & 21.13 & 0.26 \\                     
    S350 $> 3\sigma$ & 18.61 & 19.19 & -0.58 \\                           
    S250 $> 5\sigma$ only & 19.06 & 19.41 & -0.35 \\
    \hline
    \label{tab:prior_medians}
  \end{tabular}
\end{table*}

When we consider the median temperature estimates with each prior in
table \ref{tab:prior_medians}, we see that the second two samples --
which require $5\sigma$ PACS detections -- have higher median values
due to the removal of the coldest sources from the sub-sample; these
sources are generally undetected in our comparatively shallow PACS
data. The last two sub-samples in table \ref{tab:prior_medians} have
lower values of \tcold, and are mostly undetected by PACS; in
particular, the 350\,$\mu$m selection criterion preferentially picks
the colder sources in the catalogue, with SEDs peaking at longer
wavelengths.

In summary, we have used the DCE08 $15\,K < \tcold\ < 25\,K$
temperature priors for the following reasons:
\begin{itemize}
  \item Simulations and studies of the PACS-complete sample suggest that
    we can explain our data adequately using the DCE08 prior,
  \item The narrower prior also limits the impact of \tcold\ errors on
    \mdust,
  \item Detailed studies using a variety of multiwavelength data and
    modelling techniques see no evidence for substantial cold dust
    components with temperatures lower than 15\,K in the kind of dust
    rich spiral galaxies being investigated here (e.g. Dunne \&\ Eales
    2001, Popescu et al. 2002, Vlahakis, Dunne \&\ Eales 2005, Draine
    et al. 2007, Willmer et al., 2009, Bendo et al., 2010, Boselli et
    al. 2010, Kramer et al. 2010, Bernard et al. 2010).
\end{itemize}

\section{Defining the good fits}
\label{chi2_sims_appendix}

\begin{figure}
  \centering
  \includegraphics[width=0.99\columnwidth, angle=0]{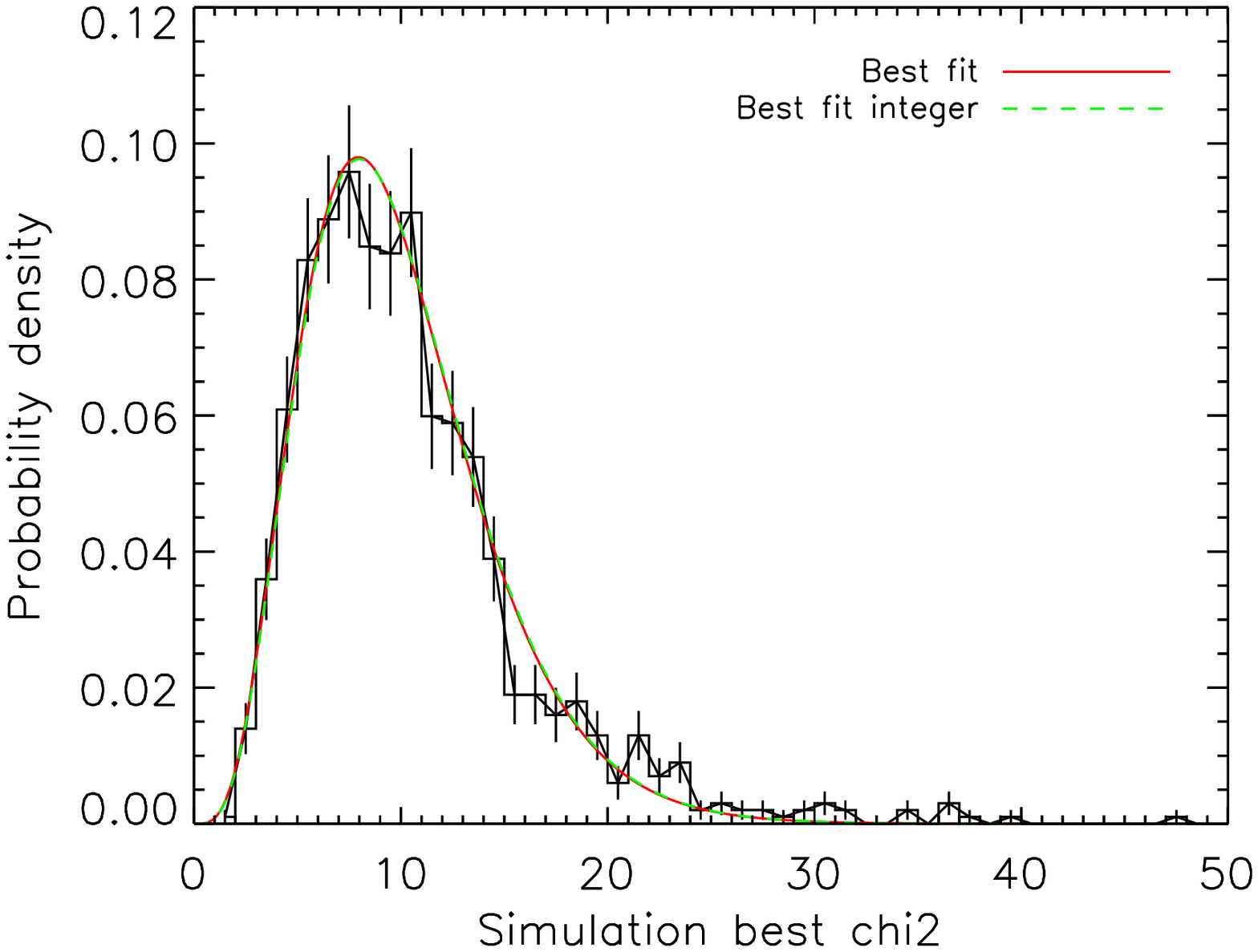}
  \caption{The distribution of $\chi^2$ values for the one thousand
    galaxy simulations referred to in the text. The shape of the
    resulting histogram is well-described by Equation \ref{chi2dist},
    and depends on only one parameter, the number of degrees of
    freedom for that particular combination of input photometry
    ($N_{\mathrm{dof}}$). In this way, $N_{\mathrm{dof}}$ can then be
    determined for each combination of photometry by simply minimising
    Equation \ref{chi2dist}. By fitting a quadratic to these results,
    we may relate $N_{\mathrm{dof}}$ and $N_{\mathrm{bands}}$. The
    resulting relationship is shown in Equation \ref{ndof_eq}. The
    combination of Equations \ref{chi2dist} \& \ref{ndof_eq} allows us
    to define 99\%\ confidence intervals and in turn determine which
    galaxies in our sample are well-described by our model.}
  \label{dofsim}
\end{figure}

We took a selection of best-fit model SEDs, and varied their
photometry according to a set of Gaussian distributions with a median
of zero and a standard deviation equal to the minimum photometric
error in each band (i.e. the values added in quadrature, as defined in
Section \ref{sec:data}). These values were chosen since they were the
dominant source of error for bright sources. We then removed a sub-set
of photometry to reflect the heterogeneity within our real data set,
and re-calculated their properties 1000 times each. The resulting
histograms of $\chi^2$ values (e.g. Figure \ref{dofsim}) enabled us to
estimate the number of degrees of freedom given that particular
sub-set of photometry, by performing a simple minimisation of Equation
\ref{chi2dist} to the derived $\chi^2$ probability density functions:

\begin{equation}
P \propto \frac{1}{2^{(N_{\mathrm{dof}}/2)}\Gamma(N_{\mathrm{dof}}/2)}\left(\chi^2\right)^{(N_{\mathrm{dof}}/2) - 1}\exp^{-\chi^2/2} 
\label{chi2dist}
\end{equation}

\noindent where $\Gamma(x)$ represents the Gamma function,
$N_{\mathrm{dof}}$ the number of degrees of freedom, and $\chi^2$ the
median $\chi^2$ value for each bin. Due to the relatively small
redshift range covered by our sample, we average each solution, and
determine that the relationship between the number of degrees of
freedom ($N_{\rm dof}$) and the number of photometric bands with
measurements ($N_{\rm bands}$) is given by Equation \ref{ndof_eq}:

\begin{equation}
\begin{split}
N_{\mathrm{dof}} \approx (-2.820 \pm 0.745) + (0.661\pm0.132) N_{\mathrm{bands}} \\+ (7.91\pm5.50 \times 10^{-3}) N_{\mathrm{bands}}^2.
\end{split}
\label{ndof_eq}
\end{equation}

\noindent With this information, we are then able to use Equation
\ref{chi2dist}, in conjunction with the number of degrees of freedom
estimate from Equation \ref{ndof_eq}, to determine a 99 per cent
confidence interval on $\chi^2$. Those galaxies outside the interval
on $\chi^2$ have less than 1 per cent chance of being consistent with
our model. In this way we may remove those ``bad fits'' from further
analysis. We derived equation \ref{ndof_eq} using galaxies with
between 6 and 19 detections, so these values constitute the bounds on
$N_{\mathrm{bands}}$ over which we believe it is valid.

Finally, in comparing our simulations to our real sample, we note that
in contrast to our simulations, the photometric errors used in our SED
fitting are not strictly Gaussian due to problems associated with
e.g. deblending or calibration issues, particularly in regions
neighbouring saturated stars, however the difference is not expected
to be large.

\section{Additional tests for bias in the fitting}
\label{bias_tests}

Due to the complexity of the energy-balance SED fitting method, we
wanted to perform additional checks for bias in the derived
parameters, and determine which parameters depend on one
another. 

\begin{figure*}
  \centering
  \subfigure[$L_{\mathrm{dust}}$]{\includegraphics[width=0.9\columnwidth]{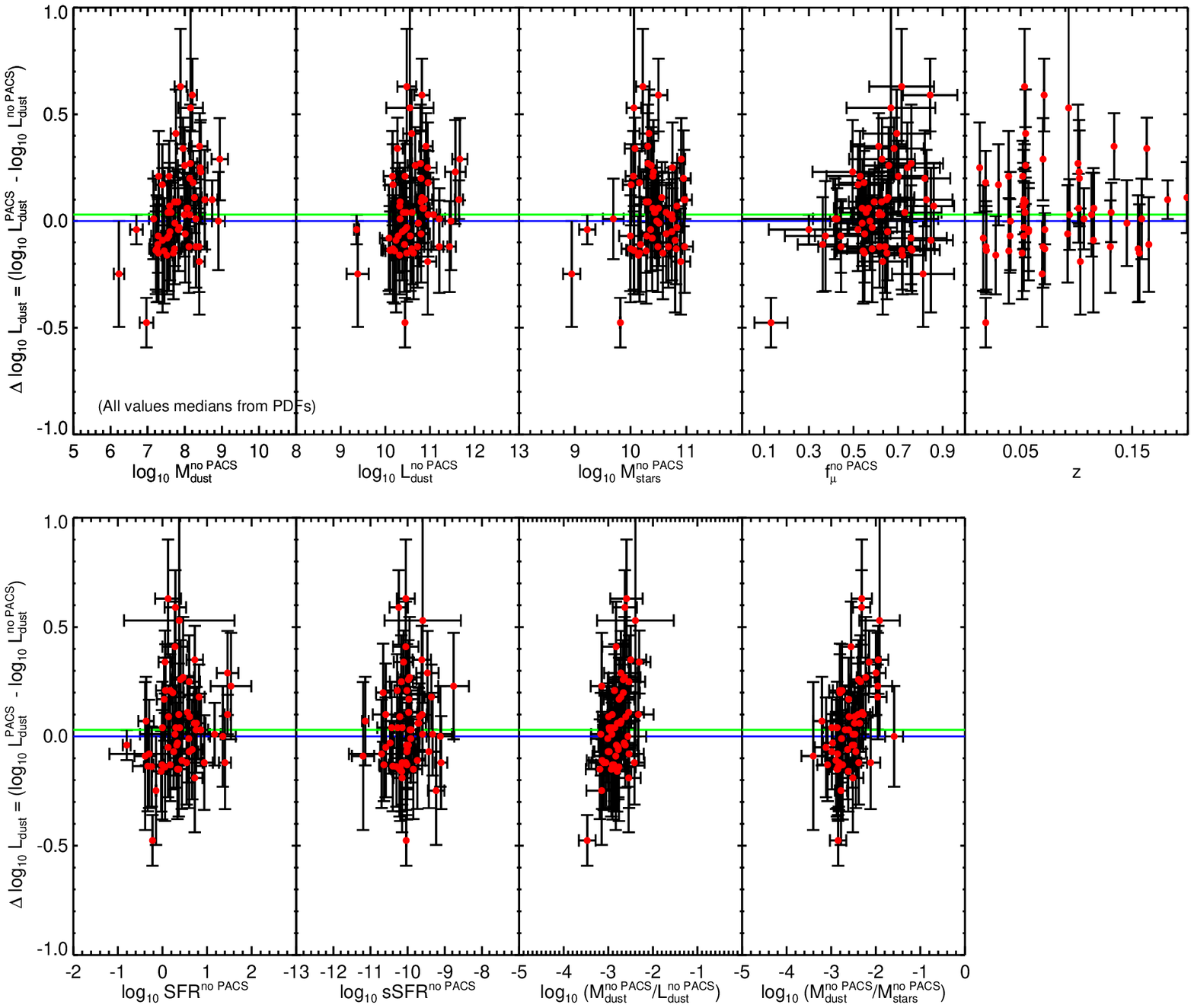}}
  \hspace{1cm}
  \subfigure[$M_{\mathrm{dust}}$]{\includegraphics[width=0.9\columnwidth]{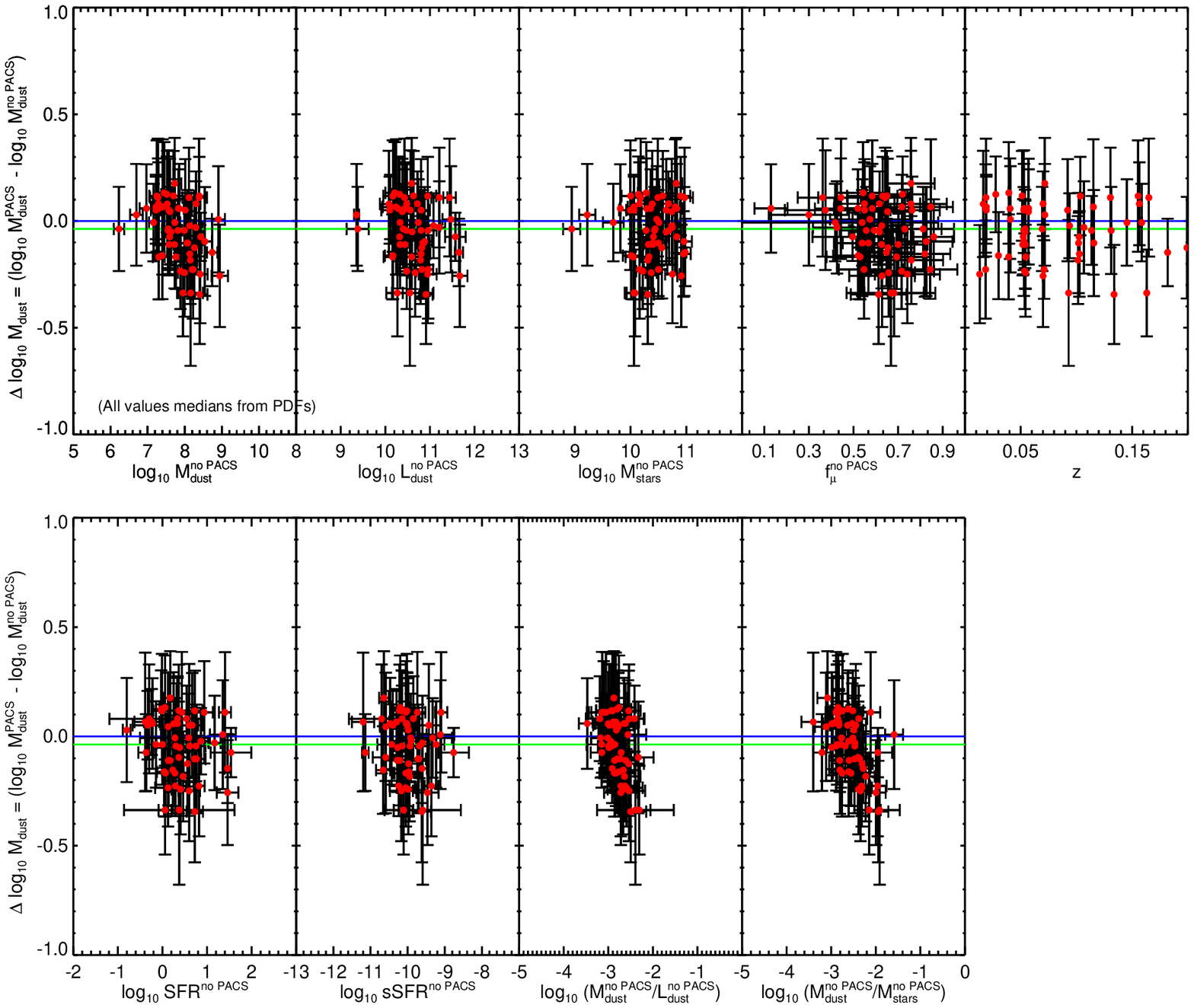}}
  \subfigure[$f_{\mu}$]{\includegraphics[width=0.9\columnwidth]{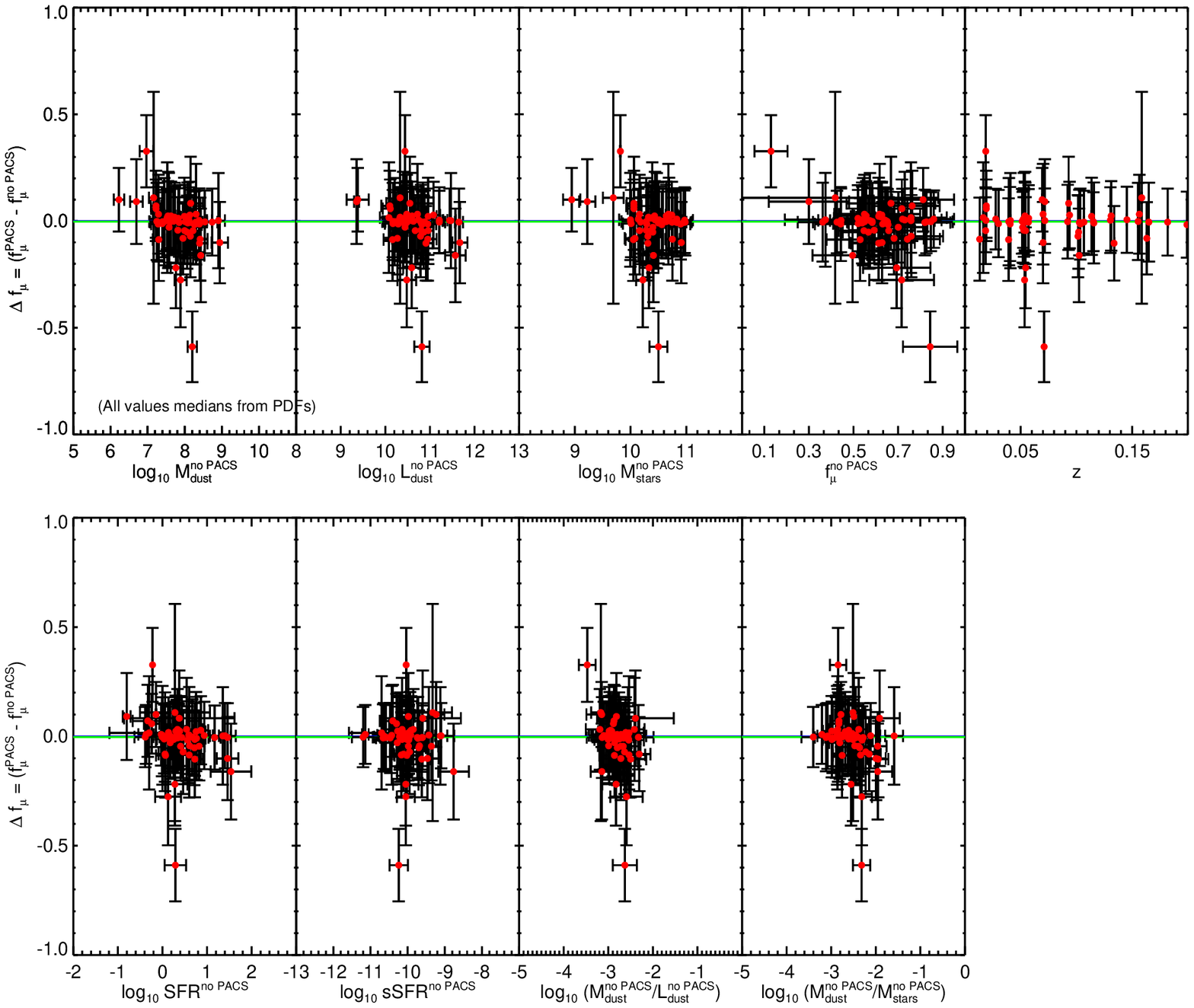}}
  \hspace{1cm}
  \subfigure[\mstars]{\includegraphics[width=0.9\columnwidth]{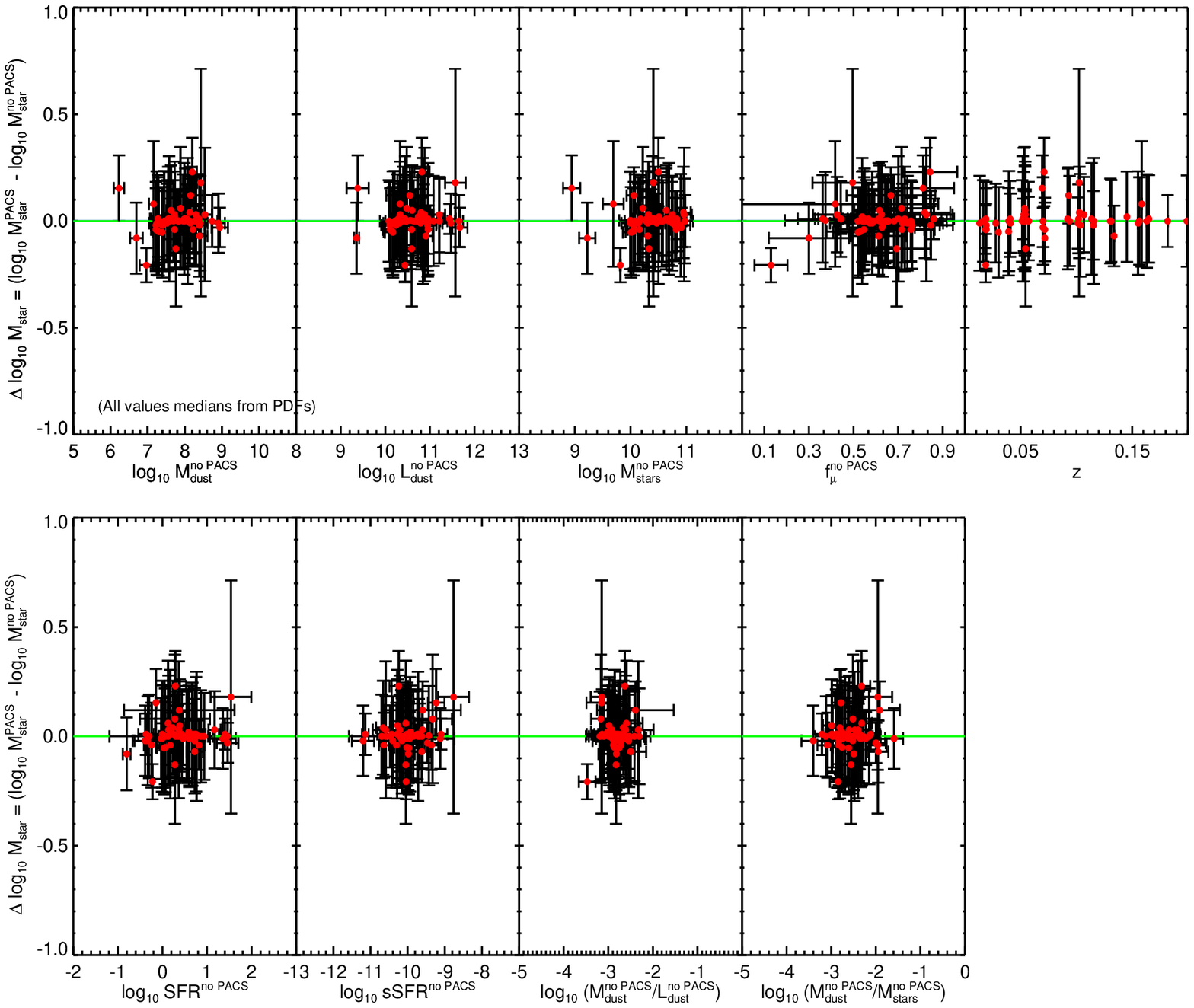}}
  \subfigure[$SFR$]{\includegraphics[width=0.9\columnwidth]{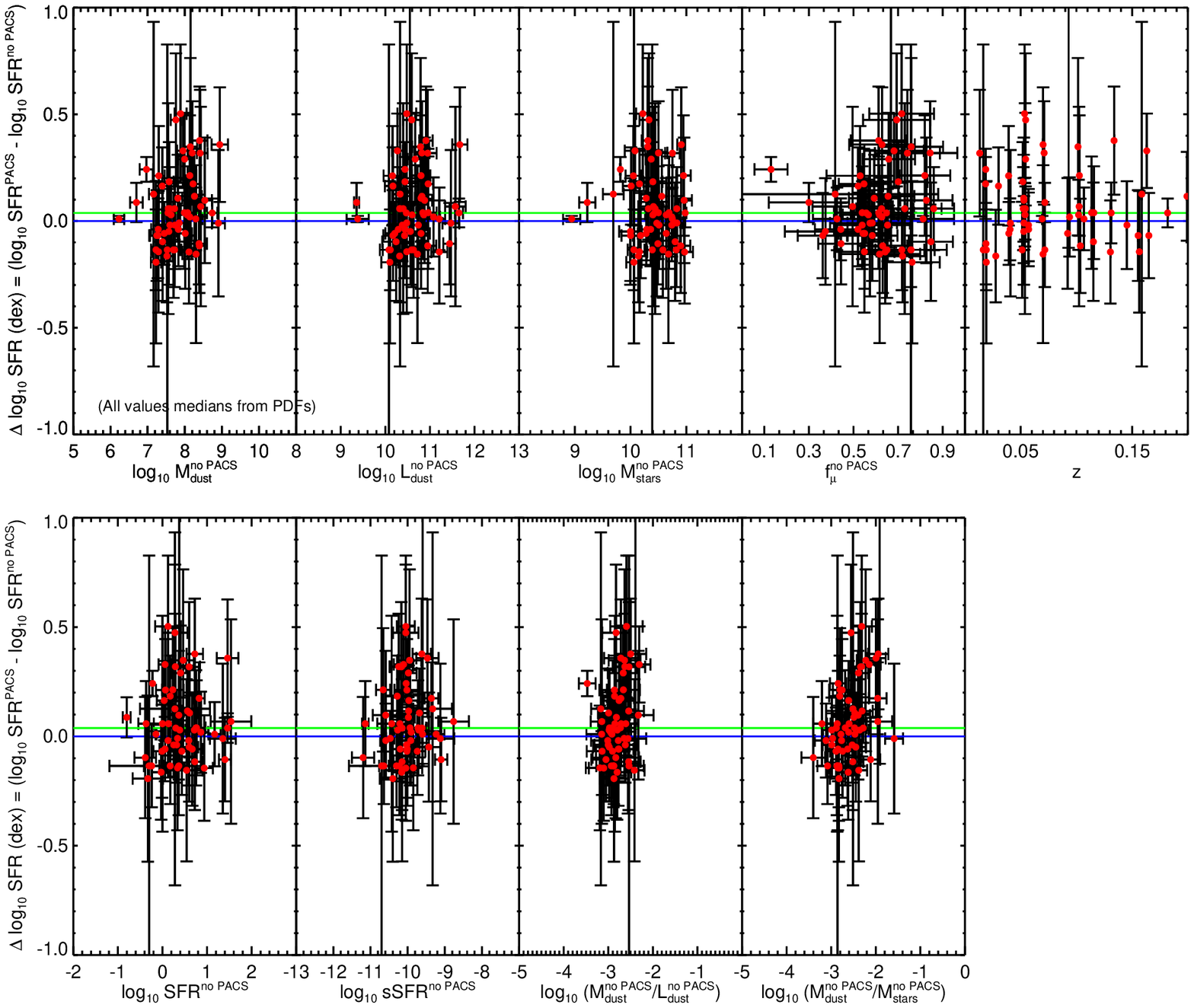}}
  \hspace{1cm}
  \subfigure[$sSFR$]{\includegraphics[width=0.9\columnwidth]{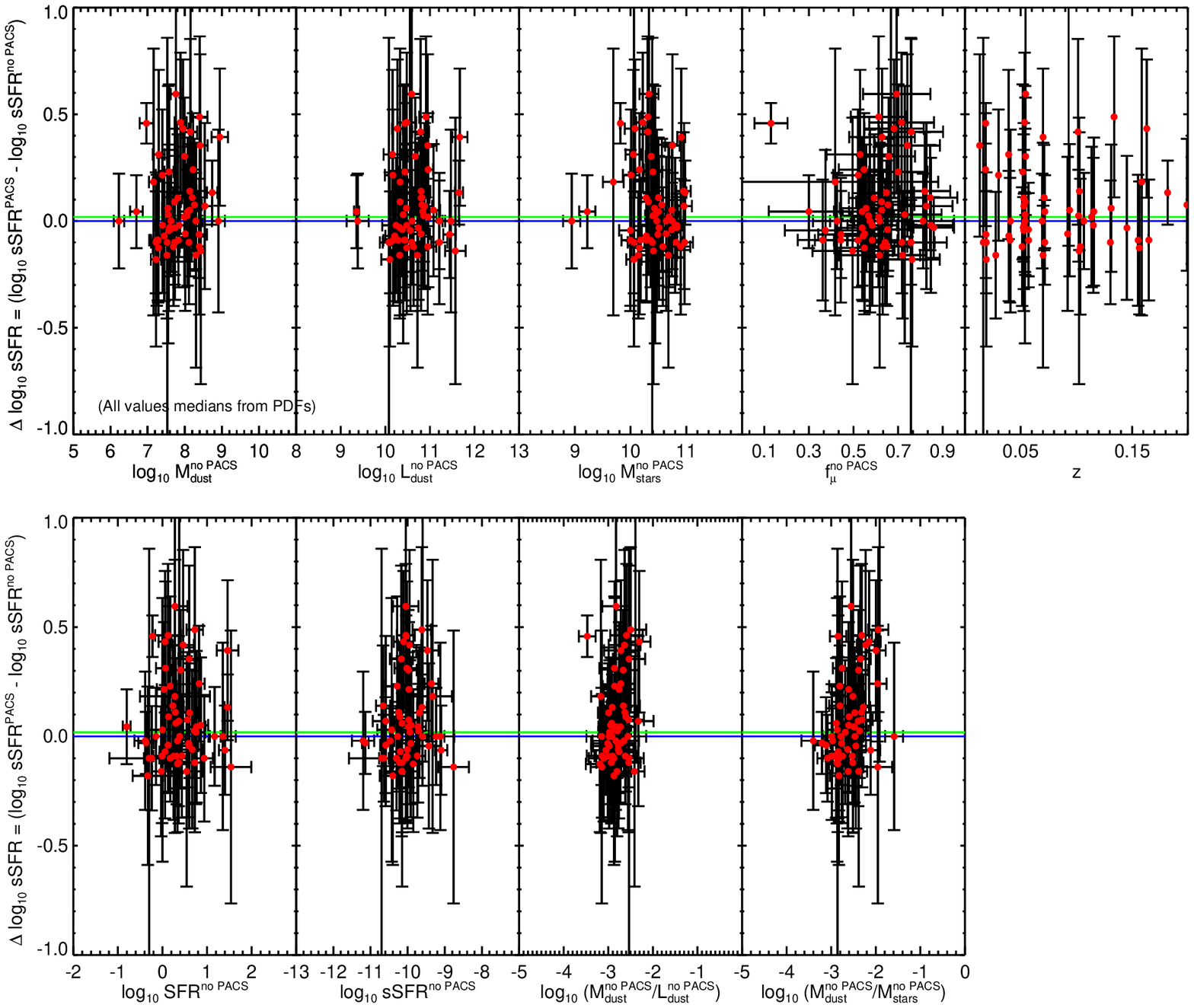}}
  \caption{Variation in (a) \ldust, (b) \mdust, (c) $f_\mu$, (d)
    \mstars, (e) SFR and (f) sSFR as a function of each of the
    parameters discussed in this paper for the PACS-complete sample
    described in Section \ref{pacscomp}. The data points are shown as
    red circles, with horizontal error bars indicating the
    uncertainties derived neglecting the PACS data, and the vertical
    position derived according to equation \ref{deltapacs_equation},
    considering the asymmetric errors on the median likelihood values
    for each parameter. The blue horizontal line indicates $\Delta Q =
    0$ whilst the green horizontal line denotes the difference between
    the median values of the stacked PDFs for the whole PACS complete
    sample when the PACS data are included and when they are omitted.}
  \label{further_bias_tests}
\end{figure*}

In turn, we calculate the variation on our key parameters (\ldust,
\mdust, $f_\mu$, \mstars, SFR and sSFR) introduced in our
``PACS-complete'' sample when the PACS data are included and when they
are omitted from the fitting, as a function of each of the derived
parameters included in this analysis. We consider the difference in
each parameter, Q (e.g. \ldust, sSFR, etc):

\begin{equation}
  \Delta Q = Q^{\mathrm{PACS}} - Q^{\mathrm{no PACS}},
  \label{deltapacs_equation}
\end{equation}

\noindent where $Q^{\mathrm{PACS}}$ and $Q^{\mathrm{no PACS}}$ are the
median likelihood estimates for a particular source including or
ignoring the PACS information in the fitting. In this way we are able
to probe for a skew in the bias between estimates of a parameter
e.g. do we over-estimate \ldust\ for particular \ldust\ when galaxies
aren't detected by PACS? Even though we have used our stacked PDFs to
probe for overall bias and found that it is minimal for all of the key
parameters that we discuss in this paper (they are detailed in table
\ref{tab:parameters}), it is still possible that $\Delta Q$ is skewed
for some parameter combinations. In figure \ref{further_bias_tests},
we plot these values along with their error bars. The individual data
points are shown as red circles, with error bars in the horizontal
direction derived in the absence of the PACS data. The blue solid line
indicates zero offset, while the green solid line indicates the median
of the stacked PDF for all galaxies in the ``PACS-complete'' sample
that we explore here.

In figure \ref{further_bias_tests}\,(a), we consider possible bias in
\ldust. We find that there is a skewed bias between $\Delta \ldust$
and \mdustnopacs; though the scatter is large, these tests suggest
that \ldust\ is underestimated at high \mdust\ and underestimated at
lower values in the absence of PACS data, though the offset averaged
across the whole range of \mdustnopacs\ is small. $\Delta \ldust$
appears broadly unbiased across all values of \ldustnopacs, while
there is little evidence for any bias with respect to \mstars, \fmu,
(specific) SFR or redshift. There is evidence for bias in \ldust\ with
respect to \mdld\ and \mdms, though these biases are dominated by the
bias in \mdust\ already discussed.

Figure \ref{further_bias_tests}\,(b) shows the variation in $\Delta
\mdust$ as a function of the same key parameters, suggesting that
there may be a tendency to overestimate \mdust\ in the absence of PACS
data at the highest dust masses probed by this study and there is a
small tendency for dust masses to be overestimated in the absence of
PACS data on average (comparing the blue and green horizontal lines in
figure \ref{further_bias_tests}). The related parameters
\mdustnopacs\slash\ldustnopacs and \mdustnopacs\slash\mstarsnopacs)
also show evidence for skew.

Figures \ref{further_bias_tests}\,(c) and (d) show the variation in
$\Delta \fmu$ and $\Delta \mstars$; it is clear that they are unbiased
with respect to the other model outputs that we use in this paper, as
probed by the ``PACS-complete'' sample. Though the spread in $\Delta
SFR$ and $\Delta sSFR$ is larger, reflected in the larger error bars
in figure \ref{further_bias_tests}\,(e) \&\ (f), it is difficult to
discern any skew or bias between these parameters and the other model
outputs.

\section{Stacking samples of SEDs}
\label{stack_calc}

\begin{figure}
  \centering
 \includegraphics[width=0.99\columnwidth]{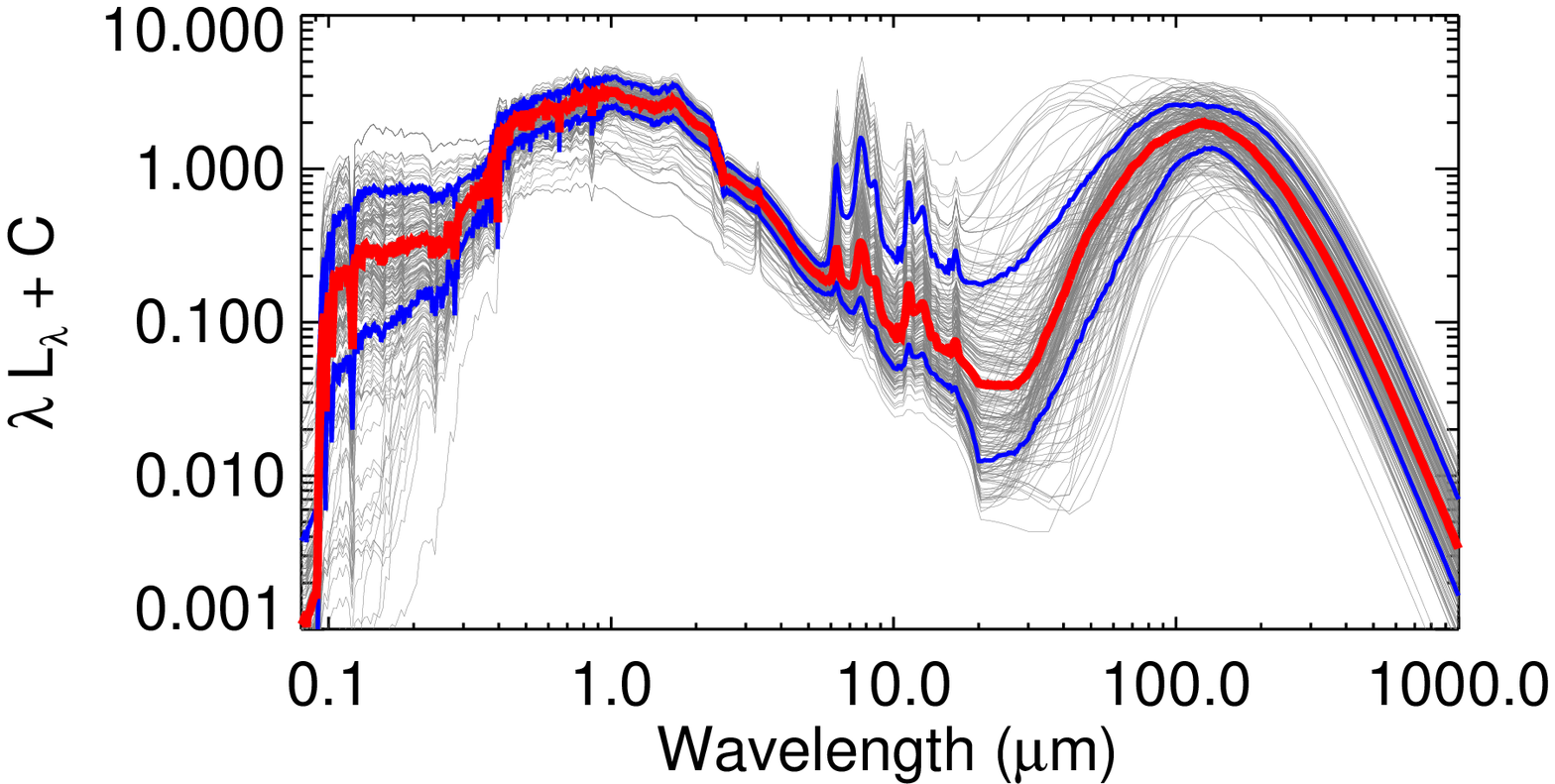}
  \caption{Individual best-fit SEDs (grey points) are first normalised
    between 0.2 and 500\,$\mu$m, and then the median (the red line)
    and 16$^{\mathrm{th}}$ and 84$^{\mathrm{th}}$ percentiles of the
    ensemble (blue lines) are calculated as a function of
    wavelength. This particular example shows the stacked SEDs of
    those 216 galaxies with good best-fit models at $z < 0.35$ and
    with $10.0 < \log_{10} (L_{\mathrm{dust}} \slash
    L_{\mathrm{solar}}) < 10.5$.}
  \label{stack_method}
\end{figure}

In order to calculate the median of an ensemble of SEDs, we first
normalise each individual SED to the mean between 0.2 and 500\,$\mu$m
(in units of $\lambda F_\lambda$), such that each template is given
equal weighting in the stack, then take the median of the ensemble of
normalised SED values in each wavelength bin. We also determine the
16$^{\mathrm{th}}$ and 84$^{\mathrm{th}}$ percentiles of the
cumulative SED-distribution as a function of wavelength. These values
provide a measure of the spread in the SEDs of the galaxies which go
into the stack, in contrast to the estimated error on the median
template shown in figure \ref{sSFR_variation_v2}, which is determined
using the median statistics method of Gott et al. (2001). Our method
of determining the range of values in our SED stacks (i.e.  the
16-84$^{\mathrm{th}}$ percentiles) is illustrated in Figure
\ref{stack_method}, in which the individual best-fit SEDs that go into
the median template (in this case for 216 galaxies with good fits, at
$z < 0.35$, and with $10.0 < \log_{10} (L_{\mathrm{dust}} \slash
L_{\mathrm{solar}}) < 10.5$) are shown in grey, while the median and
the afore-mentioned percentiles are overlaid in red and blue lines,
respectively.

Note that for the purposes of calculating these stacked templates, we
bin according to the best-fit values returned from the fitting code
rather than the medians of the PDF that we use for analysis
elsewhere. This distinction is noteworthy since we determine only the
best-fit model SEDs for each galaxy (as opposed to the full PDF at
each wavelength). Whilst there is generally excellent agreement
between the best-fit and the median-likelihood estimates of any given
parameter, the two may differ in individual cases, adding unrealistic
outliers in the stacked SEDs if median-likelihood values are used for
these purposes.

Whilst this method of stacking SEDs determines templates designed to
be representative of a typical galaxy in a given sample (or
sub-sample) of galaxies, it is important to note that we do not expect
such templates to reproduce the total cosmic spectral energy
distribution (e.g. Hill et al., 2010, Somerville et al., 2012, Driver
et al., {\it in prep}), which would require calculating the sum of the
emergent SEDs without prior normalisation. We leave further discussion
of this topic for a future publication.

\label{lastpage}
\bsp

\end{document}